%% file: main.tex
\documentclass{aa}
\usepackage[varg]{txfonts}
\usepackage{xcolor}
\usepackage[autostyle, english=british]{csquotes}
\usepackage{amsmath,amsfonts,amssymb,mathrsfs,mathtools}
\usepackage{bbold}  
\newcommand{\iu}{{i\mkern1mu}}
\usepackage{pgf}
\usepackage[utf8]{inputenc}\DeclareUnicodeCharacter{2212}{-}
\usepackage{cleveref}
\DeclareMathOperator{\tr}{tr}
\DeclareMathOperator{\lognormal}{LogNormal}
\DeclareMathOperator{\normal}{Normal}

\DeclareMathOperator{\flex}{flex}
\DeclareMathOperator{\asp}{asp}
\DeclareMathOperator{\fluc}{fluc}
\DeclareMathOperator{\avgsl}{avgsl}

\newcommand{\tflex}{\enquote{$\flex$}}
\newcommand{\tasp}{\enquote{$\asp$}}
\newcommand{\tfluc}{\enquote{$\fluc$}}
\newcommand{\tavgsl}{\enquote{$\avgsl$}}

\newcommand{\xiI}{\xi^{(I)}}
\newcommand{\xisig}{\xi^{(\sigma)}}
\newcommand{\resolve}{\texttt{resolve}}
\newcommand{\ducc}{ducc}
\newcommand{\clean}{\texttt{CLEAN}}
\newcommand{\wsclean}{\texttt{wsclean}}
\newcommand{\ssclean}{single-scale \clean}
\newcommand{\msclean}{multi-scale \clean}
\newcommand{\fluxunit}{\mathrm{Jy}/\mathrm{arcsec}^2}
\newcommand{\intd}{\,\mathrm{d}}

\begin{document}
\title{Comparison of classical and Bayesian imaging in radio interferometry}
\subtitle{Cygnus~A with \clean\ and \resolve}
\author{Philipp Arras\inst{\ref{mpa},\ref{lmu},\ref{tum}}\and Hertzog L. Bester\inst{\ref{sarao},\ref{rhodes}} \and Richard A. Perley\inst{\ref{nrao}}\and Reimar Leike\inst{\ref{mpa},\ref{lmu}} \and Oleg Smirnov\inst{\ref{rhodes},\ref{sarao}} \and Rüdiger Westermann\inst{\ref{tum}}\and Torsten A. Enßlin\inst{\ref{mpa},\ref{lmu}}} %
\institute{Max-Planck Institut für Astrophysik, Karl-Schwarzschild-Str. 1, 85748 Garching, Germany\label{mpa}\and
  Ludwig-Maximilians-Universität München (LMU), Geschwister-Scholl-Platz 1, 80539 München, Germany\label{lmu}\and
  Technische Universität München (TUM), Boltzmannstr. 3, 85748 Garching, Germany\label{tum}\and
  South African Radio Astronomy Observatory, 2 Fir Street, Black River Park, Observatory (North Gate entrance), 7925, South Africa\label{sarao}\and
  Department of Physics and Electronics, Rhodes University, Grahamstown, 6140, South Africa\label{rhodes}\and
  National Radio Astronomy Observatory, P.O. Box O, Socorro, NM 87801, USA\label{nrao}}
\date{Received <date>/ Accepted <date>}
\abstract{
  \clean, the commonly employed imaging algorithm in radio interferometry, suffers from a number of shortcomings:
  In its basic version, it does not have the concept of diffuse flux, and the common practice of convolving the \clean\ components with the \clean\ beam erases the potential for super-resolution;
  it does not output uncertainty information;
  it produces images with unphysical negative flux regions;
  and its results are highly dependent on the so-called weighting scheme as well as on any human choice of \clean\ masks for guiding the imaging.
  Here, we present the Bayesian imaging algorithm \resolve , which solves the above problems and naturally leads to super-resolution.
  We take a VLA observation of Cygnus~A at four different frequencies and image it with \ssclean, \msclean, and \resolve.
  Alongside the sky brightness distribution, \resolve\ estimates a baseline-dependent correction function for the noise budget, the Bayesian equivalent of a weighting scheme.
  We report noise correction factors between $\protect\input{stdcorr_min}$ and $\protect\input{stdcorr_max}$.
  The enhancements achieved by \resolve\ come at the cost of higher computational effort.
} 
\maketitle


\section{Introduction}
Radio interferometers provide insights into a variety of astrophysical processes that deepen our knowledge of astrophysics and cosmology in general.
A common strategy to improve radio observations is to upgrade the hardware: increase the number of antennas or their sensitivity.
This paper takes the orthogonal approach and improves one part of radio pipelines, the imaging and deconvolution step.
Interferometers do not directly measure the sky brightness distribution but rather measure its modified Fourier components.
Therefore, the step from the data to the image is non-trivial.

One of the first deconvolution algorithms, \ssclean\ \citep{hogbom1974aperture}, is still in use today.
It was developed for the computational resources of the 1970s and assumes that the sky brightness distribution consists of point sources.
The basic idea behind \ssclean\ is to transform the Fourier data into image space, find the brightest point sources in descending order, simulate a measurement of those point sources, subtract them from the data, and iterate.
Finally, the collection of point sources, called \clean\ components, is usually convolved with the so-called \clean\ beam, which is supposed to represent the intrinsic resolution of the radio interferometer.
In practice, this algorithm converges to some approximation of the actual sky brightness distribution.

The assumption that the sky consists of point sources is problematic because typical radio interferometers are also capable of capturing faint diffuse emission.
Therefore, \citet{cornwell2008}, \citet{rau2011}, and \citet{offringamsclean} extended \clean\ to using Gaussian-shaped structures as basis functions.
The resulting algorithm is called \msclean\ and is the de-facto standard for deconvolving extended structures.

There are several major reasons to rethink the \clean\ approach to imaging and deconvolution now that more computational resources are available and significant progress in Bayesian inference has been made relative to the 1970s.
First, in order to allow \clean\ to undo initial and too greedy flux assignments, \clean\ components are usually not required to be positive.
Therefore, the final sky brightness distribution is not necessarily positive and almost all maps produced from radio interferometric data contain unphysical negative-flux regions.
Second, the convolution with the \clean\ beam fundamentally limits the resolution of the image despite the fact that super-resolution is possible \citep{superresolution,convex}.
In particular, the location of bright compact sources can be determined with much higher accuracy than suggested by the \clean\ beam.
Third, the weighting scheme, which is a function that rescales the influence of each data point on the final image depending on the baseline length or proximity of other measurements, crucially influences the output image.
A prescription for setting the weighting scheme, such that the resulting image resembles the actual sky brightness distribution in the best possible way, does not exist.
Finally, \clean\ does not output reliable uncertainty information.

We intend to address the above issues by updating the Bayesian imaging algorithm \resolve\ developed in \citet{resolve18,resolve19} and originally pioneered by \citet{resolve15} and \citet{fastresolve}.
Bayesian inference is the framework of choice for this as it is the only consistent extension of Boolean logic to uncertainties via real-valued probabilities \citep{cox}.
\resolve\ is formulated in the language of information field theory \citep{ensslin09} in symbiosis with the inference algorithm Metric Gaussian Variational Inference \citep[MGVI,][]{mgvi}.
It combines the imaging and deconvolution steps of the \clean\ approach.
Indeed, \resolve\ significantly improves the resolution of the image as super-resolution is built in.

Bayesian imaging in radio astronomy is not new.
Maximum entropy imaging was the most prominent of the first such algorithms based on the minimalistic prior assumption that photons could arrive from all directions and no intrinsic emission structures can be assumed a priori \citep{cornwellmaxent, memgullskilling}.
While this has been proven to be particularly successful for imaging diffuse emission, \citet[Section 3.2.2]{resolve16} demonstrate that \resolve\ can outperform maximum entropy imaging.
The reasons for this include that the latter does not assume any correlations between pixels a priori or a brightness distribution for each pixel with an exponential cut-off for high values.

Related approaches include \citet{optimalimaging} and \citet{sutter2014}, who use Bayesian inference as well.
These approaches are, however, limited to Gaussian priors and relatively few pixels because Gibbs sampling is used.

Another approach to deconvolution has leveraged convex optimization theory, and in particular, the relatively new field of compressive sensing \citep{csoriginal}.
Originally formulated as the SARA (sparsity averaging) reconstruction algorithm \citep{sara}, this has produced approaches such as PURIFY \citep{purify} and HyperSARA \citep{hypersara}.
These methods have demonstrated good performance on extended emission, and in particular, on the data we use for this study \citep{convex}.
This class of algorithms can be thought of as yielding maximum-a-posterior point estimates of the sky under a sparsity prior, however recent work by \citet{uncertaintycs} shows a way to incorporate uncertainty estimates into the approach.
These uncertainty estimates are not an uncertainty map for the whole sky brightness distribution but rather a hypothesis test to assess the discovery significance of single sources.
This approach is based on the assumption that the functionals that need to be optimized are (log-)convex and has been demonstrated to work on large data sets.
One of our main insights is that noise inference is needed (at least for the data sets that we have analysed) because otherwise the noise statistics of the data are not correct.
Uncertainties that are derived from incorrect error bars on the data cannot be reliable.
In our understanding noise inference would render the optimization problem non-convex.
\citet{cai2018uncertaintyI} propose a hybrid approach, where compressive sensing is combined with Markov Chain Monte Carlo sampling.

This paper is structured as follows: \Cref{sec:measurementmodel} describes the underlying data model common to the compared imaging algorithms.
\Cref{sec:resolve} defines the novel \resolve\ algorithm and specifies the prior assumptions and \cref{sec:traditional} recapitulates the \ssclean\ and \msclean\ algorithms.
All three algorithms are compared in \cref{sec:comparison} by applying them to the same four data sets.

\section{Measurement model}
\label{sec:measurementmodel}
Astrophysical signals undergo a variety of transformations as they travel from their source to where they are observed on Earth.
We restrict ourselves to an ideal, unpolarized phase tracking interferometer, in which case the measurement process obeys \citep[for example][]{book}:
\begin{align}
  \begin{multlined}[t]
    \label{eq:nfft1}
    d_{uvw} = n_{uvw} +\\
    \iint_{\{(l, m)\in\mathbb R^2 \mid l^2+m^2< 1\}} \frac{\mathfrak a(l, m) \, I(l,m)}{\sqrt{1-l^2-m^2}}\, e^{2\pi \iu [ul+vm+w(1-\sqrt{1-l^2-m^2})]} \intd (l, m),
  \end{multlined}
\end{align}
where $d_{uvw}$ represents the data taken by the interferometer (commonly referred to as visibilities), $n_{uvw}$ represents an additive noise realization, $\mathfrak a (l, m)$ is the antenna sensitivity pattern and $I(l, m)$ the true sky brightness distribution.
The data space coordinates $(u, v, w)$ record the relative positions of antenna pairs as the Earth rotates under the frame of the sky.
The coordinates $(l, m, \sqrt{1-l^2-m^2})$ denote the positions of points on the celestial sphere.
The integral goes over the half of the sphere that is above the horizon.
If the array is arranged such that $w \rightarrow 0$ or if the field of view is very small ($l^2+m^2 \rightarrow 1$), \cref{eq:nfft1} reduces to a two-dimensional Fourier transform of the apparent sky $\mathfrak a (l,m)\,I(l,m)$.
This assumption, referred to as the coplanar array approximation, is discussed further in \cref{sec:traditional}.

In practice the integral in \cref{eq:nfft1} is discretized to allow numerical evaluation.
Then, the measurement model simplifies to:
\begin{align}
 d &= R I + n,\label{eq:discretemeasmodel}
\end{align}
where $R \in \mathrm{Lin}_{\mathbb R}(\mathbb R^N, \mathbb C^M)$ is a discretization of \cref{eq:nfft1}, which maps a discretized image $I \in \mathbb{R}^N$ to visibilities in $\mathbb{C}^M$, and $n\in\mathbb C^M$ is the noise present in the observation.
Both \resolve\ and \wsclean\ use the software library \ducc\footnote{\url{https://gitlab.mpcdf.mpg.de/mtr/ducc}} (Distinctly Useful Code Collection) for evaluating the integral.

Since visibilities consist of an average of a large number of products of antenna voltages, it can be assumed, by the central limit theorem, that the noise is Gaussian with diagonal covariance $N$: $n \curvearrowleft \mathscr G (n, N)$.
Thus, the likelihood probability density is given by:
\begin{align}
 \mathcal P(d \mid I, N)  = \mathscr G (d-RI, N) := \frac{1}{\sqrt{|2\pi N|}}\, e^{-\frac12 (d-RI)^\dagger N^{-1}(d-RI)},
\end{align}
where ${}^\dagger$ denotes the complex conjugate transpose.
For better readability, but also because it is the quantity that needs to be implemented for \resolve , we define the information Hamiltonian $\mathcal H (d \mid I, N) := -\log P(d \mid I, N)$ \citep{ensslin09}.
Then,
\begin{align}
  \mathcal H (d \mid I, N) = \frac12 (d - RI)^\dagger N^{-1} (d - RI)+h(N),\label{eq:simplefid}
\end{align}
where $h(N)$ is a normalization term constant in $I$.
Many traditional imaging algorithms employ this expression without $h(N)$ as the data fidelity term that ought to be minimized.

We conclude this section with two comments.
First, note that \cref{eq:simplefid} stores all information about the measurement device and the data at hand.
No specific assumptions about the data processing have been made yet.
Therefore, \cref{eq:simplefid} is the starting point of both \resolve\ and \clean.
We call the process of turning \cref{eq:simplefid} into an image \enquote{imaging} and do not differentiate between imaging and \enquote{deconvolution}.
Second, the process of recovering the true sky brightness distribution from the measured visibilities is an inverse problem.
In \cref{eq:discretemeasmodel}, the sky $I$ cannot be computed uniquely from $d$ and $N$ alone because the Fourier space coverage (commonly called uv-coverage) is not complete and because of the presence of noise.
We may know the noise level $N$ but we never know the noise realization $n$.
This is why turning data into the quantity of interest, in our case $I$, is a non-trivial task.
The appearance of uncertainties is a direct consequence of the non-invertibility of $R$ and the presence of $n$.

\section{Resolve}
\label{sec:resolve}

\resolve\ is a Bayesian imaging algorithm for radio interferometers.
It is formulated in the language of information field theory \citep{ensslin09, ensslin11, ensslin18} and was first presented in \citet{resolve15} and then upgraded in \citet{resolve16}, \citet{fastresolve}, and \citet{resolve18}.
\citet{resolve19} added antenna-based direction-independent calibration to \resolve\ such that calibration and imaging can be performed simultaneously.
This paper presents another \resolve\ feature for the first time: automatic data weighting.
Additionally, the diffuse sky model is updated to a special case of the model presented in \citet{ifteht}.
The implementation is free software\footnote{\url{https://gitlab.mpcdf.mpg.de/ift/resolve}}.

\subsection{Inference scheme}
\resolve\ views radio interferometric imaging as a Bayesian inference problem:
it combines a likelihood and a prior probability density to a posterior probability density.
We generalize the likelihood to depend on general model parameters $\xi$ (previously $I$ and $N$).
The likelihood contains all information about the measurement process and the noise.
In contrast, the prior $\mathcal P(\xi)$ is a probability density that assigns to every possible value of the model parameters $\xi$ a probability that represents the knowledge on the model parameters before having looked at the data.
These two quantities are combined with a normalization factor $\mathcal P (d)$ to Bayes' theorem:
\begin{align}
  \label{eq:bayes}
  \mathcal P(\xi\mid d) = \frac{\mathcal P(d\mid \xi) \,\mathcal P(\xi)}{\mathcal P(d)}.
\end{align}
$\mathcal P(\xi\mid d)$ gives the probability for all configurations of the model parameters after having looked at the data.

\resolve\ uses Bayes' theorem together with the re-parameterization trick \citep{reparatrick}:
It is always possible to transform the inference problem such that the prior density is a standard normal distribution: $\mathcal P (\xi) = \mathscr{G}(\xi,\mathbb 1)$.
In this approach, all prior knowledge is formally encoded in the likelihood.
Put differently, the task of defining the inference problem is to write down a function that takes standard normal samples as input, transforms them into sensible samples of the quantity of interest with their assumed prior statistics and finally computes the actual likelihood.

For our imaging purposes $\xi$ is a roughly 10~million-dimensional vector.
Exactly representing non-trivial high-dimensional probability densities on computers is virtually impossible.
Therefore, approximation schemes need to be employed.
For the application at hand, we choose the Metric Gaussian Variational Inference \citep[MGVI,][]{mgvi} implementation in NIFTy \citep{nifty5} because it strikes a balance between computational affordability and expressiveness in the sense that it is able to capture off-diagonal elements of the posterior uncertainty covariance matrix.

\subsection{On weighting schemes}
\label{sec:weighting}
\clean\ assumes a certain weighting scheme that induces changes in the noise level.
A weighting scheme is necessary for two reasons:
It can be used to reweight by the density of the uv-coverage to make it effectively uniform, which \clean\ needs to perform best (see \cref{sec:traditional}).
\resolve\ does not need this kind of correction because it is based on forward modelling and Bayesian statistics:
A more densely sampled region in uv-space leads to more information in this region and not to inconsistencies in the inference.

Additionally, there exist weighting schemes that further reweight the visibilities based on the baseline length.
This weighting represents the tradeoff between sensitivity (up-weight short baselines) and resolution (uniform weighting).
Depending on the application \clean\ users need to choose between those extremes themselves.

Moreover, we find that short baselines are subject to higher systematic noise.
For the data sets at hand, this systematic noise is up to a factor of $\input{stdcorr_max}$ higher than the thermal noise level  (see \cref{fig:stdcorr}).
If the noise variance of the visibilities were correct, that value would be 1.
To \clean\ higher systematic noise is indistinguishable from non-uniform sampling; to a Bayesian algorithm, which takes the uncertainty information of the input data seriously, it makes a crucial difference.
Therefore, the advanced version of \resolve\ presented here assumes that the thermal measurement uncertainties need to be rescaled by a factor that depends only on the baseline length and which is correlated with respect to that coordinate.
This correction function (or Bayesian weighting scheme) is learned from the data alongside the actual image.
The details on this approach are described in the next section.

\subsection{Assumptions and data model}
\label{sec:model}
To specify \resolve, the standardized likelihood $\mathcal P (d \mid  \xi)$ in \cref{eq:bayes} needs to be defined.
In addition to the thermal noise level $\sigma_{\mathrm{th}}$, which is generated by the antenna receivers, calibrated visibilities may be subject to systematic effects.
In order to account for these the thermal variance is multiplied by a correction factor $\alpha$, which is unknown and assumed to depend on the baseline length:
\begin{align}
  \label{eq:noisemodel}
\sigma(\xisig) &= \sigma_{\mathrm{th}} \cdot \alpha(\xisig),
\end{align}
where $\xisig$ refers to the part of $\xi$ that parameterizes $\sigma$.
Consequently the noise standard deviation $\sigma$ itself becomes a variable part of the inference.
The sky brightness distribution $I$ is variable as well (meaning that it depends on $\xi$) and the simulated data $s$ are given by:
\begin{align}
s(\xiI)&= \int \frac{\mathfrak a \cdot I(\xiI)}{\sqrt{1-l^2-m^2}}\, e^{2\pi\iu[ul+vm+w(1-\sqrt{1-l^2-m^2})]}\intd (l, m) ,\label{eq:nfft}
\end{align}
where $\xiI$ refers to the part of $\xi$ that parameterizes $I$ and $I(\xiI)$ is the discretized sky brightness distribution in units Jy/$\mathrm{arcsec}^2$.

The remaining task is to specify $I(\xiI)$ and $\alpha(\xisig)$.
For the sky brightness distribution we assume two additive components: A point source component modelled with a pixel-wise inverse gamma prior \citep{d3po} and a component for diffuse emission.
A priori we assume the diffuse emission to be log-normal distributed with unknown homogeneous and isotropic correlation structure.
This is motivated by the expectation that emission varies over several magnitudes.
Furthermore, we assume that the noise correction function $\alpha$ is log-normal distributed since it needs to be strictly positive and also may vary strongly.

Let $F^{(n)}(\xi)$ be a function that maps standard normal distributed parameters $\xi$ on a $n$-dimensional Gaussian random field with periodic boundary conditions and homogeneous and isotropic correlation structure \citep{ensslin18}.
The specific form of $F^{(n)}(\xi)$ is explained in \cref{sec:cf}.
Then:
\begin{align}
 I (\xiI) &=\exp F^{(2)}(\xiI) + (\mathrm{CDF}^{-1}_{\mathrm{InvGamma}} \circ \mathrm{CDF}_{\mathrm{Normal}})(\xiI), \label{eq:skymodel}\\
 \alpha(\xisig) &= (C\circ\exp)\left[ F^{(1)}(\xisig)\right] ,
\end{align}
where $\circ$ denotes function composition, $\mathrm{CDF}_{\mathrm{Normal}}$ and $\mathrm{CDF}^{-1}_{\mathrm{InvGamma}}$ refer to the cumulative density function of the standard normal distribution and the inverse cumulative density function of the Inverse Gamma distribution, respectively, and $C$ is a cropping operator that returns only the first half of the (one-dimensional) log-normal field.
This is necessary because $\alpha$ is not a periodic quantity and we use fast Fourier Transforms, which assume periodicity.
While the diffuse component of the sky brightness distribution is not periodic either, it is not necessary to apply zero-padding there since the flux is expected to vanish at the image boundaries.
The point sources are restricted to the locations a priori known to contain point sources.

All in all, the likelihood density is given by:
\begin{align}
  \mathcal P (d \mid \sigma(\xisig), s(\xiI)) &= |2\pi \widehat{\sigma^2}|^{-1} e^{-\frac12 (s - d)^\dagger \widehat{\sigma^{-2}}(s-d)},\label{eq:likelihood}\\
    \mathcal H (d \mid \sigma(\xisig), s(\xiI)) &= \frac12 (s-d)^\dagger \widehat{\sigma^{-2}} (s-d) + 2 \sum_i\log \sigma_i  + c, \label{eq:likelihood-hamiltonian}
\end{align}
where $\widehat{x}$ denotes a diagonal matrix with $x$ on its diagonal and $c$ is a normalization constant.
The sum goes over all data points and the dependency of $\sigma$ and $s$ on $\xi$ is left implicit.
The normalization factor in \cref{eq:likelihood} is chosen such that \cref{eq:likelihood} is normalized if $d$ is viewed as combination of two sets of real random variables:
\begin{align}
  d =\Re(d) + i\Im(d), \quad \int \mathcal P(d\mid \xi) \intd \Re(d) \intd \Im(d)  = 1.
\end{align}
The following two subsections (\cref{sec:cf,sec:sampling}) describe the technical details of the \resolve\ sky model and the sampling procedure.
\Cref{sec:traditional} describes the technical details of \ssclean\ and \msclean .
Non-technical readers may safely skip directly to \cref{sec:traditional} or even \cref{sec:comparison}.

\subsection{Correlated field model with unknown correlation structure}
\label{sec:cf}
The following section closely follows \citet[Methods section]{ifteht}, which derives the correlated field model in a more general context.
For reasons of clarity and comprehensibility, we repeat the derivation here for the specific case at hand and adopted to the notation used here.
The main reason for the complexity of the model below is that for modelling diffuse emission neither a specific correlation kernel nor a parametric form for the kernel shall be assumed.
Rather, our goal is to make the correlation kernel part of the inference as well.
This reduces the risk of biasing the end result by choosing a specific kernel as prior.

In order to simplify the notation we drop the indices $(I)$ and $(\sigma)$ for this section and write: $F^{(n)} = F^{(n)}(\xi)$.
Still the model $F^{(n)}$ is used for both the correction function $\alpha$ and the diffuse component of the sky brightness distribution while the domains are one-dimensional and two-dimensional, respectively.
In the following, standard normal variables will appear in various places.
Therefore, we write $\xi = (\xi_0, \xi_1, \ldots)$ and $\xi_{>n} = (\xi_{n+1}, \xi_{n+2}, \ldots)$ where each $\xi_i$ is a collection of standard normal variables.

The task is to write down a function that takes a standard normal random variable $\xi$ as input and returns a realization of a correlated field with unknown homogeneous and isotropic correlation structure.
This means that the two-point correlation function depends on the distance between the sampling points only:
\begin{align}
S = \left\langle F^{(n)}(\xi)(x) \, F^{(n)}(\xi)(y)\right\rangle_{\mathscr G (\xi, \mathbb 1)} = f(|x-y|),
\end{align}
where $\big\langle x \big\rangle_P$ denote the expectation value of $x$ over he distribution $P$.
For homogeneous and isotropic processes the Wiener-Khintchin theorem \citep{Wiener, Khintchin} states that the two-point correlation function of the process is diagonal in Fourier space.
Let the $n$-dimensional discrete Fourier transform be the map $\mathcal F^{(n)}: X_h\to X$ where $X$ is a regular grid space with shape $(N_1, \ldots, N_n)$ and pixel sizes $(\Delta x_1, \ldots, \Delta x_n)$ and $X_h$ its harmonic counterpart: It has the same shape and pixel sizes  $((N_1 \Delta x_1)^{-1}, \ldots, (N_n \Delta x_n)^{-1})$.
Let us define:
\begin{align}
 F^{(n)}(\xi) = \mathrm{offset} + \mathcal F^{(n)} \left( \mathrm{vol} \cdot A(\xi_{>0}) \cdot \xi_0 \right),
\end{align}
where $\mathrm{offset}$ is the (known) mean of the Gaussian random field, $\hat A \hat A^\dagger = S$ in Fourier basis, $\mathrm{vol}=\prod_i N_i\Delta x_i$ is the total volume of the space and $\xi$ is a standard normal random field.
The volume factors in the Fourier transform are defined such that the zero mode in Fourier space is the integral over position space:
\begin{align}
x_{0\cdots0} = \sum_{i_1=0}^{N_1}\cdots\sum_{i_n=0}^{N_n}\left( \Delta x_1\cdots\Delta x_n\cdot\mathcal F^{(n)}(x) \right)
\end{align}
for all $n$-dim fields $x$.
Then the set $\big\{F^{(n)}(\xi) \mid \xi \curvearrowleft \mathscr G(\xi, \mathbb 1)\big\}$ is a collection of correlated fields with unknown correlation structure, meaning that $A$ still depends on $\xi$.
$\xi_0$ is defined on that space as well and \enquote{$\cdot$} denotes pixel-wise multiplication.

If we could derive a sensible form of the correlation structure $A$ for both the diffuse emission and the correction function a priori, we could insert it here and infer only $\xi$.
However, we are not aware of a method to set the correlation structure by hand without introducing any biases for a given data set.
Therefore, we let the data inform the correlation structure $A$ as well and set a prior on $A$.
This approach may be viewed as a hyper parameter search integrated into the inference itself.
In the following we will see that even the parameters needed to model $A$ are inferred from the data.
So it is really a nested hyper parameter search.

The presented model has five hyper parameters.
In order to emulate a hyper parameter search, we do not set them directly but rather make them part of the inference and let the algorithm tune them itself.
The hyper parameters which are necessarily positive are modelled with a log-normal prior as generated from standard normal variables $\xi_i$ via:
\begin{align}
  \lognormal (\xi_i; \mathfrak m, \mathfrak s) &:= \exp \left( \mathfrak m + \tilde{\mathfrak s} \, \xi_i -\tfrac12 \tilde{\mathfrak s}^2\right),\\
  \tilde{\mathfrak s} &:= \sqrt{\log \left(1+\left(\tfrac{\mathfrak s}{\mathfrak m}\right)^2\right)}\,,
\end{align}
where $\mathfrak m$ and $\mathfrak s$ refer to mean and standard deviation of the log-normal distribution; the ones that can be positive or negative have a Gaussian prior and are denoted by $\normal (\xi_i; \mathfrak m, \mathfrak s) := \mathfrak m+ \mathfrak s\, \xi_i$.
The values for $\mathfrak m$ and $\mathfrak s$ as well as for the other hyper parameters are summarized in \cref{tab:hpres}.

The zero mode controls the overall diffuse flux scale.
Its standard deviation $A_0$ is a positive quantity and we choose it to be log-normal distributed a priori:
\begin{align}
  \label{eq:lognormalexample}
A_{\vec 0}(\xi_1) = \lognormal (\xi_1; \mathfrak m_1, \mathfrak s_1).
\end{align}

The non-zero modes $\vec k \neq 0$ control the fluctuations of the random process.
In order to be able to set a prior on the total fluctuations, we define:
\begin{align}
 A_{\vec k}(\xi_{>1})  = \sqrt{\frac{p_{\vec k}(\xi_{>2})}{\sum_{\vec k} p_{\vec k}(\xi_{>2})}}\cdot \fluc (\xi_2), \quad \text{for } \vec k \neq 0,
\end{align}
where $p_{\vec k}$ is the model for the power spectrum of $F^{(n)}$ up to the multiplicative term \tfluc{}.
By this definition we ensure that \tfluc{} is the point-wise standard deviation of the final process: $\langle s_x s_x \rangle = \fluc^2$ for all $x$ after having subtracted the contribution from $A_{\vec 0}$.
\tfluc{} is strictly positive and we model it with a log-normal prior: $\fluc = \lognormal (\xi_2; \mathfrak m_2, \mathfrak s_2)$.

The remaining piece is the actual form of $p_{\vec k}$ for $\vec k \neq 0$.
The prior knowledge we want to encode into this model is:
First, diffuse emission is correlated, meaning that falling power spectra and specifically $p_{|\vec k|}\sim |\vec k|^{-s}, s > 0$ shall be preferred.
And second, periodically repeating patterns in the sky brightness distribution are not expected or equivalently strong peaks in the power spectrum shall be penalized.
In order to define $p_{\vec k}$ in a non-parametric fashion and to represent the above power law property, we choose to transform $p_{\vec k}$ into double-logarithmic space in which power laws become affine linear functions:
\begin{align}
p_{\vec k} = e^{a_t}, \quad\text{with } t = \log |\vec k|, \, \vec k \neq \vec 0.
\end{align}
We choose to model $a_t$ as an integrated Wiener process, that is a general continuous random process:
\begin{align}
\partial_t^2 a_t = \eta_t,
\end{align}
where $\eta_t$ is Gaussian distributed.
In this form the process is not Markovian and is not suited to be evaluated as a forward model.
Therefore, we track the derivatives $b_t$ of $a_t$ as degrees of freedom themselves:
\begin{align}
  \label{eq:wienerprocess}
  \partial_t \begin{pmatrix}a_t \\b_t \end{pmatrix} + \begin{pmatrix}0&-1\\0&0\end{pmatrix} \begin{pmatrix}a_t\\b_t \end{pmatrix}
  = \begin{pmatrix} \sqrt{\asp}\flex \xi_3 \\ \flex \xi_4 \end{pmatrix},
\end{align}
where the specific form of the variances on the right-hand side of the equation will be interpreted below.
Subsequently, we will call \tasp{} asperity and \tflex{} flexibility.
The solution to \cref{eq:wienerprocess} for $b_t$ is a Wiener process.
Therefore, $a_t$ is an integrated Wiener process for $\asp = 0$.
$\asp > 0$ leads to an additional (not integrated) Wiener process on $a_t$.
The solution to \cref{eq:wienerprocess} is:
\begin{align}
  b_{t_n} &= b_{t_{n-1}}  + \flex \sqrt{\Delta t_n}\, \xi_4,\\
  a_{t_n} &= a_{t_{n-1}} + \frac{\Delta t_n}{2} (b_{t_n} + b_{t_{n-1}}) + \flex \sqrt{\frac1{12} \Delta t_n^3 + \asp \Delta t_n}\, \xi_3,
\end{align}
where $t_n$ is the $n$th (discretized) value of $t$ and $\Delta t_n = t_n - t_{n-1}$.
This formulation allows us to compute samples of the process $a_t$ from standard normal inputs $\xi_3$ and $ \xi_4$.
\tflex{} and \tasp{} are both positive quantities and are modelled with lognormal priors: $\flex = \lognormal (\xi_5; \mathfrak m_5, \mathfrak s_5)$ and $\asp = \lognormal (\xi_6; \mathfrak m_6, \mathfrak s_6)$.
As can be seen from \cref{eq:wienerprocess} \tflex{} controls the overall variance of the integrated Wiener process.
The model is set up such that it produces power spectra that can deviate from a power law.
\tasp{} determines the relative strength between the Wiener and the integrated Wiener process.
The limit $\asp \to 0$ is well-defined.
In this case, $a_t$ is a pure integrated Wiener process and $\asp > 0$ adds non-smooth parts to it.
More intuitively, this means that vanishing \tasp{} lead to effectively turn off the non-smooth part of the power spectrum model.
Then, the generated power spectra can be differentiated twice on double-logarithmic scale.
A non-vanishing \tasp{} gives the model the possibility to add small non-smooth structures on top of the smooth power spectrum.
Since \tasp{} is also variable during the inference process, we choose not to set it to zero a priori since the algorithm can do it itself if needed.

Finally, we modify the model such that it is possible to set a prior on the average slope of the integrated Wiener process.
This is necessary to encode a preference for falling spectra.
To this end, the difference between the first and the last pixel of the integrated Wiener process is replaced by a linear component whose slope is \tavgsl{}:
\begin{align}
  \tilde a_{t_i} = a_{t_i} - a_{t_n} \cdot \frac{t_i - t_1}{t_n-t_1} + (t_i-t_1) \cdot \avgsl, \quad \forall i \in\{1, \ldots, n\}.
\end{align}
The slope is modelled with a Gaussian prior: $\avgsl = \normal (\xi_7; \mathfrak m_7, \mathfrak s_7)$.

In summary, this defines a model that is able to generate Gaussian random fields of arbitrary dimension with unknown correlation structure.
The random field is assumed to have homogeneous and isotropic correlation structure.
The power spectrum itself is modelled in double-logarithmic space as a mixture of a Wiener process and an integrated Wiener process with the possibility of specifying the overall slope of the process.
This model is used in its one-dimensional version for the weighting scheme field $\alpha$ and in its two-dimensional version for the diffuse component of the sky brightness distribution $I$.

\subsection{Sampling with variable noise covariance}\label{sec:sampling}
To find approximate posterior samples, \resolve\ employs the MGVI algorithm \citep{mgvi}.
This algorithm performs a natural gradient descent to find the minimum of:
\begin{align}
    E(\overline{\xi}) = \frac{1}{N}\sum_{i=1}^N \mathcal H(d\mid \xi=\overline\xi +\xi_i) + \frac{1}{2}\overline{\xi}^\dagger\overline{\xi} ,
  \label{eq:objfunc}
\end{align}
where $\overline{\xi}$ is the latent posterior mean and $\xi_i$ are samples that represent the uncertainty of the posterior.
They are drawn as zero centred Gaussian random samples with the inverse Bayesian Fisher metric as covariance:
\begin{align}
    \xi_i \curvearrowleft \mathscr G \bigg(\xi \;\Big|\; 0, \Big[\mathbb{1} + \nabla_\xi (\sigma, s)^\dagger\big\vert_{\overline{\xi}}\, F_{\sigma, s}\,  \nabla_\xi (\sigma, s)\big\vert_{\overline{\xi}}\Big]^{-1}\bigg),
\end{align}
where $\nabla_\xi(\sigma, s)\big\vert_{\overline{\xi}}$ is the Jacobian of $s$ and $\sigma$ as a function of $\xi$ evaluated at the latent mean $\overline{\xi}$, and $F$ is the Fisher information metric of the likelihood in terms of the visibility $s$ and the noise standard deviation $\sigma$.
These samples from this inverse metric can be drawn without the need of inverting explicit matrices, by using the conjugate gradient algorithm. We refer to \citet[discussion around eq.~(58)]{mgvi} for a detailed description.

For the computation of the Fisher metric of a complex Gaussian distribution, the real and imaginary parts of the visibility $s$ are treated individually in order to avoid ambiguities related to complex versus real random variables.
Using \cref{eq:likelihood-hamiltonian} we arrive at:
\begin{align}
    F_{\sigma, s} &= \left\langle
      \begin{pmatrix}
        \nabla_\sigma \mathcal H(d\mid \sigma, s)\\
        \nabla_{\Re (s)} H(d\mid \sigma, s)\\
        \nabla_{\Im (s)} H(d\mid \sigma, s)
    \end{pmatrix}
      \begin{pmatrix}
        \nabla_\sigma \mathcal H(d\mid \sigma, s)\\
        \nabla_{\Re (s)} H(d\mid \sigma, s)\\
        \nabla_{\Im (s)} H(d\mid \sigma, s)
    \end{pmatrix}^T
    \right\rangle_{P(d\mid \sigma, \xi)}\nonumber\\
    &=\begin{pmatrix}
        4\sigma^{-2} & 0 & 0\\
        0 & \sigma^{-2} & 0 \\
        0 &  0 &\sigma^{-2}
    \end{pmatrix}. \label{eq:variable-covariance-gaussian-fisher}
\end{align}
To draw random variates with this covariance we use normal random variates and multiply them with the square root of the diagonal of the matrix in \cref{eq:variable-covariance-gaussian-fisher}.
In the NIFTy package implementing these operations, this Fisher metric is given as a function of $\sigma^{-2}$ instead, that can be obtained from \cref{eq:variable-covariance-gaussian-fisher} by applying the Jacobian $\frac{\partial \sigma}{\partial \sigma^{-2}}$:
\begin{align}
    F_{\sigma^{-2}, s} &= \begin{pmatrix} \left(\frac{\partial \sigma}{\partial \sigma^{-2}}\right)^T 4\sigma^{-2} \left(\frac{\partial \sigma}{\partial \sigma^{-2}}\right)& 0 & 0\\ 0 & \sigma^{-2} & 0 \\
        0 &  0 &\sigma^{-2} \end{pmatrix}\nonumber\\
    &= \begin{pmatrix} \sigma^{4} & 0 & 0\\ 0 & \sigma^{-2} & 0 \\
        0 &  0 &\sigma^{-2} \end{pmatrix}.
\end{align}
For computational speed, the real and imaginary parts of the visibilities are combined into complex floating point numbers where possible.

\section{Traditional \clean\ imaging algorithms}\label{sec:traditional}

\subsection{\expandafter\MakeUppercase\ssclean}\label{sec:ssclean}
This section outlines the main ideas behind the \clean\ algorithm.
First, the most basic variant of \clean\ \citep{hogbom1974aperture} is described followed by a discussion of additional approximations that make it more efficient \citep{clarkclean1980} and a more sophisticated version of the algorithm that overcomes coplanar array approximation \citep{cottonschwab}.

At its heart, \clean\ is an optimization algorithm that seeks to minimize \cref{eq:simplefid}.
But since this problem is ill-posed (the operator $R^\dagger N^{-1} R$ occurring in \cref{eq:simplefid} is not invertible), a unique minimum does not exist.
For a patch of sky consisting purely of point sources, one could seek the smallest number of points that would result in the dirty image when convolved with the PSF.

A practical solution, as formalized by \citet{hogbom1974aperture}, involves starting from an empty sky model and then iteratively adding components to it until the residual image appears noise-like.
More precisely, noting that the residual image equates to the dirty image at the outset, we proceed by finding the brightest pixel in the residual image.
Then, using the intuition that the dirty image is the true image convolved by the PSF, we centre the PSF at the current brightest pixel, multiply it by the flux value in the pixel and subtract some fraction of it from the residual image.
At the same time, the model image is updated by adding in the same fraction of the pixel value at the location of the pixel.
This procedure is iterated until a satisfactory solution is found, for example when the residual appears noise-like or its brightest pixel is lower than some predetermined value.
This solution loosely corresponds to the smallest number of point sources necessary to explain the data.
The one tunable parameter in the algorithm is the fraction of the flux of the point source that is added to the model at a time.
This parameter is called loop gain.

This surprisingly simple procedure is so effective that it is still the most commonly used deconvolution algorithm in radio astronomy.
However, it relies on the approximation
\begin{align}
  \label{eq:hogbom}
R^\dagger N^{-1} R \approx I^{PSF} *,
\end{align}
where $*$ denotes convolution and $I^{PSF}$ is an image of the point spread function (PSF), namely the result of applying $R^\dagger N^{-1} R$ to an image that has only a unit pixel at its centre.
In \cref{eq:hogbom}, equality only holds when the coplanar array approximation is valid\footnote{The PSF is direction-dependent when the array is non-coplanar.}.
This leads to two alternate forms of the derivative of the likelihood Hamiltonian:
\begin{align}
  \nabla_I \mathcal H(d \mid  I, N) &= R^\dagger N^{-1}\left(d - RI\right) \approx I^D - I^{PSF} * I,\label{eq:likgrad}
\end{align}
where the latter approximation is exact if the coplanar array approximation is valid and the primary beam structure is negligible or ignored.
For the maximum likelihood solution, set the right hand side of \cref{eq:likgrad} to zero.
This leads to the classic notion that the dirty image is the image convolved by the PSF:
\begin{align}
  I^D &= I^{PSF} * I.\label{eq:dirtyconvolve}
\end{align}
Especially if the number of image pixels is much smaller than the number of data points, this allows computation of the gradients in \cref{eq:likgrad} very efficiently.
The reason for this is that the operator $I^{PSF} * $ can be implemented efficiently using the fast Fourier transform (FFT), whereas $R^\dagger N^{-1} R$ requires a combination of convolutional gridding (including possible $w$-term corrections) and the FFT.

The key to the speed of the \clean\ algorithm comes from the intuition provided by \cref{eq:dirtyconvolve}.
During model building the convolution is not performed explicitly, rather the PSF is centred on the location of the current pixel and subtracted from the residual pixel-wise.
Since point sources can be located right at the edge of the image, the PSF image needs to be twice the size in both dimensions of the residual image.
To save memory and computational time, \citet{clarkclean1980} approximated the PSF by a smaller version and restricted the regions in which PSF side lobes are subtracted.
This is possible since the PSF side lobes typically fall off fairly rapidly, especially for arrays with good uv-overage.
However, it is paid for by artefacts being added to the model if the approximation is not done carefully.
For this reason the Clark approximation is often used in combination with a \clean\ mask\footnote{\clean\ masks are not only used to limit deconvolution artefacts but also to preclude possible calibration artefacts, a topic that is beyond the scope of the current discussion.}, the region in which real emission is expected.
Outside the mask boundaries the algorithm is not allowed to allocate components.
However, even with a mask, such aggressive image space approximations inevitably lead to artefacts.
Thus, to prevent artefacts from accumulating, the residual has to be computed by subtracting the model convolved with the full PSF from the dirty image.
This step, which uses an FFT-based convolution, was termed the major cycle to distinguish it from the less accurate but much faster approximate computation of the residual termed the minor cycle.
\citet{cottonschwab} generalized this idea to use the full measurement operator instead of an FFT-based convolution leading to a different and more robust form of major cycle.

A major cycle corresponds to an exact evaluation of the gradient using the first of the two expressions for the gradient in \cref{eq:likgrad}.
It removes artefacts stemming from incomplete subtraction of PSF side lobes by subtracting the model correctly in visibility space.
In addition, by incorporating w-projection \citet{wprojection} or w-stacking \citet{wstacking} techniques into the implementation of the measurement operator, it is possible to compute the gradient without utilising the coplanar array approximation.
Since computing the gradient exactly is an expensive operation, it should preferably be done as few times as possible.
Högbom \clean\ can be used in combination with the Clark approximation to add multiple components to the model while keeping track of the approximate gradient.
This is called the minor cycle.
Eventually, the current model is confronted with the full data using the exact expression for the gradient and the procedure is repeated until some convergence criteria are met.
Since new regions of emission are uncovered as the corrupting effects of the brightest sources are removed, dynamic masking strategies, in which the mask is adapted from one major cycle to the next, are often employed.

The criterion at which to stop the minor cycle and perform another exact evaluation of the gradient affects both the computational cost and the quality of the final result.
Careful user input is often required to balance the tradeoff between these two factors.
Because of the convolutional nature of the problem, the level of artefacts introduced by exploiting image space approximations is proportional to the brightest pixel in the residual image.
Thus, running the minor cycle for too long adds artefacts to the model.
In principle it is possible to correct for these artefacts in subsequent iterations, but in practice this is potentially unstable.
As convergence criterion for the minor loop, a parameter called major loop gain or peak factor is defined:
Iterate minor loops until the residual has decreased by the peak factor.
A sensible choice depends on the field of view and the degree of non-coplanarity of the array.
Typical values are around 0.15.

In AIPS, the software we used for our \ssclean\ maps, a new major cycle $i+1$ starts if the flux of the next clean component is smaller than $m_i(1 + a_i)$, a current map specific reference flux $m_i$ times a cycle-dependent factor $1+a_i$, which is stirred according to the following heuristic.
The starting value for this factor, $a_0$, depends on the ratio $\rho =\frac{r_0 - m_0}{m_0}$ where $r_i$ and $m_i$ are the peak and lowest flux of the absolute residual image in the $i$th major cycle, respectively, and is defined as:
\begin{align}
  a_0 = \left\{
  \begin{array}{lcl}
    0.05 \cdot \rho &:&  \rho \geq 3\\
    0.02 \cdot \rho &:&  1 \leq \rho < 3\\
    0.01 \cdot \rho &:&  \rho < 1.
  \end{array}
  \right.
\end{align}
Then, $a$ increases at each iteration: $a_{i+1} = a_i + n_i^{-1} \left(\frac{m_i}{r_i}\right)^f$ where $n_i$ is the current number of \clean\ components and $f$ is a free parameter.
Larger $f$s let $a_i$ decrease more slowly.

Especially if extended emission is present, model images produced by \clean\ are so far from realistic representatives of the true sky that astronomers cannot work with them directly.
They are the best fit to the data under the implicit prior imposed by \clean\ but fail miserably at capturing extended source morphology or frequency spectra.
Therefore, the results produced by \clean\ are interpreted with the help of the so-called restored image.
The first step in creating the restored image is to convolve the model image with the \clean\ beam, a Gaussian that approximates the primary lobe of the PSF.
This represents the intrinsic resolution of the instrument that is assumed to be constant across the image.
Next, in an attempt to account for any undeconvolved flux and set the noise floor for the observation, the residual image is added to the model convolved with the PSF.
The noise floor, which is taken to be the RMS of the resulting image in regions devoid of structure, is then supposed to give an estimate of the uncertainty in each pixel.

All in all, careful user input is required to successfully use \clean\ for imaging.
Fortunately the tunable parameters are actually quite easy to set once the user has developed some intuition for them.
However, the model images produced by \ssclean\ are completely unphysical when there are extended sources in the field.
In extreme cases, \ssclean\ fails to fully deconvolve the faint diffuse emission in the field and can lead to imaging artefacts.
A possible explanation for this is that, at each iteration, \ssclean\ tries to minimize the objective function by interpolating residual visibility amplitudes with a constant function.
This limitation has been partially addressed by the multi-scale variants of the \clean\ algorithm.

\subsection{\expandafter\MakeUppercase \msclean}
\expandafter\MakeUppercase\msclean\ \citep{cornwell2008,rau2011,offringamsclean} is an extension of \ssclean\ that imposes sparsity in a dictionary of functions, as opposed to just the delta function.
Most implementations use a pre-determined number of either circular Gaussian components or the tapered quadratic function \citep{cornwell2008} in addition to the delta function.
While this model is still not a physical representation of the sky, diffuse structures within the field of view are more faithfully represented.
Most \msclean\ implementations share the major and minor cycle structure of Cotton-Schwab \clean\ with the major cycle implemented in exactly the same way.
However, the minor cycle differs between the many variants of \msclean.
The implementation used for the current comparison is described in detail in \citet{offringamsclean} and implemented in the \wsclean\ software package \citep{wstacking}.

The starting point for \wsclean's multi-scale algorithm is to select the size of the scale kernels.
While this can be specified manually, \wsclean\ also provides a feature to determine them automatically from the uv-coverage of the observation.
In this case, the first scale always corresponds to the delta function kernel scale.
The second scale is then selected as the full width window of the tapered quadratic function that is four times larger than the smallest theoretical scale in the image (determined from the maximum baseline).
The size of the corresponding Gaussian scale kernels is set to approximately match the extent of the tapered quadratic function.
As noted in \citet{offringamsclean}, the factor of four was empirically determined to work well in practice.
If smaller scales are used, point sources are sometimes represented with this scale instead of the delta scale.
Each subsequent scale then has double the width of the previous one and scales are added until they no longer fit into the image or until some predetermined maximum size is reached.

Once the scales have been selected, the algorithm identifies the dominant scale at each iteration.
This is achieved by convolving the residual image with each Gaussian scale kernel and comparing the peaks in the resulting convolved images subject to a scale bias function (conceptually similar to matched filtering).
The scale bias function (see \citet{offringamsclean} for full details) can be used to balance the selection of large and small scales.
It introduces a tunable parameter to the algorithm, viz.\ the scale bias $\beta$.
With the dominant scale identified, the model is updated with a component corresponding to this scale at the location of the maximum in the convolved residual image.
As with \ssclean, the model is not updated with the full flux in the pixel but only some fraction thereof.
The exact fraction is scale-dependent (see again \citet{offringamsclean} for details).
To keep track of the approximate residual, the PSF convolved with the scale kernel multiplied by this same fraction is subtracted from the residual image.

The additional convolutions required to determine the dominant scale at each iteration introduce an additional computational cost compared to \ssclean.
For this reason, \wsclean\ provides the option of running an additional sub-minor loop that fixes the dominant scale until the peak in the scale convolved image decreases by some pre-specified fraction (or for a fixed number of iterations).
This significantly decreases the computational cost of the algorithm but it is still more expensive than \ssclean.
While we will not delve into the exact details of how the sub-minor loop is implemented, we will note that it introduces yet another tunable parameter to the algorithm that is similar to the peak factor of Cotton-Schwab \clean.
This parameter, called multiscale-gain in \wsclean, determines how long a specific scale should be \clean ed before re-determining the dominant scale in the approximate residual.
Importantly, the sub-minor loop also makes use of a Clark-like approximation to restrict regions in which peak finding and PSF subtraction should be performed.
This improves both the speed and the quality of the reconstructed images.

While we have not discussed all the details behind the \msclean\ implementation in \wsclean, our discussion should make it clear that it introduces additional tunable parameters to the algorithm.
Most of the time the algorithm performs reasonably well with these parameters left to their defaults.
However, some degree of tuning and manual inspection is sometimes required, especially for fields with complicated morphologies.

\subsection{Motivation to improve CLEAN}
\label{sec:problems}
Classical radio interferometric imaging suffers from a variety of problems.
Two of these problems stand out in particular: the lack of reliable uncertainty estimates and the unphysical nature of model images produced by \clean.
As we discuss below, \clean\ forces astronomers to conflate these two issues in a way that makes it very difficult to derive robust scientific conclusions in the sense that it is not guaranteed that two observers would convert the same data set into the same sky image and that meaningful statistical uncertainty information would be provided by the algorithm.

Astronomers need to account for uncertainties in both flux and position and these two notions of uncertainty are correlated in a non-trivial way that is determined by both the uv-coverage and the signal-to-noise ratio of the observation.
However, model images produced by \clean\ are not representative of the true flux distribution of the sky and come without any uncertainty estimates.
This can be attributed to the fact that \clean\ is not based on statistical theory but rather is a heuristic that tries to represent flux in form of pre-determined basis functions (delta peaks, Gaussians) via flux-greedy algorithms.
As a result, astronomers turn to the restored image (see \cref{sec:ssclean}) instead of relying directly on the model produced by \clean.
Compared to the model image, the restored image has two favourable qualities viz.\ it accounts for the (assumed constant) intrinsic instrumental resolution and it displays structures in the image relative to the noise floor of the observation.
These two aspects are supposed to roughly account for uncertainties in position and flux respectively.
However, besides the fact that adding the residuals back in introduces structures in the image that are not real, and that the restored image has inconsistent units\footnote{The residual has different units from the model convolved by the \clean\ beam.}, this is completely unsatisfactory from a statistical point of view.
Firstly, the restored image completely neglects the correlation between uncertainties in flux and position, information that is crucial to determine whether a discovery is real or not.
In fact, since the act of convolving the model image by the \clean\ beam assumes that the resolution is constant across the image, whereas it is known that super-resolution of high signal-to-noise structures is possible, the restored image paints a rather pessimistic picture of the capabilities of radio interferometers.
Secondly, both the \enquote{noise in the image} and the size of the clean beam depend on the weighting scheme that has been used.
It is difficult to attach any degree of confidence to the results since the weighting scheme is a free parameter of \clean.
\citet[Figure~1 and 2]{convex}) shows the impact of different weighting schemes on the final image.
This limitation is borne out quite explicitly in the data set chosen for the current comparison in \cref{sec:comparison}.
Furthermore, since \clean\ outputs images that contain regions with unphysical negative flux\footnote{Negative flux is also an artefact of discretising the measurement operator \cref{eq:discretemeasmodel} since the response of a point source situated exactly in between two pixels is a $\mbox{sinc}$ function.}, astronomers need to assess for themselves which parts of the image to trust in the first place.
The above limitations provide opportunities for speculative scientific conclusions that cannot be backed up by statistically rigorous arguments.
They also make it impossible to quantitatively compare images from radio interferometers processed by \clean\ to for example astrophysical simulations.

In addition to the above, \clean\ relies on user input that involves the careful construction of masks, selecting an appropriate weighting scheme and setting hyper-parameters such as loop gains and stopping criteria etc.
This results in an effective prior:
It is known that \clean\ imposes some measure of sparsity in the chosen dictionary of functions, but it is unclear how to write down the explicit form of the effective prior.
The problem is exacerbated by \clean\ using a form of backward modelling that does not perform well when there are very little data available or when the uv-coverage is highly non-uniform, as is the case for typical VLBI observations.
Thus, the way that \clean\ is implemented is fundamentally incompatible with Bayesian inference making it impossible to infer, or indeed marginalize over, optimal values for the parameters it requires.
This is clearly problematic as far as scientific rigour is concerned.

This illustrates that the notions of uncertainty, resolution and sensitivity are tightly coupled concepts when interpreting images produced by radio interferometers.
As such it is not sufficient to apply a post-processing step such as making the restored image to derive scientific conclusions from radio maps.
In fact, doing so potentially limits the usefulness of interferometric data because it eliminates the possibility of super-resolution at the outset.
This is a result of incorrect prior specification and not properly accounting for the interaction between the data fidelity and the prior term during imaging.
Obtaining sensible posterior estimates requires combining the linear Fourier measurement taken by the interferometer with a prior that respects the physics of the underlying problem, such as enforcing positivity in the spatial domain for example.
To this end, \resolve\ approximates the posterior with MGVI, an algorithm that can track non-trivial cross-correlations.
Instead of providing a point estimate with associated error bars, MGVI provides samples from the approximate posterior that can then be used to compute expectation values of any derived quantities while accounting for cross correlations between parameters.

In summary, the absence of proper uncertainty information, potential negativity of flux, the arbitrariness of the weighting scheme, problems with little data and non-uniform uv-coverage and loss of resolution by convolving with the \clean\ beam illustrate the necessity to improve beyond the \clean-based algorithms.

\section{Comparison of results from \resolve\ and \clean}

\label{sec:comparison}
Here we compare the performance of the three imaging approaches presented in \cref{sec:resolve,sec:traditional}.
To this end we use VLA observations of Cygnus~A that have been flagged and calibrated with standard methods.
For more details on the data reduction process refer to \citet{datapaper}.
We use single-channel data sets at the frequencies 2052, 4811, 8427 and 13360~MHz.
The \clean\ maps have been converted from the unit Jy/beam to $\fluxunit$ by multiplication with the half-width-half-maximum area of the \clean\ beam.
All data and the results of the three different methods are archived \citet{zenodo}\footnote{\url{https://doi.org/10.5281/zenodo.4267057}}.

\subsection{Configuration}

\begin{table}
  \centering
  \begin{tabular}{lrrrr}
    \hline\hline
    & $\alpha$ mean & $\alpha$ sd &$I$ mean & $I$ sd\\\hline
    Offset&0&---& 21 & ---\\
    {[1]} Zero mode variance & 2 & 2 & 1 & 0.1\\
    {[2]} Fluctuations & 2 & 2 & 5 & 1 \\
    {[5]} Flexibility & 1.2 & 0.4 & 1.2 & 0.4\\
    {[6]} Asperity & 0.2 & 0.2 & 0.2 & 0.2 \\
    {[7]} Average slope & -2 & 0.5 & -2 & 0.5\\\hline
  \end{tabular}
  \caption{Hyper parameters for \resolve\ runs.
  The numbers in the brackets refer to the index of the excitation vector $\xi$ to which the specified mean $\mathfrak m$ and standard deviation $\mathfrak s$ belong, see for example \cref{eq:lognormalexample}.}
  \label{tab:hpres}
\end{table}

All values for the hyper parameters of \resolve\ are summarized in \cref{tab:hpres}.
The \resolve\ parameters separate into those for the sky brightness distribution and those for the Bayesian weighting scheme.
For the latter, they are chosen such that the model has much flexibility to adopt to the exact situation.
Because $\alpha$ provides a multiplicative correction to the noise levels, the offset is set to zero (which becomes one, that is to say no correction, after exponentiation).
The zero mode standard deviation is set to a high value because the overall noise level might be completely different.
The fluctuations also have a large standard deviation such that the algorithm can easily tune that parameter.
A value of 2 means that we expect the correction function $\alpha$ to vary within one standard deviation two e-folds up and down.
The flexibility and asperity parameters of the power spectrum \tflex{} and \tasp{} are set such that the algorithm can pick up non-trivial values but not too extreme ones here.
The average slope of the power spectrum is chosen to vary around -2.
In other words, the Bayesian weighting scheme $\alpha$ depends in a differentiable fashion on the baseline length a priori.
A relatively high a priori standard deviation of 0.4 enables the algorithm to tune the slope to the appropriate value.
The most important aspect of the hyper parameter setting is that the resulting prior has enough variance to capture the actual Bayesian weighting scheme and sky brightness distribution.
As discussed above the model is set up in such a way that it can adjust its hyper parameters on its own.
All parameters discussed in this section are really hyper parameters of that hyper parameter search.
For the sky brightness distribution we know a priori that typical flux values in regions with emission vary on scales of $10^8$ and $10^{12}$~Jy/sr.
Therefore a sensible offset for the Gaussian field is $\log(10^9) \approx 20$.
A priori we let that value vary two e-folds up and down in one standard deviation which means that within three standard deviations typical flux values between $\approx 10^6$ and $\approx 10^{11}$~Jy/sr can be reached.
However, as always we make the standard deviations themselves a parameter and choose 2 for the standard deviation of the standard deviation of the zero mode which makes virtually all offsets possible.
As positions for the point sources modelled with an inverse-gamma prior (see \cref{eq:skymodel}) we assume a point source at the phase centre and a second one located at $(0.7, -0.44)$~arcsec relative to the phase centre \citep[Cygnus~A-2,][]{discovery}.

Apart from the hyper parameters we need to specify the minimization procedure for \resolve\ \citep{mgvi}.
In order to arrive at a sensible starting position for the actual inference we proceed in the following steps:
First, compute the maximum-a-posterior solution assuming the error bars provided by the telescope.
This means that we set $\alpha = 1$ in \cref{eq:noisemodel}.
Second, use five mirrored parameter samples $\xi$, as generated by MGVI, to approximate the Metric Gaussian Kullback-Leibler divergence and solve the inference problem with respect to $\xi^{(\sigma)}$ only.
In other words, we find a good weighting scheme $\alpha$ conditional to the sky brightness distribution found before.
Third, solve the MGVI inference problem for the sky brightness distribution conditional to the found weighting scheme using five mirrored samples.
Fourth, solve the full inference problem for the sky brightness distribution and the Bayesian weighting scheme simultaneously.
Fifth, terminate after the second iteration.
Sixth, flag all data points that are more than $6\sigma$ away from the model data taking the Bayesian weighting scheme into account and restart from the first step.

In all cases, we approximate the Metric Gaussian Kullback-Leibler divergence using five mirrored samples.
These samples are drawn with the help of conjugate gradient runs (see \cref{sec:sampling}).
These conjugate gradients are declared converged when the conjugate gradient energy does not change by more than 0.1 three times in a row.
As an upper limit for the maximum number of conjugate gradient steps we choose 2000.
Not iterating the conjugate gradient algorithm until convergence, which is not computationally feasible, does not introduce biases in the inference but rather increases the posterior variance as discussed in \citet{mgvi}.

\begin{table}
  \centering
  \begin{tabular}{lr}
    \hline\hline
    Parameter & Value\\\hline
    j & 20\\
    size & 4096 3072\\
    padding& 2.0 \\
    scale&  0.04asec \\
    weight & briggs 0\\
    gain & 0.1 \\
    mgain & 0.8 \\
    niter & 1000000 \\
    nmiter & 10 \\
    multiscale-gain & 0.1 \\
    auto-mask & 2.0 \\\hline
  \end{tabular}
  \caption{Common hyper parameters for \msclean\ runs.
    The parameters that differ for the four runs are described in the main text.
    Additionally, the options \texttt{multiscale}, \texttt{no-small-inversion}, \texttt{use-wgridder}, \texttt{local-rms} have been used.}
  \label{tab:hpms}
\end{table}

The \msclean\ results produced for the current comparison were obtained by first doing an imaging run with uniform weighting down to a fairly low threshold and using \wsclean's auto-masking feature.
The resulting images were used to define an external mask containing the most prominent features.
A second imaging run down to a deeper threshold was then performed using Briggs weighting with a robustness factor of -1.
These images were then used to refine the mask and to flag obvious outliers in the data.
The outliers were identified by computing whitened residual visibilities and flagging all data points with whitened residual visibility amplitudes larger than five time the global average.
On average this resulted in about 1\% of the data being flagged which is more than expected from the noise statistics.
This could indicate that a small amount of bad data slipped through the initial pre-processing steps (flagging and calibration).
The final imaging run was then performed using the refined mask and Briggs weighting with a robustness factor of zero.
While the procedure could be refined further, we found that doing so results in diminishing returns in terms of improving the final result.

The \wsclean\ settings reported in \cref{tab:hpms} are common to all the data sets for the final \msclean\ imaging run.
The image size was set so that the PSF for the 13 GHz data set has just more than five pixels across the FWHM of the primary lobe, a rule of thumb that is commonly employed to set the required pixel sizes for an observation.
Twenty threads are employed to approximately match the computational resources given to resolve.
In addition to auto-masking, which is set to kick in when the peak of the residual is approximately twice the value of the RMS in the image, a manual FITS mask was supplied using the \texttt{fits-mask} option.
The masks for the different data sets are shown in \cref{fig:msmasks}.
In all cases the scales were automatically selected.
The only parameter that differs between data sets is the threshold at which to stop \clean ing, specified through the \texttt{threshold} parameter in \wsclean .
These were set to 0.002, 0.0007, 0.0003, and 0.0002 for the 2, 4, 8, and 13 GHz data sets, respectively, which approximately matches the noise floor in the final restored images.
A value of zero for the Briggs robustness factor was chosen as it usually gives a fairly good tradeoff between sensitivity and resolution.
However, as discussed in \cref{sec:problems}, the need to specify the weighting scheme manually is one of the main limitations of \clean .
This is especially evident in the 8 GHz observation where the Cygnus~A-2 is just visible using a robustness factor of zero whereas it is clearly visible in the images with a robustness factor on minus one.
Cygnus~A-2 is completely lost when using natural weighting, which is where the interferometer is most sensitive to faint diffuse structures.
For \ssclean, the default settings as implemented in AIPS are used.

\subsection{Analysis of results}

\begin{figure*}
  \centering
  \input{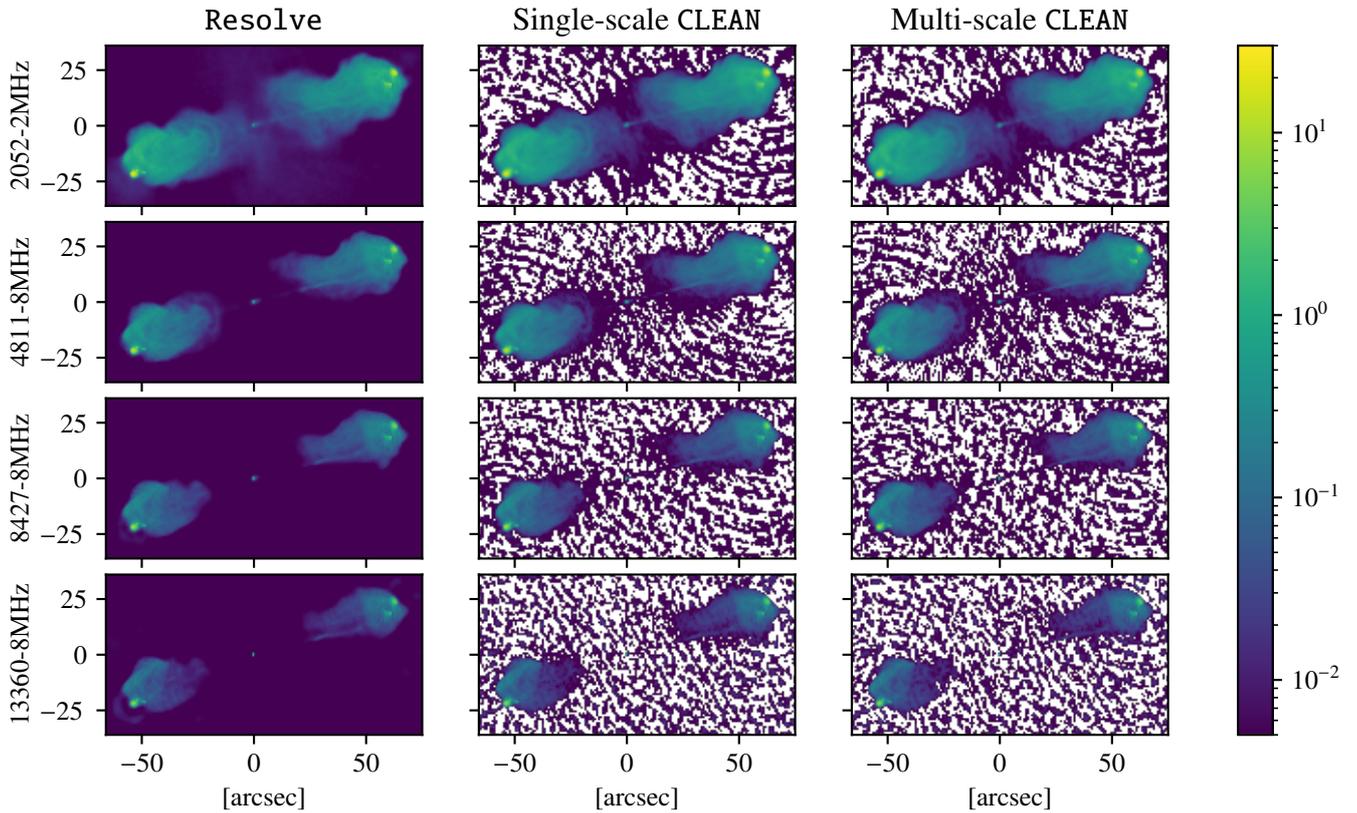}
  \caption{Overview of imaging results. The first column shows the \resolve\ posterior mean, the middle and last column show \ssclean\ \msclean\ results, respectively.
    The colour bar has units $\fluxunit$.
    Negative flux regions are displayed in white. See also different scaled version in \cref{fig:all2}.}
  \label{fig:all}
\end{figure*}

\Cref{fig:all} shows a summary of the results of the twelve runs: four frequencies imaged with three different algorithms.
The units of the \clean\ images have been converted to $\fluxunit$ (by dividing the \clean\ output in Jy/beam by the beam area $\frac{\pi}{4\log 2}\cdot$\texttt{BMAJ}$\cdot$\texttt{BMIN}).
Then the pixel values of all images can be directly compared to each other.
As discussed above, the output of \resolve\ is not a single image but rather a collection of posterior samples.
For the purpose of comparison we display the pixel-wise posterior mean.

\begin{table}
  \input{points}
  \caption{\resolve\ point source fluxes.
    Source 0 refers to the central source Cygnus~A and Source 1 to the fainter secondary source Cygnus~A-2.
  The standard deviation is computed from the \resolve\ posterior samples and does not account for calibration uncertainties and other effects, see main text.}
\label{tab:points}
\end{table}

\begin{figure*}
  \centering
  \input{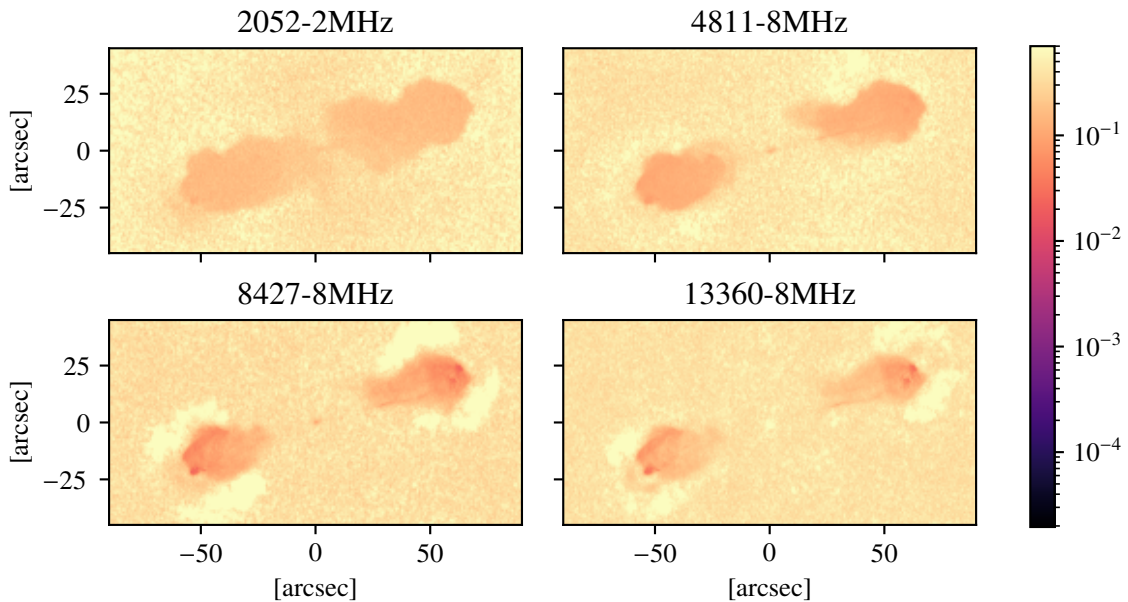}
  \caption{Relative pixel-wise posterior uncertainty of \resolve\ runs.
    All plots are clipped to 0.7 from above and the two pixels with point sources are ignored in determining the colour bar.
    Their uncertainty is reported in \cref{tab:points}.
  }
  \label{fig:uncertaintylog}
\end{figure*}

\Cref{fig:all} shows that the \resolve\ maps do not feature any negative flux regions.
Since this was a strict prior assumption for the algorithm, this is the expected result.
The \ssclean\ and the \msclean\ have many negative flux regions where no (bright) sources are located.
Otherwise, the results of these two algorithms are similar.
Additionally, \cref{fig:uncertaintylog,fig:uncertaintylinear} show the pixel-wise posterior uncertainty of the \resolve\ runs.
These figures do not contain the whole uncertainty information which is stored in the posterior samples.
The posterior distribution for each pixel is not Gaussian and therefore the higher moments are non-trivial.
Additionally, the cross-correlation between the pixels cannot be recovered from the pixel-wise posterior uncertainty.

\begin{figure*}
  \centering
  \input{bigg0.pgf}
  \caption{Zoomed-in version of the \ssclean\ reconstruction of the $13.36$~GHz data set focusing on the western lobe and rotated anti-clockwise by 90 degrees.
    The colour bar is the same as in \cref{fig:all}.
    Negative flux regions have been set to lower limit of the colour map.}
  \label{fig:biggssclean}
\end{figure*}
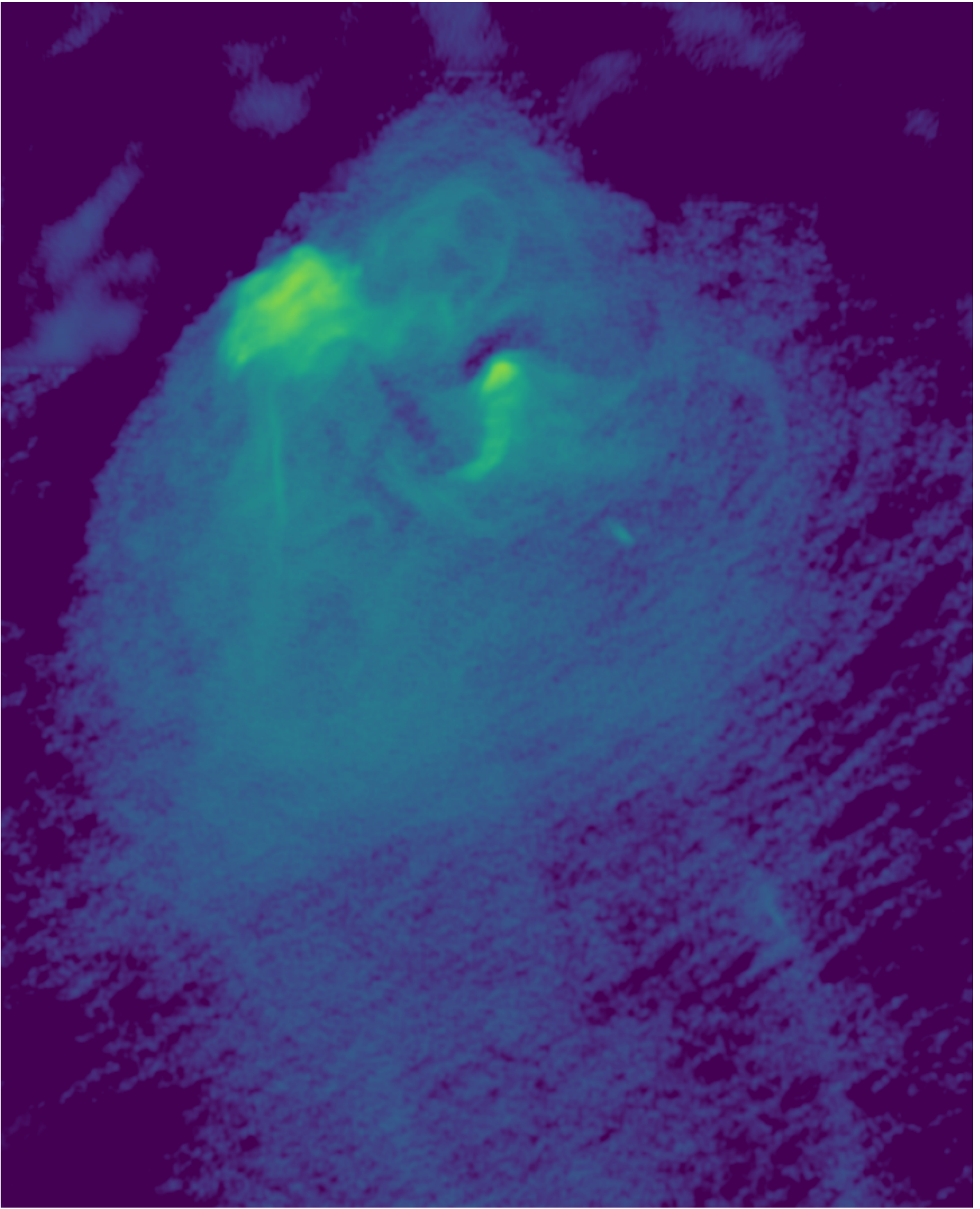

\begin{figure*}
  \centering
  \input{bigg1.pgf}
  \caption{Same as \cref{fig:biggssclean}, but with \msclean\ reconstruction.}
  \label{fig:biggmsclean}
\end{figure*}
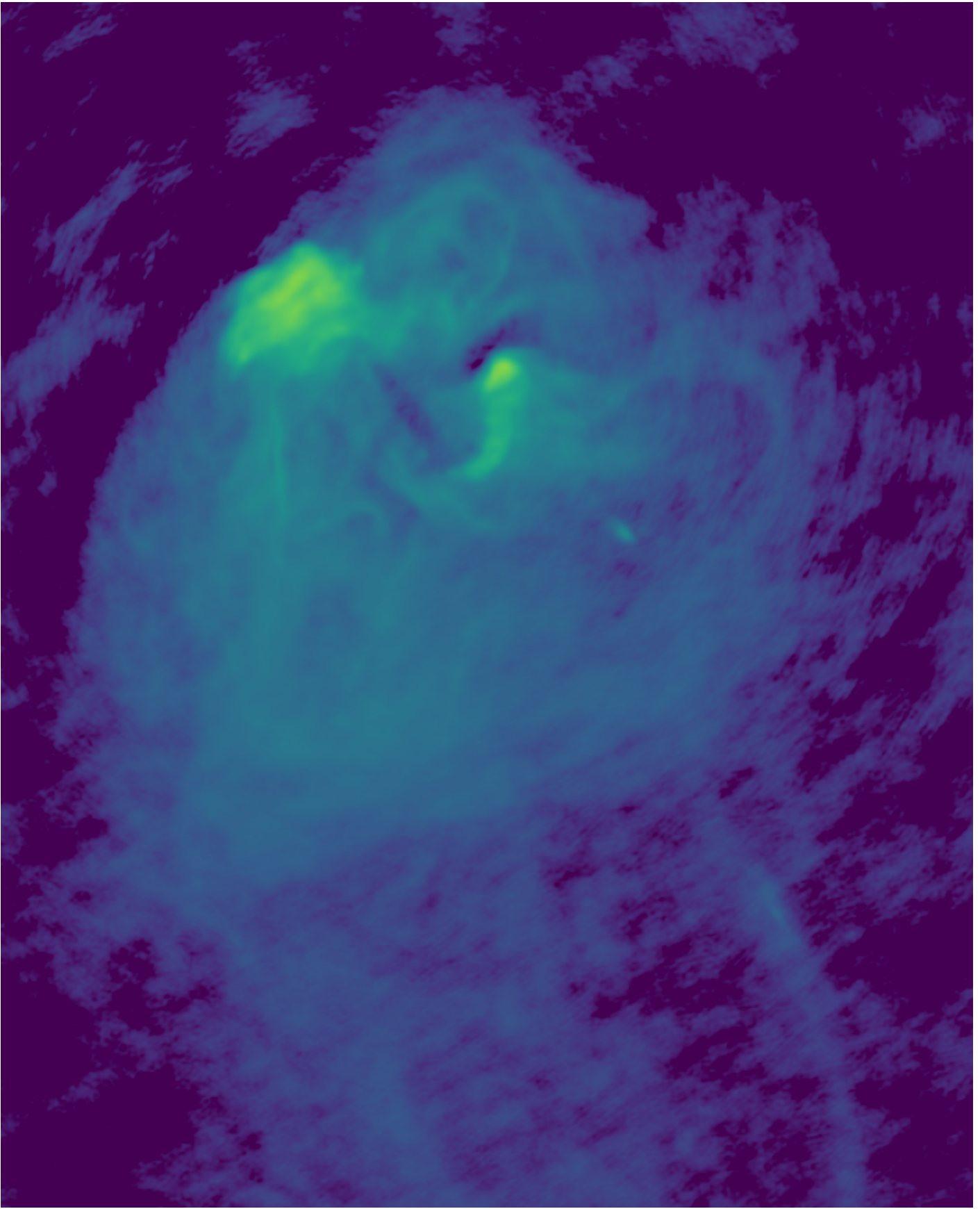

\begin{figure*}
  \centering
  \input{bigg2.pgf}
  \caption{Same as \cref{fig:biggssclean}, but with \resolve\ posterior mean.}
  \label{fig:biggresolve}
\end{figure*}
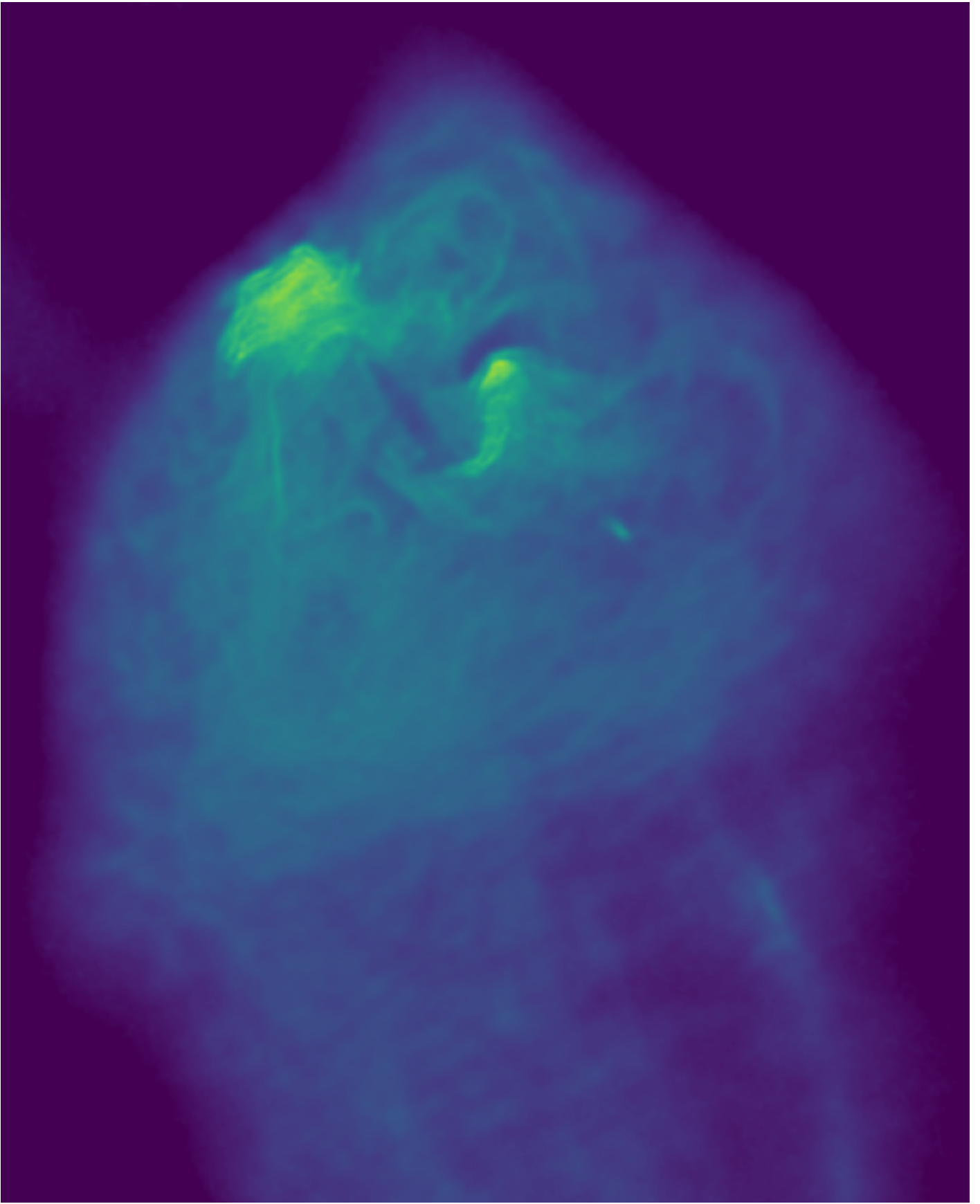

\begin{figure*}
  \centering
  \input{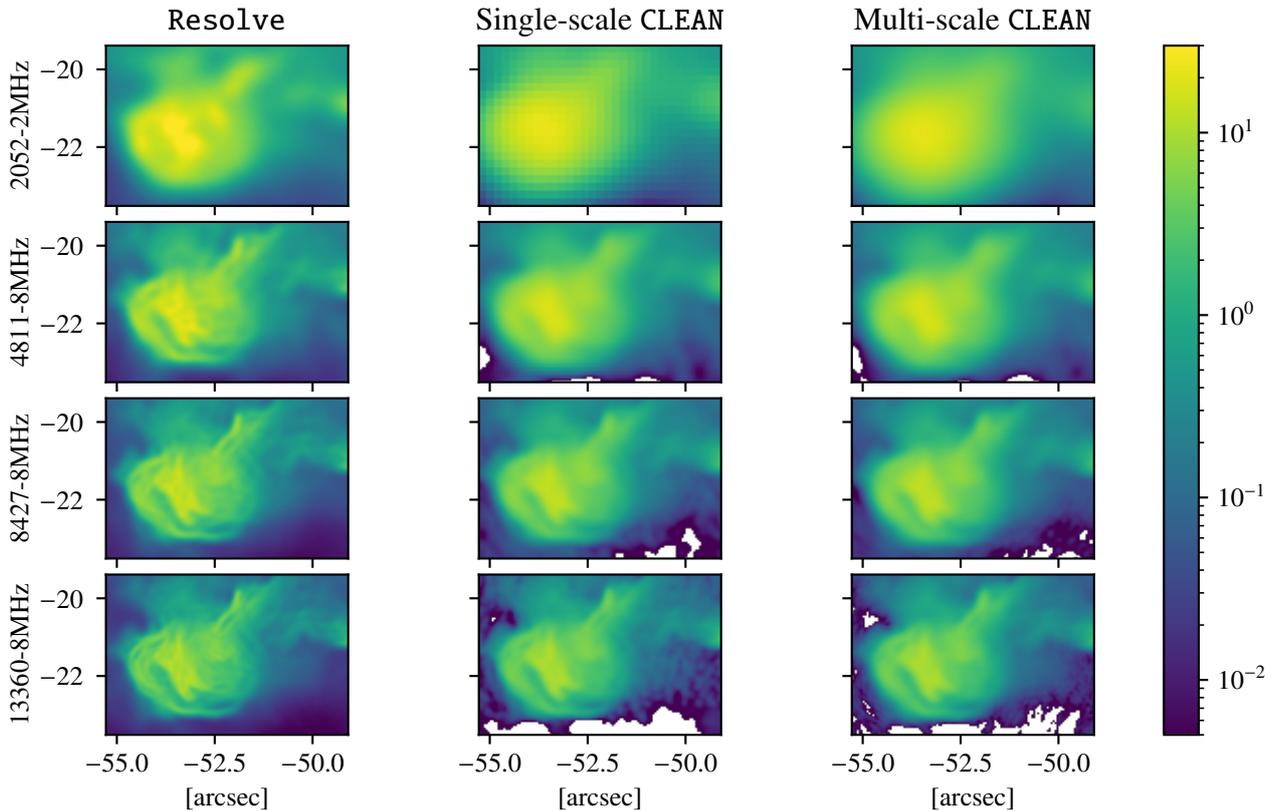}
  \caption{Overview of imaging results. Zoomed-in version of \cref{fig:all} focusing on the eastern hot spot.}
  \label{fig:zoomed}
\end{figure*}

In order to investigate the results further, \crefrange{fig:biggssclean}{fig:biggresolve} show the western lobe of the $13.36$~GHz observation only and \cref{fig:zoomed} shows the bottom left hot spot of all observations.
In the \clean\ results it can be seen that the resolution improves significantly when going to higher frequencies.
This is due to the natural increase of an interferometer: the higher the observation frequency, the higher the intrinsic resolution.
The same is true for the \resolve\ maps.
However, \resolve\ also achieves higher resolution than \clean\ at lower frequencies.
By eye, the resolution of the \resolve\ $4.8$~GHz map is comparable to the \clean\ $13.4$~GHz map.
This phenomenon is called super-resolution and is possible by the non-trivial interaction between likelihood and prior:
By adding the constraint that the sky brightness distribution is positive, information about Fourier modes which correspond to baselines longer than the actual maximum baseline can be inferred from the data.
The high resolution features that turn up at lower frequencies can be validated at the higher frequency \clean\ maps.
This is possible because the synchrotron radiation which is responsible for the emission has a very broad frequency spectrum.
Unless there are internal or external absorption effects which are not believed to be happening here, there cannot be major differences in the brightness over frequency ratios of a few.
Additionally, it can be observed that the ripples in the fainter regions next to the hotspot which are present in both \clean\ reconstructions are not present in the \resolve\ one.
This is rooted in the fact that \resolve\ can take the noise level properly into account and let the prior smooth within the regions which are less informed by the data because the flux level is lower.

\begin{figure*}
  \centering
  \input{overview_contour_13360_leftverysmall.pgf}
  \caption{Comparison of \msclean\ (blue contour lines, grey regions: negative flux regions) and four \resolve\ posterior samples (red) at 13.4~GHz.
  }
  \label{fig:smallcontour}
\end{figure*}
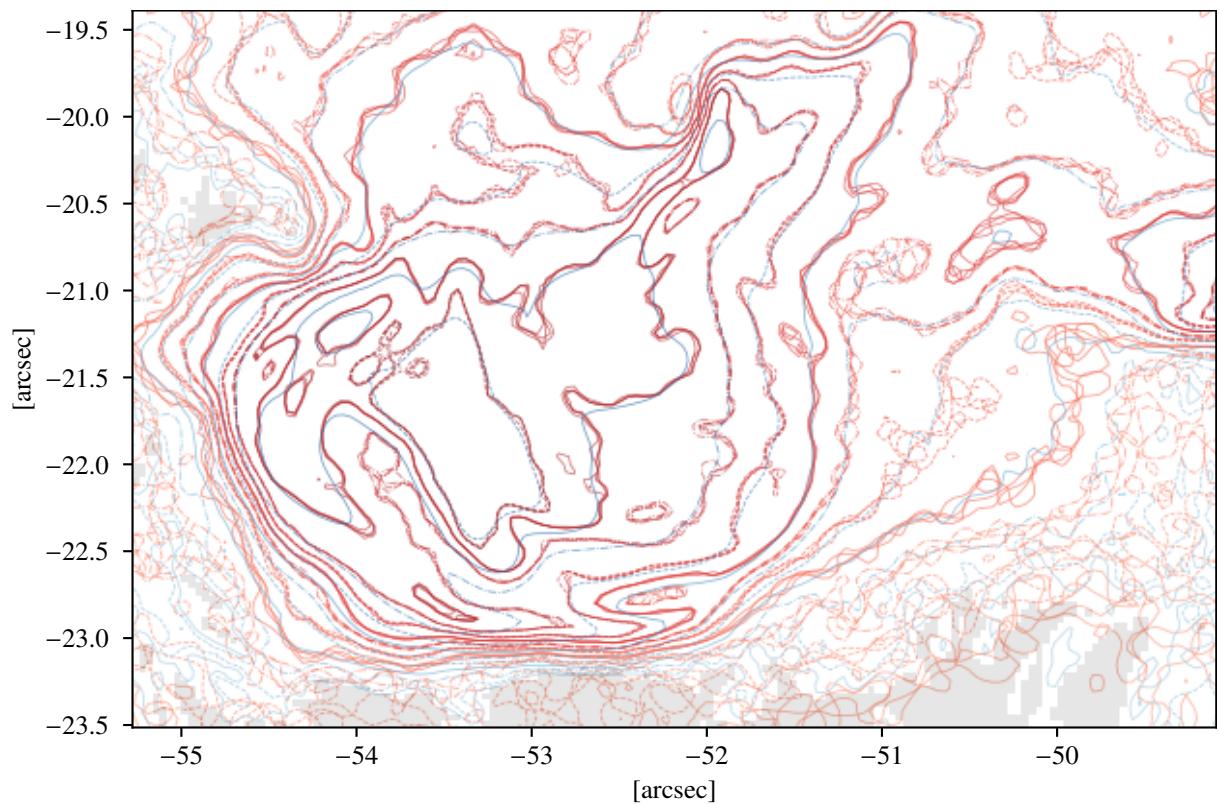

\Cref{fig:smallcontour} shows a direct comparison of the \msclean\ result and posterior samples of \resolve.
It can be observed that the \resolve\ samples significantly deviate from the \msclean\ map.
In addition, it becomes apparent that \resolve\ assigns significant flux in regions which have negative flux in the \ssclean\ result.

\begin{figure*}
  \centering
  \input{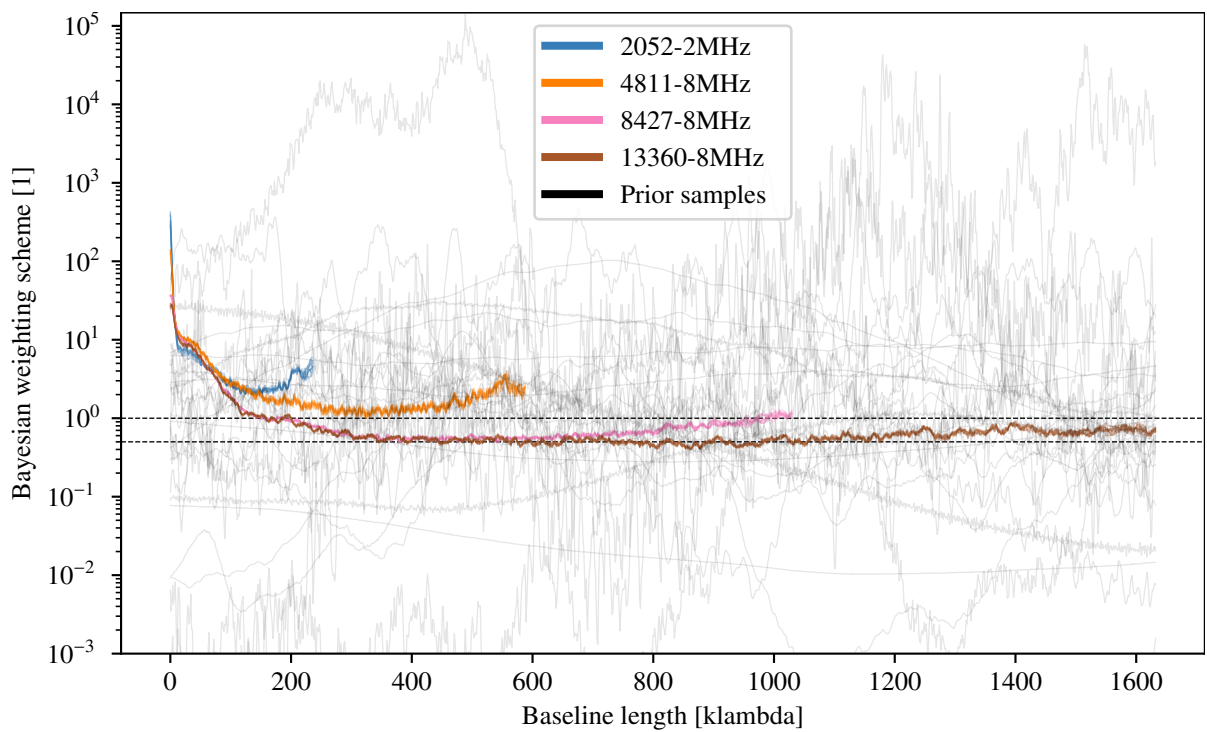}
  \caption{Posterior samples of the Bayesian weighting scheme $\alpha$ and prior samples for the $13.36$~GHz data set.
    The dashed lines are located at values 0.5 and 1.
    The latter corresponds to no correction at all.
    The light grey lines are prior samples that illustrate the flexibility of the a priori assumed Bayesian weighting schemes.}
  \label{fig:stdcorr}
\end{figure*}

\Cref{fig:stdcorr} displays posterior samples of the Bayesian weighting scheme.
It can be observed that the prior samples have higher variance and show a huge variety of correlation structures.
This shows that the prior is agnostic enough not to bias the result in a specific direction.
Generally, the correction factor decreases with baseline length.
Its minimum and maximum values are $\input{stdcorr_min}$ and $\input{stdcorr_max}$, respectively, across all four data sets and all posterior samples.
That means that the actual noise level of some visibilities is $\input{stdcorr_max}$ times higher than promised by the SIGMA column of the measurement set.
For medium to long baseline lengths the correction factor takes values between $\approx 0.5$ and $\approx 1$.
A relative factor of 0.5 could originate from different conventions regarding the covariance of a complex Gaussian probability density.
For the 2~GHz data set the correction factor remains at values $\approx 8$ even at longer baseline lengths.
So this data set seems to have an overall higher noise level than specified.
For long baseline lengths the noise level increases consistently.
This effect may be explained by inconsistencies in the data due to pointing errors.
Especially at high frequencies, Cygnus~A has comparable angular size to the primary beam.
Particularly near the zenith (Cygnus~A transits 8~degrees from the zenith), the VLA antennas do not point accurately.
The errors induces by this cannot be modelled by antenna-based calibration solutions.
Therefore, pointing errors introduce inconsistencies in the data.
An additional source of inconsistencies in the data might be inconsistent calibration solutions which have been introduced in the data during the self-calibration procedure in which negative components in the sky brightness distribution have been used.
An approach similar to \citet{resolve19} may be able to compute consistent calibration solutions in the first place.

In the following, we briefly discuss some of the materials that can be found in \cref{sec:supplementary}.
\Cref{fig:residualmaps} displays residual maps as they are computed by \wsclean.
Residual maps are defined by the r.h.s.\ of \cref{eq:likgrad} divided by $\tr N^{-1}$.
It is uncommon to plot the residual image based on the restored image in the \clean\ framework.
However, if the science-ready image is considered to be the restored image, it is vitally important to actually compute the residuals from it and not from a different image.
It can be observed that the \msclean\ model image fits the data very well whereas the restored \msclean\ image performs significantly worse.

From a signal reconstruction point of view, these residual maps have to be taken with a grain of salt since for example a non-uniform uv-coverage biases the visual appearance of the maps and overfitting cannot be detected.
Therefore, \cref{fig:histogramresolvewgts,fig:histogramcleanwgts} show histograms in data space for all three methods of the (posterior) residuals weighted with the \resolve\ weights $\sigma(\xisig)$ and the \wsclean\ imaging weights, respectively.
For better comparison, the residuals for the \msclean\ model image are included.
These histograms show how consistent the final images and the original data are.
For this comparison the error bars on the data are needed.
As stated above the error bars which come with the data and represent the thermal noise cannot be trusted.
Therefore, we compute the noise-weighted residuals based on the error bars which \resolve\ infers on-the-fly and the error bars (also called weighting scheme) which \wsclean\ uses for our \msclean\ reconstructions.
If the assumed data model is able to represent the true sky brightness distribution and its measurement the noise-weighted residuals should be standard-normal distributed.
This expected distribution is indicated in \cref{fig:histogramresolvewgts,fig:histogramcleanwgts} with dashed black lines.
\Cref{tab:chisq} provides the reduced $\chi^2$ values for all histograms in \cref{fig:histogramresolvewgts,fig:histogramcleanwgts}.
If the noise-weighted residuals are standard-normal distributed, $\chi^2_\mathrm{reduced}=1$.
The reduced $\chi^2$ values of the \resolve\ posterior with Bayesian weighting are all close to 1.
This means that the error bars indeed can be rescaled by a baseline-length-dependent factor and that \resolve\ is successful in doing so.
The \msclean\ model image overfits the data according to the \wsclean\ weighting scheme but achieves values close to 1 using the Bayesian weighting scheme as well.
In contrast the reduced $\chi^2$ values for the restored images produced by \ssclean\ and \msclean\ exceed all sensible values for both weighting schemes.
One may argue that an image which comes with reduced $\chi^2$ values of $>100$ does not have much in common with the original data.
All in all, the residuals show that the \resolve\ and the \clean\ reconstructions differ significantly already on the data level.

For inspecting low flux areas \cref{fig:all2} displays a saturated version of \cref{fig:all} and \cref{fig:contour_2052_all} compares the \msclean\ result with the \resolve\ posterior mean for the $2.4$~GHz data set.
It can be observed that all three algorithms pick up the faint emission.
For \resolve, the three higher frequency data reconstructions exhibit regions next to the main lobes which are very faint.
It looks like \resolve\ tries to make these regions negative which is not possible due to the prior.
For the $13.4$~GHz data set, even the central regions features such a dip.
All this can be explained by inconsistencies described above as well.

\Cref{tab:points} summarizes the fluxes of the two point sources including their posterior standard deviation.
Most probably, the provided uncertainty underestimates the true uncertainty for several reasons:
First, these uncertainties are conditional to the knowledge that two point sources are located at the given positions.
Therefore, the information needed to determine the position of the point sources is not included in the error bars.
Second, inconsistencies in the data induced by the calibration can lead to underestimating posterior variance because contradictory data points pull with strong force in opposite directions in the likelihood during the inference.
This results in too little posterior variance.
Third, MGVI only provides an lower bound on the true uncertainty but still its estimates are found to be largely sensible as shown in \citet{mgvi}.

Generally, it can be observed that the posterior standard deviation decreases with increasing frequency.
This is expected since interferometers with effectively longer baselines are more sensitive to point sources.
Our results from \cref{tab:points} can be compared to \citet[Table~1]{discovery}.
At $8.5$~GHz \citet{discovery} reports $1368$~mJy for the central source and $(4.15\pm 0.35)$~mJy for Cygnus~A-2.
At 13~GHz they report $1440$~mJy and $(4.86\pm 0.17)$~mJy.
These measurements have been taken in July 2015 whereas our measurements are from November 30 and December 5, 2015.
The comparison is still valid since \citet{discovery} showed that the sources are not significantly variable on the scale of one year.
We can observe that all flux values are in the right ballpark and the fluxes of Cygnus~A-2 agree within $2\sigma$.
The fluxes for the central source cannot be compared well because \citet{discovery} do not provide uncertainties on it.
However, taking only the \resolve\ uncertainties into account, the flux values differ significantly.
For the lower two frequencies no data are available in \citet{discovery} because the sources are not resolved by \clean.
The \resolve\ results give the posterior knowledge on the secondary source given its position.
In this way, statements about the flux of Cygnus~A-2 at low frequencies can be made even though it is not resolved.
Thus, we can claim the discovery of Cygnus~A-2 given its position on a $3\sigma$ and $7\sigma$ level for the $2.1$ and $4.8$~GHz observations, respectively.

\subsection{Computational aspects}
Each \resolve\ run needs $\approx 500\,000$ evaluations of the response and $\approx 400\,000$ evaluations of its adjoint.
That makes the response part of the imaging algorithm a factor of $\approx 50\,000$ more expensive compared to \clean\ approaches.
The good news is that the implementation of the radio response \cref{eq:nfft} in the package \ducc\ scales well with the number of data points and that the response calls can be parallelized over the sum in \cref{eq:objfunc}.

The \resolve\ runs have been performed on a single node with five MPI tasks, each of which needs $\approx 2.2~\mathrm{GB}$ main memory.
Each MPI task uses four threads for the parallelization of the radio response and the Fast Fourier Transforms.
The wall time for each \resolve\ run is between 80 and 90~h.

\expandafter\MakeUppercase \ssclean\ takes below 30 minutes for imaging each channel on a modern laptop.
Thus, \resolve\ is approximately 180 times slower that \ssclean\ here.
This comparison does not include that the \resolve\ had five times the number of CPUs available.

\expandafter\MakeUppercase \msclean\ takes about 2 hours during the final round of imaging on the 13 GHz data set.
This number does not account for the time taken during the initial rounds of imaging used to tune the hyper parameters and construct the mask which can be a time-consuming process.
However, it should be kept in mind that \clean\ scales much better when the dimensionality of the image is much smaller than that of the data, which is not the case here.
This is because \clean\ only requires about 10--30 applications of the full measurement operator and its adjoint, even including all preprocessing steps.
Taking 90~min for the average \msclean\ run, \resolve\ is 60 times slower than \msclean.

\section{Conclusions}
\label{sec:conclusions}
This paper compares the output of two algorithms traditionally applied in the radio interferometry community (\ssclean\ and \msclean) with a Bayesian approach to imaging called \resolve.
We demonstrate that \resolve\ overcomes a variety of problems present in traditional imaging:
The sky brightness distribution is a strictly positive quantity, the algorithm quantifies the uncertainty on the sky brightness distribution, and the weighting scheme is determined non-parametrically.
Additionally, \resolve\ provides varying resolution depending on the position on the sky into account, which enables super-resolution.
We find that \ssclean\ and \msclean\ give similar results.
In contrast, \resolve\ produces images with higher resolution: The 4.8~GHz map has comparable resolution to the 13.4~GHz \clean\ maps.
These advantages are at the cost of additional computational time, in our cases $\approx 90$~h wall time on a single node.

Future work may extend \resolve\ to multi-frequency reconstructions where the correlation structure in frequency axis is taken into account as well in order to increase resolution.
Additionally, direction-independent and antenna-based calibration may be integrated into \resolve .
Finally, the prior on the sky brightness distribution may be extended to deal with polarization data as well.

\begin{acknowledgements}
We thank Vincent Eberle and Simon Ding for feedback on drafts of the manuscript, Philipp Frank for his work on the correlated field model in NIFTy, Eric Greisen for explanations regarding AIPS, George Heald, Jakob Knollmüller, Wasim Raja, and Shane O'Sullivan for discussions, explanations, and feedback, and Martin Reinecke for his work on the software packages NIFTy and ducc and comments on an early draft of the manuscript.
P.~Arras acknowledges financial support by the German Federal Ministry of Education and Research (BMBF) under grant 05A17PB1 (Verbundprojekt D-MeerKAT).
The research of O.~Smirnov is supported by the South African Research Chairs Initiative of the Department of Science and Technology and National Research Foundation.
The National Radio Astronomy Observatory is a facility of the National Science Foundations operated under cooperative agreement by Associated Unversities, Inc.
\end{acknowledgements}

\bibliographystyle{aa}
\bibliography{bib.bib}

\appendix

\section{Supplementary material}\label{sec:supplementary}

\begin{figure*}
  \centering
  \input{overview_all_masks.pgf}
  \caption{Masks used for \msclean\ runs.}\label{fig:msmasks}
\end{figure*}

\begin{figure*}
  \centering
  \input{uncertainty0.pgf}
  \caption{Relative pixel-wise posterior uncertainty of \resolve\ runs on linear scale.
    The two pixels with point sources are ignored in determining the colour bar.}\label{fig:uncertaintylinear}
\end{figure*}
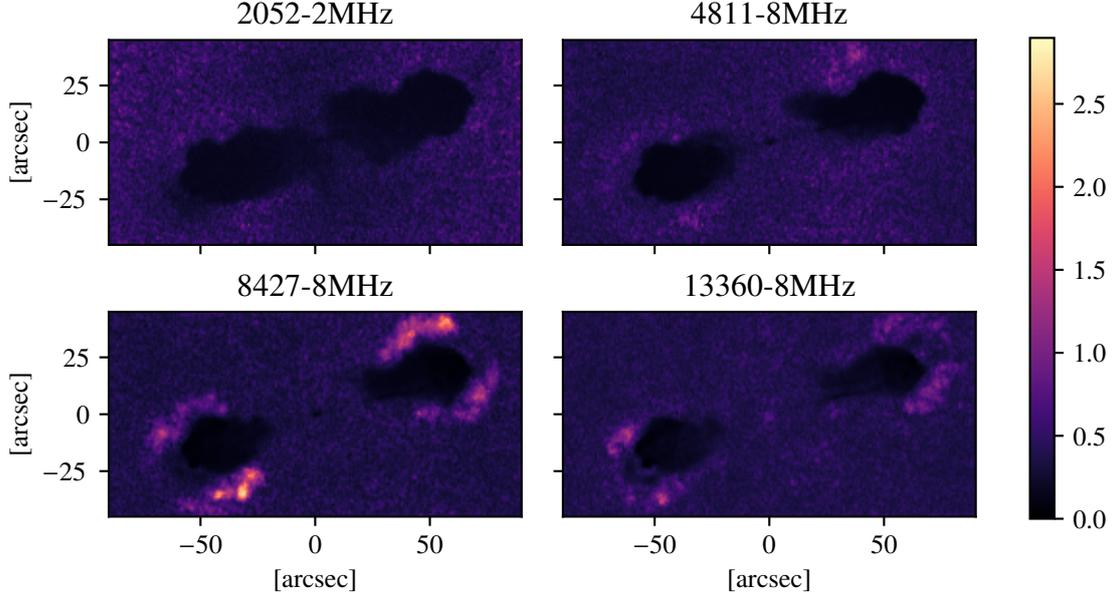

\begin{figure*}
  \centering
  \input{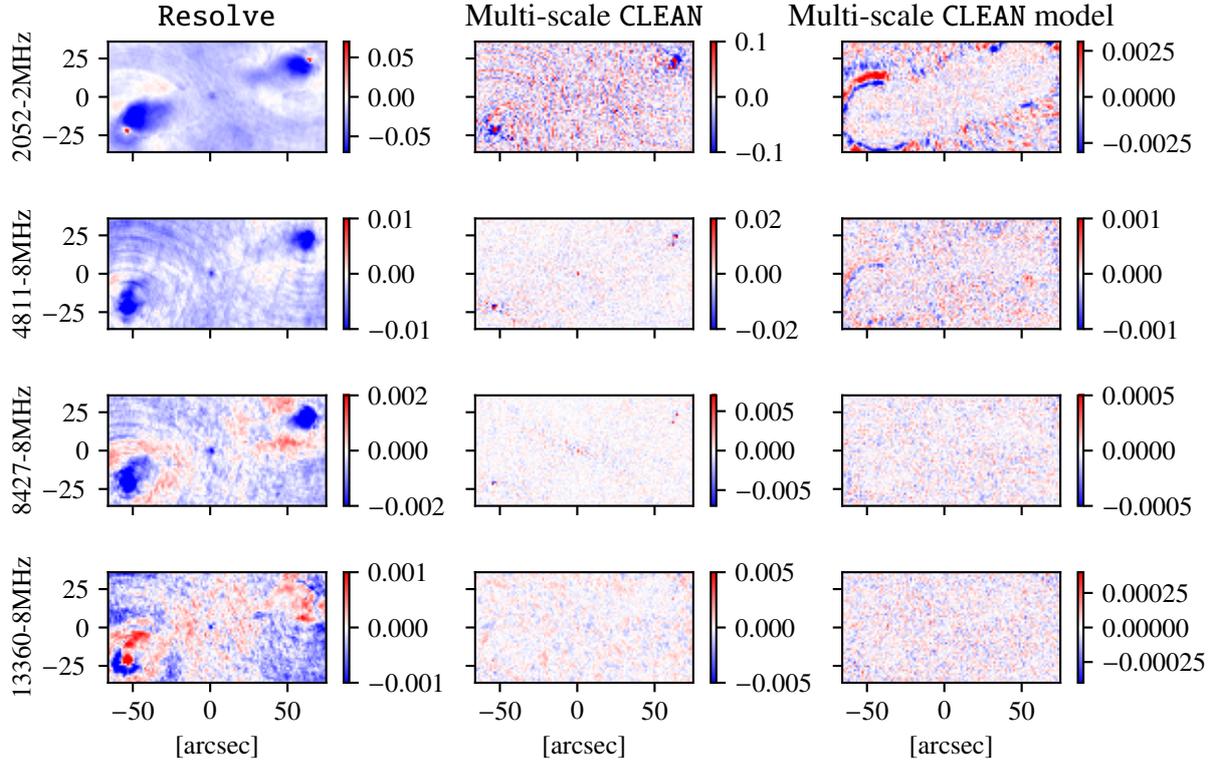}
  \caption{Residual maps.
    The first and second column display residual maps computed with the Bayesian weights.
    The third column displays the residual map for the \msclean\ model image with \wsclean\ weighting.
    All colour bars have the unit Jy and are defined to be symmetric around zero with maximum five times the median of the absolute values of each image individually.
    The sign of the residual maps is determined by the r.h.s.\ of \cref{eq:likgrad}.
  }
  \label{fig:residualmaps}
\end{figure*}

\begin{figure*}
  \centering
  \input{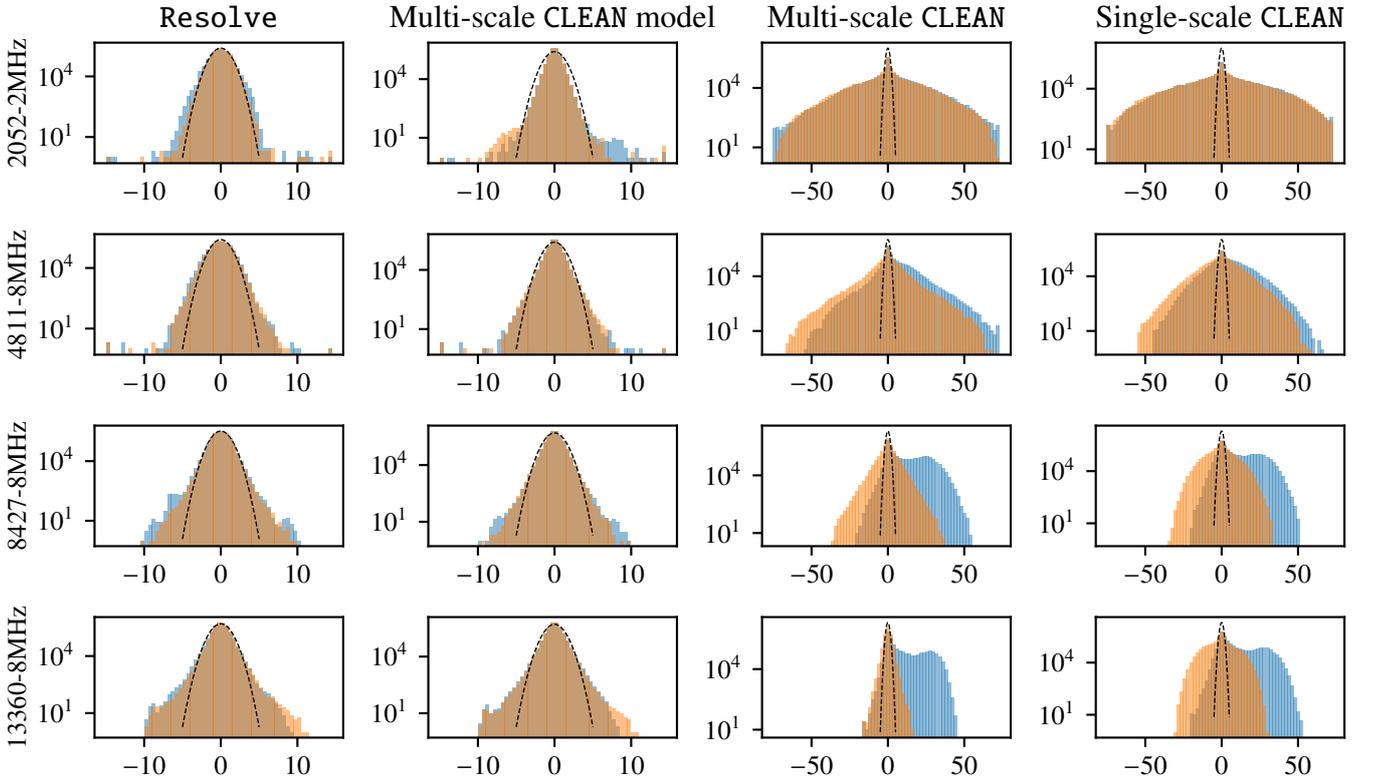}
  \caption{Histogram of (posterior) residuals weighted with $\sigma(\xisig)$, i.e.\ both the thermal noise and the Bayesian weighting scheme.
    Blue and orange bars denote real and imaginary parts, respectively.
    The black dotted line displays a standard normal Gaussian distribution scaled to the number of data points.
    For \msclean\ the residuals for both the model and restored image are shown.
    Histgram counts outside the displayed range are shown in the left- and rightmost bin.
  }\label{fig:histogramresolvewgts}
\end{figure*}

\begin{figure*}
  \centering
  \input{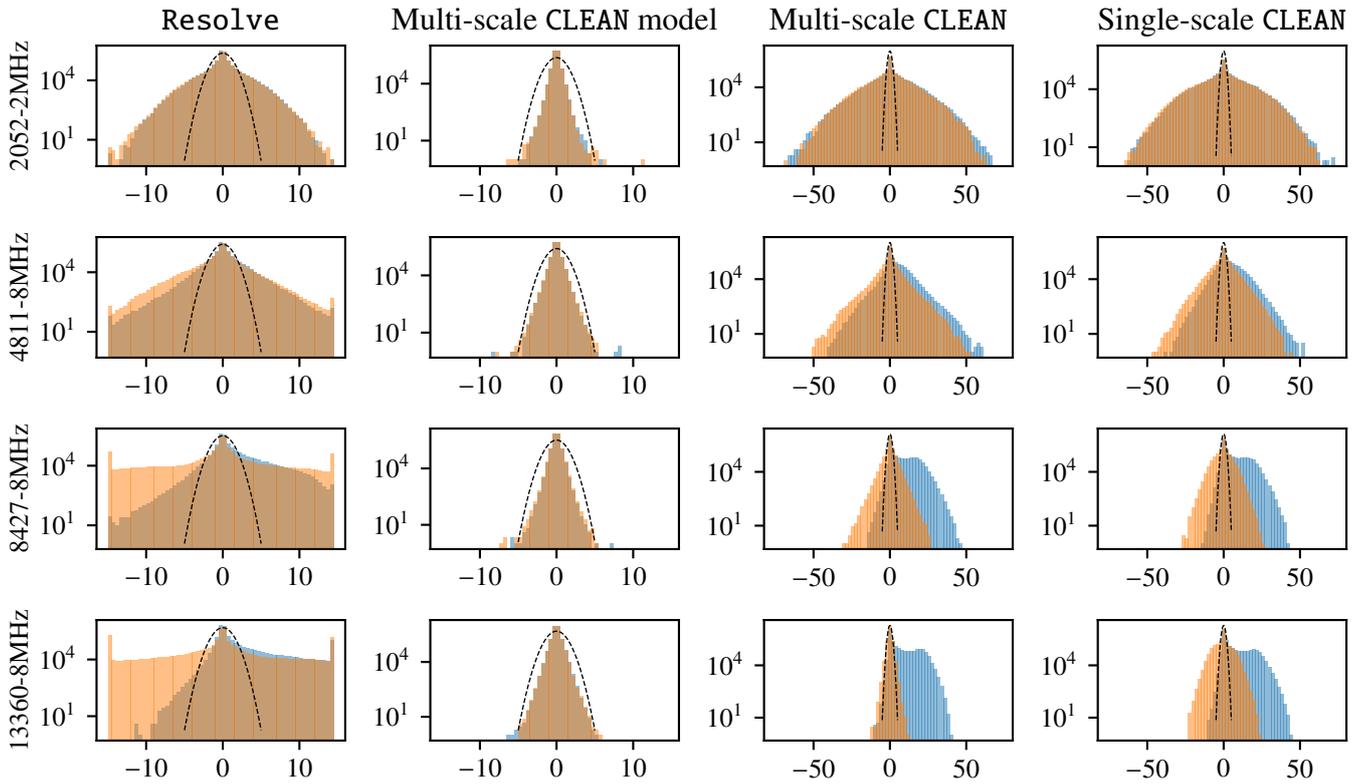}
  \caption{Histogram of noise-weighted (posterior) residuals weighted with \wsclean\ weighting scheme, i.e.\ both the thermal noise and the imaging weighting scheme employed by \wsclean.
    This weighting scheme has been used for the \msclean\ reconstruction.
    The histograms are plotted analogously to \cref{fig:histogramresolvewgts}.
  }\label{fig:histogramcleanwgts}
\end{figure*}

\begin{table*}
  \centering
  \input{chisq}
  \caption{Reduced $\chi^2$ values of all reconstructions weighted with the Bayesian $\sigma(\xisig)$ and the \wsclean\ weighting scheme.
    The first and the second value of each table entry correspond to the reduced $\chi^2$ value of the real and imaginary part of the residual, respectively.
    The latter has been used for the \msclean\ reconstruction.
    These $\chi^2$ values are in direct correspondence to the histograms displayed in \cref{fig:histogramresolvewgts,fig:histogramcleanwgts}.
    Some values are greyed out in order to emphasize the weighting that has been applied for the \resolve\ and the \msclean\ reconstruction.}
  \label{tab:chisq}
\end{table*}

\begin{figure*}
  \centering
  \input{overview_all_2.pgf}
  \caption{Same As \cref{fig:all}, but with a saturated colour bar.
    The colour bar has units $\fluxunit$.}\label{fig:all2}
\end{figure*}

\begin{figure*}
  \centering
  \input{overview_contour_2052_all.pgf}
  \caption{Comparison \msclean\ (blue, negative regions grey), \resolve\ posterior mean (orange), 2052 MHz, contour lines have multiplicative distances of $\sqrt{2}$.}\label{fig:contour_2052_all}
\end{figure*}
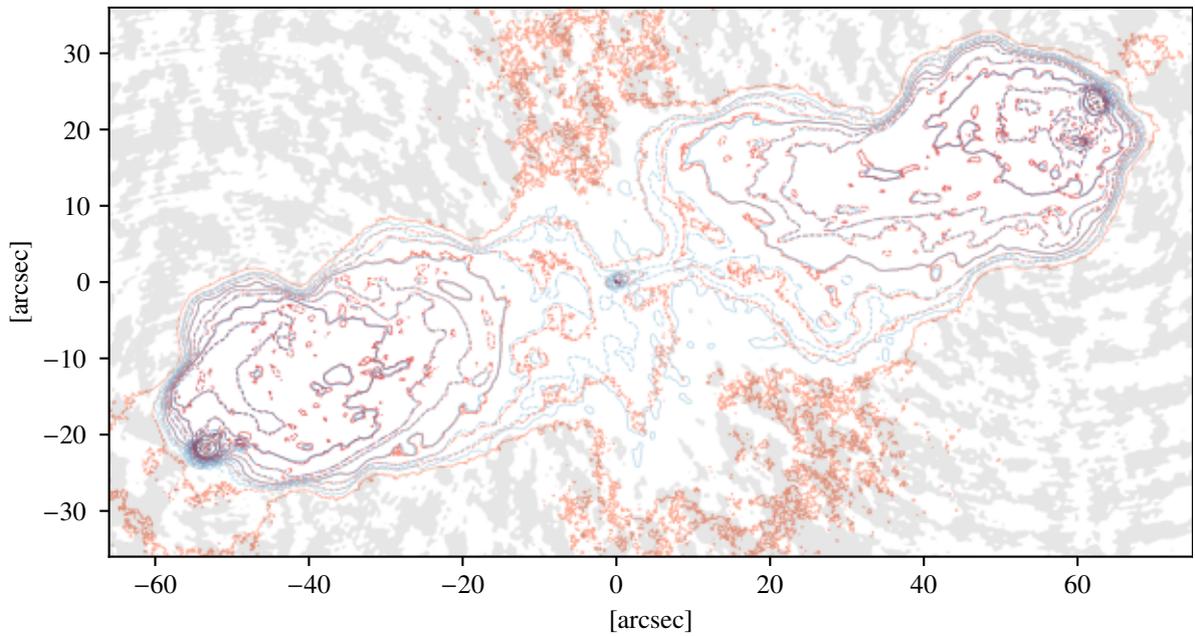

\begin{figure*}
  \centering
  \input{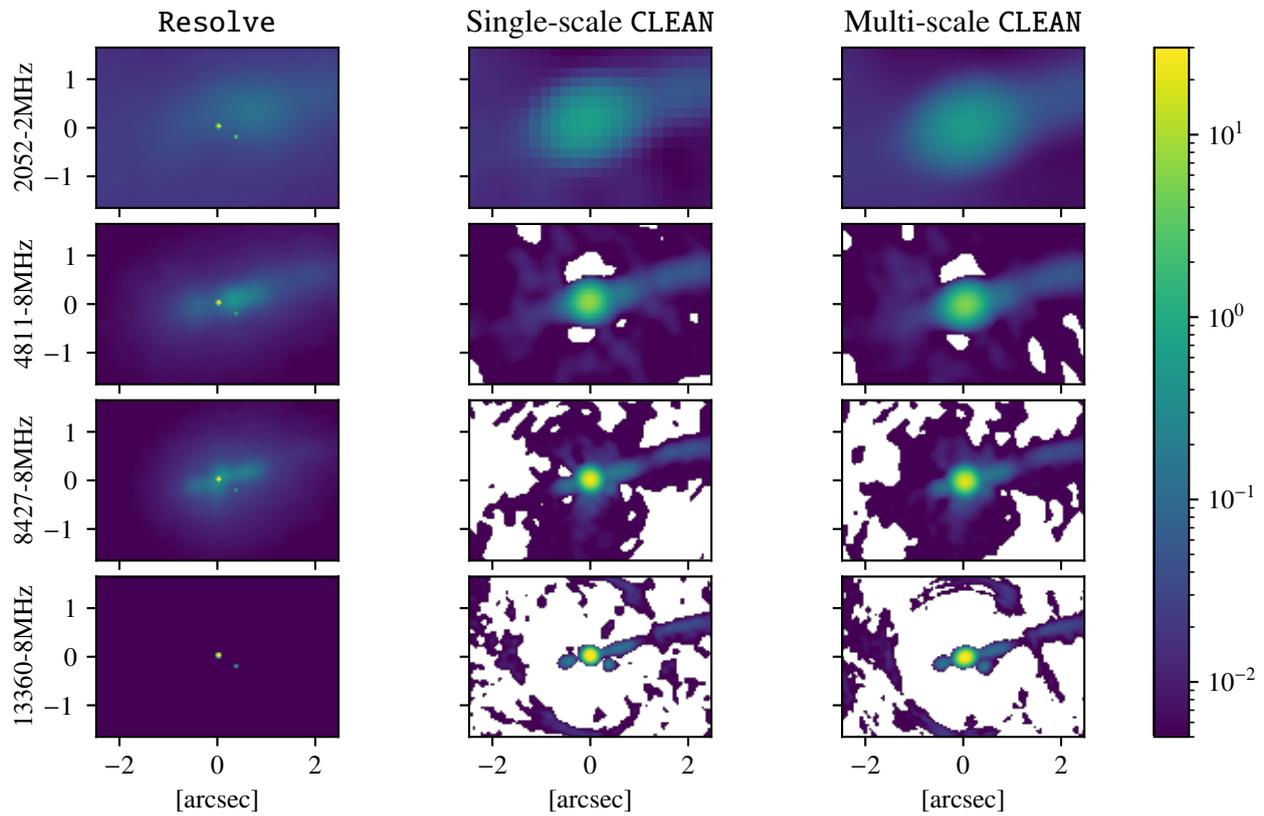}
  \caption{Overview of imaging results zoomed in to central source.
    The top row shows the \resolve\ posterior mean, the middle and last row show \ssclean\ \msclean\ results, respectively.
    The colour bar has units Jy/arcsec.
    Negative flux regions are displayed in white.}\label{fig:center}
\end{figure*}

\end{document}

%% file: stdcorr_min.tex
0.4

%% file: stdcorr_max.tex
429

%% file: points.tex
\begin{tabular}{lrr}
\hline\hline Frequency [GHz] & Source 0 [mJy] & Source 1 [mJy]\\\hline
2.052 & $585 \pm 7$ & $17 \pm 3$ \\ 4.811 & $1166.3 \pm 0.9$ & $5.5 \pm 0.8$ \\ 8.427 & $1440.4 \pm 0.7$ & $3.5 \pm 0.2$ \\ 13.36 & $1601.49 \pm 0.03$ & $4.5 \pm 0.1$ \\ \hline \end{tabular}

%% file: bigg0.pgf
\begingroup%
\makeatletter%
\begin{pgfpicture}%
\pgfpathrectangle{\pgfpointorigin}{\pgfqpoint{7.242283in}{9.052854in}}%
\pgfusepath{use as bounding box, clip}%
\begin{pgfscope}%
\pgfsetbuttcap%
\pgfsetmiterjoin%
\definecolor{currentfill}{rgb}{1.000000,1.000000,1.000000}%
\pgfsetfillcolor{currentfill}%
\pgfsetlinewidth{0.000000pt}%
\definecolor{currentstroke}{rgb}{1.000000,1.000000,1.000000}%
\pgfsetstrokecolor{currentstroke}%
\pgfsetdash{}{0pt}%
\pgfpathmoveto{\pgfqpoint{0.000000in}{0.000000in}}%
\pgfpathlineto{\pgfqpoint{7.242283in}{0.000000in}}%
\pgfpathlineto{\pgfqpoint{7.242283in}{9.052854in}}%
\pgfpathlineto{\pgfqpoint{0.000000in}{9.052854in}}%
\pgfpathclose%
\pgfusepath{fill}%
\end{pgfscope}%
\begin{pgfscope}%
\pgfpathrectangle{\pgfqpoint{0.000000in}{0.000000in}}{\pgfqpoint{7.242283in}{9.052854in}}%
\pgfusepath{clip}%
\pgfsys@transformshift{0.000000in}{0.000000in}%
\pgftext[left,bottom]{\includegraphics[interpolate=true,width=7.250000in,height=9.060000in]{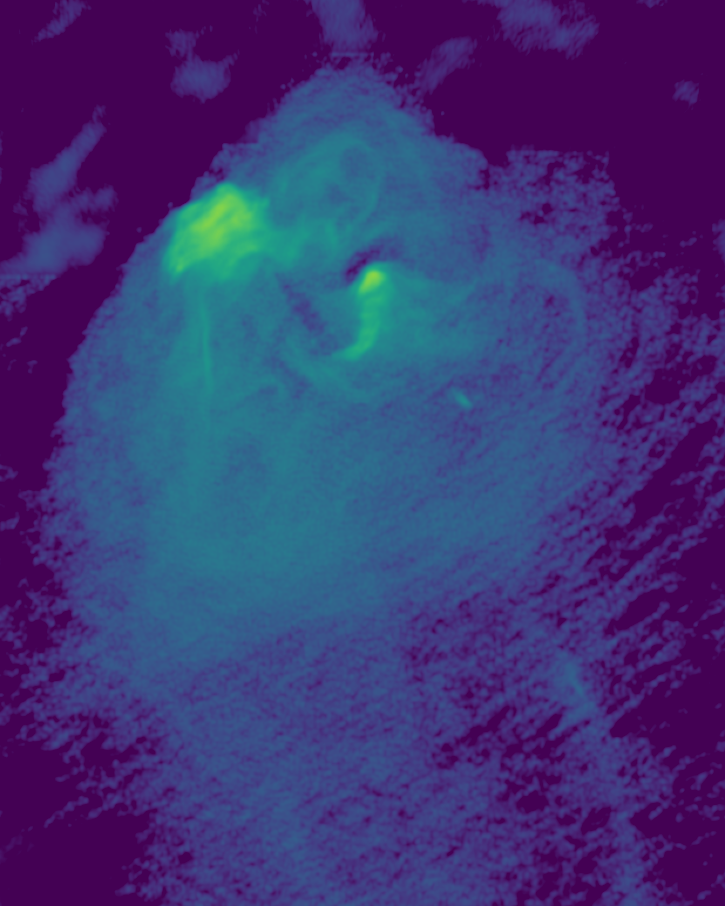}}%
\end{pgfscope}%
\end{pgfpicture}%
\makeatother%
\endgroup%

%% file: bigg1.pgf
\begingroup%
\makeatletter%
\begin{pgfpicture}%
\pgfpathrectangle{\pgfpointorigin}{\pgfqpoint{7.242283in}{9.052854in}}%
\pgfusepath{use as bounding box, clip}%
\begin{pgfscope}%
\pgfsetbuttcap%
\pgfsetmiterjoin%
\definecolor{currentfill}{rgb}{1.000000,1.000000,1.000000}%
\pgfsetfillcolor{currentfill}%
\pgfsetlinewidth{0.000000pt}%
\definecolor{currentstroke}{rgb}{1.000000,1.000000,1.000000}%
\pgfsetstrokecolor{currentstroke}%
\pgfsetdash{}{0pt}%
\pgfpathmoveto{\pgfqpoint{0.000000in}{0.000000in}}%
\pgfpathlineto{\pgfqpoint{7.242283in}{0.000000in}}%
\pgfpathlineto{\pgfqpoint{7.242283in}{9.052854in}}%
\pgfpathlineto{\pgfqpoint{0.000000in}{9.052854in}}%
\pgfpathclose%
\pgfusepath{fill}%
\end{pgfscope}%
\begin{pgfscope}%
\pgfpathrectangle{\pgfqpoint{0.000000in}{0.000000in}}{\pgfqpoint{7.242283in}{9.052854in}}%
\pgfusepath{clip}%
\pgfsys@transformshift{0.000000in}{0.000000in}%
\pgftext[left,bottom]{\includegraphics[interpolate=true,width=7.250000in,height=9.060000in]{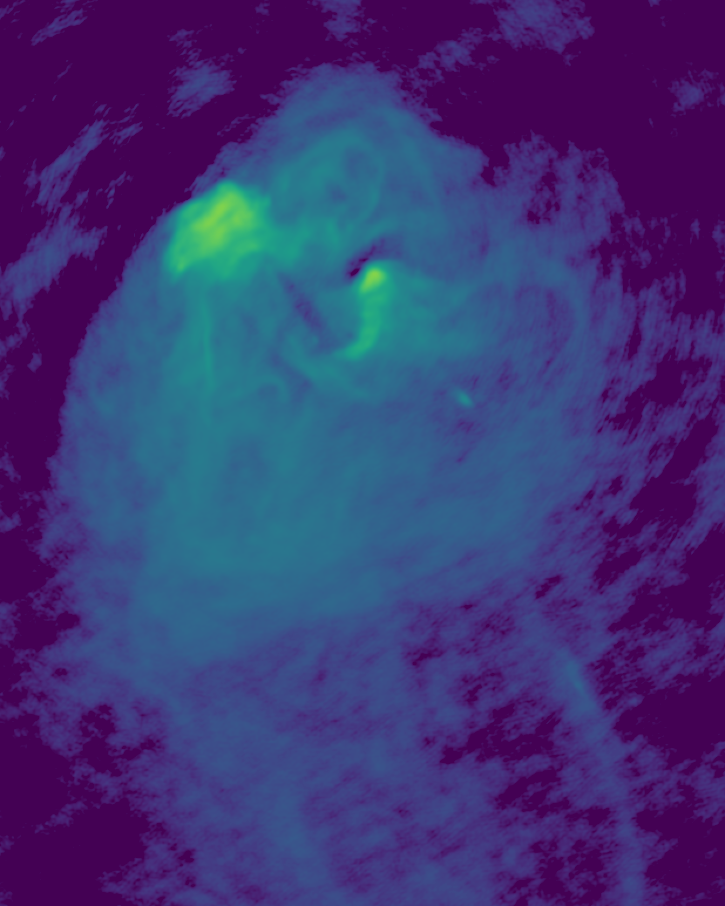}}%
\end{pgfscope}%
\end{pgfpicture}%
\makeatother%
\endgroup%

%% file: bigg2.pgf
\begingroup%
\makeatletter%
\begin{pgfpicture}%
\pgfpathrectangle{\pgfpointorigin}{\pgfqpoint{7.242283in}{9.052854in}}%
\pgfusepath{use as bounding box, clip}%
\begin{pgfscope}%
\pgfsetbuttcap%
\pgfsetmiterjoin%
\definecolor{currentfill}{rgb}{1.000000,1.000000,1.000000}%
\pgfsetfillcolor{currentfill}%
\pgfsetlinewidth{0.000000pt}%
\definecolor{currentstroke}{rgb}{1.000000,1.000000,1.000000}%
\pgfsetstrokecolor{currentstroke}%
\pgfsetdash{}{0pt}%
\pgfpathmoveto{\pgfqpoint{0.000000in}{0.000000in}}%
\pgfpathlineto{\pgfqpoint{7.242283in}{0.000000in}}%
\pgfpathlineto{\pgfqpoint{7.242283in}{9.052854in}}%
\pgfpathlineto{\pgfqpoint{0.000000in}{9.052854in}}%
\pgfpathclose%
\pgfusepath{fill}%
\end{pgfscope}%
\begin{pgfscope}%
\pgfpathrectangle{\pgfqpoint{0.000000in}{0.000000in}}{\pgfqpoint{7.242283in}{9.052854in}}%
\pgfusepath{clip}%
\pgfsys@transformshift{0.000000in}{0.000000in}%
\pgftext[left,bottom]{\includegraphics[interpolate=true,width=7.250000in,height=9.060000in]{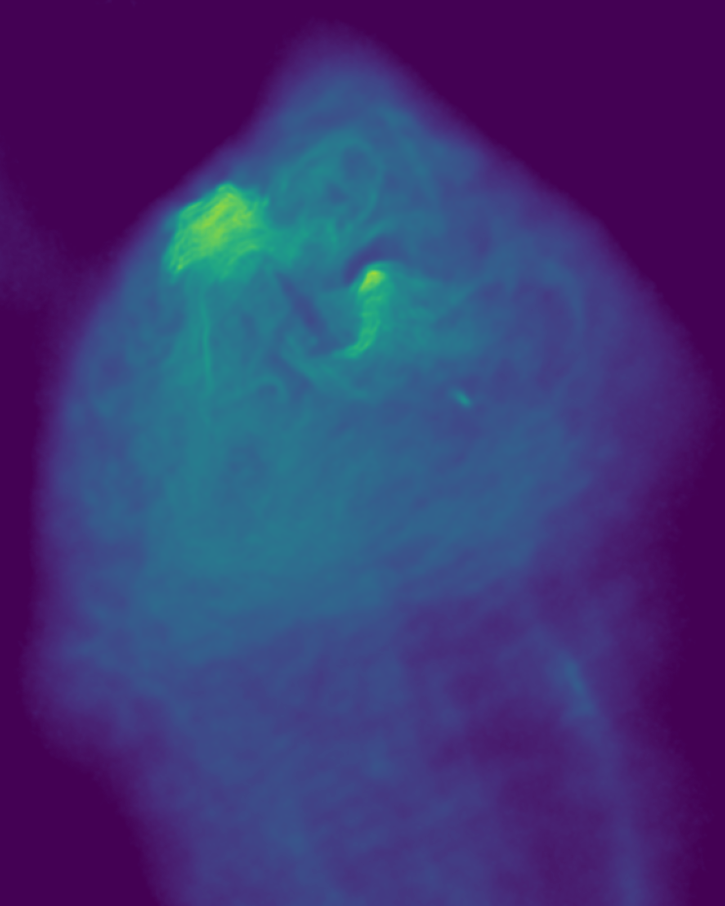}}%
\end{pgfscope}%
\end{pgfpicture}%
\makeatother%
\endgroup%

%% file: overview_contour_13360_leftverysmall.pgf
\begingroup%
\makeatletter%
\begin{pgfpicture}%
\pgfpathrectangle{\pgfpointorigin}{\pgfqpoint{6.260779in}{4.157905in}}%
\pgfusepath{use as bounding box, clip}%
\begin{pgfscope}%
\pgfsetbuttcap%
\pgfsetmiterjoin%
\definecolor{currentfill}{rgb}{1.000000,1.000000,1.000000}%
\pgfsetfillcolor{currentfill}%
\pgfsetlinewidth{0.000000pt}%
\definecolor{currentstroke}{rgb}{1.000000,1.000000,1.000000}%
\pgfsetstrokecolor{currentstroke}%
\pgfsetdash{}{0pt}%
\pgfpathmoveto{\pgfqpoint{0.000000in}{0.000000in}}%
\pgfpathlineto{\pgfqpoint{6.260779in}{0.000000in}}%
\pgfpathlineto{\pgfqpoint{6.260779in}{4.157905in}}%
\pgfpathlineto{\pgfqpoint{0.000000in}{4.157905in}}%
\pgfpathclose%
\pgfusepath{fill}%
\end{pgfscope}%
\begin{pgfscope}%
\pgfsys@transformshift{0.650000in}{0.427905in}%
\pgftext[left,bottom]{\includegraphics[interpolate=true,width=5.610000in,height=3.730000in]{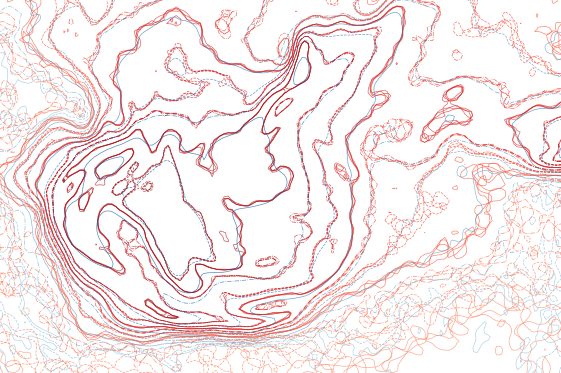}}%
\end{pgfscope}%
\begin{pgfscope}%
\pgfpathrectangle{\pgfqpoint{0.646606in}{0.415123in}}{\pgfqpoint{5.614173in}{3.742782in}}%
\pgfusepath{clip}%
\pgfsys@transformshift{0.646606in}{0.415123in}%
\pgftext[left,bottom]{\includegraphics[interpolate=true,width=5.620000in,height=3.750000in]{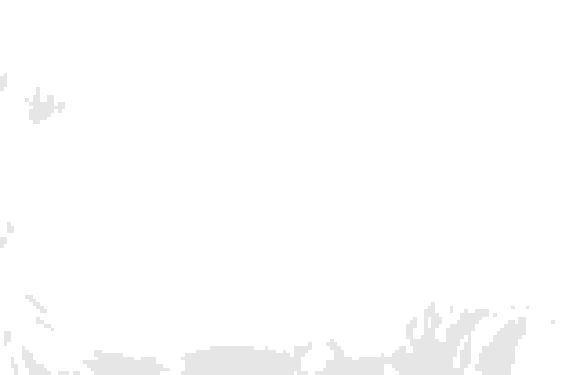}}%
\end{pgfscope}%
\begin{pgfscope}%
\pgfsetbuttcap%
\pgfsetroundjoin%
\definecolor{currentfill}{rgb}{0.000000,0.000000,0.000000}%
\pgfsetfillcolor{currentfill}%
\pgfsetlinewidth{0.803000pt}%
\definecolor{currentstroke}{rgb}{0.000000,0.000000,0.000000}%
\pgfsetstrokecolor{currentstroke}%
\pgfsetdash{}{0pt}%
\pgfsys@defobject{currentmarker}{\pgfqpoint{0.000000in}{-0.048611in}}{\pgfqpoint{0.000000in}{0.000000in}}{%
\pgfpathmoveto{\pgfqpoint{0.000000in}{0.000000in}}%
\pgfpathlineto{\pgfqpoint{0.000000in}{-0.048611in}}%
\pgfusepath{stroke,fill}%
}%
\begin{pgfscope}%
\pgfsys@transformshift{0.899707in}{0.415123in}%
\pgfsys@useobject{currentmarker}{}%
\end{pgfscope}%
\end{pgfscope}%
\begin{pgfscope}%
\definecolor{textcolor}{rgb}{0.000000,0.000000,0.000000}%
\pgfsetstrokecolor{textcolor}%
\pgfsetfillcolor{textcolor}%
\pgftext[x=0.899707in,y=0.317901in,,top]{\color{textcolor}\rmfamily\fontsize{10.000000}{12.000000}\selectfont \(\displaystyle {-55}\)}%
\end{pgfscope}%
\begin{pgfscope}%
\pgfsetbuttcap%
\pgfsetroundjoin%
\definecolor{currentfill}{rgb}{0.000000,0.000000,0.000000}%
\pgfsetfillcolor{currentfill}%
\pgfsetlinewidth{0.803000pt}%
\definecolor{currentstroke}{rgb}{0.000000,0.000000,0.000000}%
\pgfsetstrokecolor{currentstroke}%
\pgfsetdash{}{0pt}%
\pgfsys@defobject{currentmarker}{\pgfqpoint{0.000000in}{-0.048611in}}{\pgfqpoint{0.000000in}{0.000000in}}{%
\pgfpathmoveto{\pgfqpoint{0.000000in}{0.000000in}}%
\pgfpathlineto{\pgfqpoint{0.000000in}{-0.048611in}}%
\pgfusepath{stroke,fill}%
}%
\begin{pgfscope}%
\pgfsys@transformshift{1.806983in}{0.415123in}%
\pgfsys@useobject{currentmarker}{}%
\end{pgfscope}%
\end{pgfscope}%
\begin{pgfscope}%
\definecolor{textcolor}{rgb}{0.000000,0.000000,0.000000}%
\pgfsetstrokecolor{textcolor}%
\pgfsetfillcolor{textcolor}%
\pgftext[x=1.806983in,y=0.317901in,,top]{\color{textcolor}\rmfamily\fontsize{10.000000}{12.000000}\selectfont \(\displaystyle {-54}\)}%
\end{pgfscope}%
\begin{pgfscope}%
\pgfsetbuttcap%
\pgfsetroundjoin%
\definecolor{currentfill}{rgb}{0.000000,0.000000,0.000000}%
\pgfsetfillcolor{currentfill}%
\pgfsetlinewidth{0.803000pt}%
\definecolor{currentstroke}{rgb}{0.000000,0.000000,0.000000}%
\pgfsetstrokecolor{currentstroke}%
\pgfsetdash{}{0pt}%
\pgfsys@defobject{currentmarker}{\pgfqpoint{0.000000in}{-0.048611in}}{\pgfqpoint{0.000000in}{0.000000in}}{%
\pgfpathmoveto{\pgfqpoint{0.000000in}{0.000000in}}%
\pgfpathlineto{\pgfqpoint{0.000000in}{-0.048611in}}%
\pgfusepath{stroke,fill}%
}%
\begin{pgfscope}%
\pgfsys@transformshift{2.714259in}{0.415123in}%
\pgfsys@useobject{currentmarker}{}%
\end{pgfscope}%
\end{pgfscope}%
\begin{pgfscope}%
\definecolor{textcolor}{rgb}{0.000000,0.000000,0.000000}%
\pgfsetstrokecolor{textcolor}%
\pgfsetfillcolor{textcolor}%
\pgftext[x=2.714259in,y=0.317901in,,top]{\color{textcolor}\rmfamily\fontsize{10.000000}{12.000000}\selectfont \(\displaystyle {-53}\)}%
\end{pgfscope}%
\begin{pgfscope}%
\pgfsetbuttcap%
\pgfsetroundjoin%
\definecolor{currentfill}{rgb}{0.000000,0.000000,0.000000}%
\pgfsetfillcolor{currentfill}%
\pgfsetlinewidth{0.803000pt}%
\definecolor{currentstroke}{rgb}{0.000000,0.000000,0.000000}%
\pgfsetstrokecolor{currentstroke}%
\pgfsetdash{}{0pt}%
\pgfsys@defobject{currentmarker}{\pgfqpoint{0.000000in}{-0.048611in}}{\pgfqpoint{0.000000in}{0.000000in}}{%
\pgfpathmoveto{\pgfqpoint{0.000000in}{0.000000in}}%
\pgfpathlineto{\pgfqpoint{0.000000in}{-0.048611in}}%
\pgfusepath{stroke,fill}%
}%
\begin{pgfscope}%
\pgfsys@transformshift{3.621535in}{0.415123in}%
\pgfsys@useobject{currentmarker}{}%
\end{pgfscope}%
\end{pgfscope}%
\begin{pgfscope}%
\definecolor{textcolor}{rgb}{0.000000,0.000000,0.000000}%
\pgfsetstrokecolor{textcolor}%
\pgfsetfillcolor{textcolor}%
\pgftext[x=3.621535in,y=0.317901in,,top]{\color{textcolor}\rmfamily\fontsize{10.000000}{12.000000}\selectfont \(\displaystyle {-52}\)}%
\end{pgfscope}%
\begin{pgfscope}%
\pgfsetbuttcap%
\pgfsetroundjoin%
\definecolor{currentfill}{rgb}{0.000000,0.000000,0.000000}%
\pgfsetfillcolor{currentfill}%
\pgfsetlinewidth{0.803000pt}%
\definecolor{currentstroke}{rgb}{0.000000,0.000000,0.000000}%
\pgfsetstrokecolor{currentstroke}%
\pgfsetdash{}{0pt}%
\pgfsys@defobject{currentmarker}{\pgfqpoint{0.000000in}{-0.048611in}}{\pgfqpoint{0.000000in}{0.000000in}}{%
\pgfpathmoveto{\pgfqpoint{0.000000in}{0.000000in}}%
\pgfpathlineto{\pgfqpoint{0.000000in}{-0.048611in}}%
\pgfusepath{stroke,fill}%
}%
\begin{pgfscope}%
\pgfsys@transformshift{4.528811in}{0.415123in}%
\pgfsys@useobject{currentmarker}{}%
\end{pgfscope}%
\end{pgfscope}%
\begin{pgfscope}%
\definecolor{textcolor}{rgb}{0.000000,0.000000,0.000000}%
\pgfsetstrokecolor{textcolor}%
\pgfsetfillcolor{textcolor}%
\pgftext[x=4.528811in,y=0.317901in,,top]{\color{textcolor}\rmfamily\fontsize{10.000000}{12.000000}\selectfont \(\displaystyle {-51}\)}%
\end{pgfscope}%
\begin{pgfscope}%
\pgfsetbuttcap%
\pgfsetroundjoin%
\definecolor{currentfill}{rgb}{0.000000,0.000000,0.000000}%
\pgfsetfillcolor{currentfill}%
\pgfsetlinewidth{0.803000pt}%
\definecolor{currentstroke}{rgb}{0.000000,0.000000,0.000000}%
\pgfsetstrokecolor{currentstroke}%
\pgfsetdash{}{0pt}%
\pgfsys@defobject{currentmarker}{\pgfqpoint{0.000000in}{-0.048611in}}{\pgfqpoint{0.000000in}{0.000000in}}{%
\pgfpathmoveto{\pgfqpoint{0.000000in}{0.000000in}}%
\pgfpathlineto{\pgfqpoint{0.000000in}{-0.048611in}}%
\pgfusepath{stroke,fill}%
}%
\begin{pgfscope}%
\pgfsys@transformshift{5.436087in}{0.415123in}%
\pgfsys@useobject{currentmarker}{}%
\end{pgfscope}%
\end{pgfscope}%
\begin{pgfscope}%
\definecolor{textcolor}{rgb}{0.000000,0.000000,0.000000}%
\pgfsetstrokecolor{textcolor}%
\pgfsetfillcolor{textcolor}%
\pgftext[x=5.436087in,y=0.317901in,,top]{\color{textcolor}\rmfamily\fontsize{10.000000}{12.000000}\selectfont \(\displaystyle {-50}\)}%
\end{pgfscope}%
\begin{pgfscope}%
\definecolor{textcolor}{rgb}{0.000000,0.000000,0.000000}%
\pgfsetstrokecolor{textcolor}%
\pgfsetfillcolor{textcolor}%
\pgftext[x=3.453693in,y=0.138889in,,top]{\color{textcolor}\rmfamily\fontsize{10.000000}{12.000000}\selectfont [arcsec]}%
\end{pgfscope}%
\begin{pgfscope}%
\pgfsetbuttcap%
\pgfsetroundjoin%
\definecolor{currentfill}{rgb}{0.000000,0.000000,0.000000}%
\pgfsetfillcolor{currentfill}%
\pgfsetlinewidth{0.803000pt}%
\definecolor{currentstroke}{rgb}{0.000000,0.000000,0.000000}%
\pgfsetstrokecolor{currentstroke}%
\pgfsetdash{}{0pt}%
\pgfsys@defobject{currentmarker}{\pgfqpoint{-0.048611in}{0.000000in}}{\pgfqpoint{-0.000000in}{0.000000in}}{%
\pgfpathmoveto{\pgfqpoint{-0.000000in}{0.000000in}}%
\pgfpathlineto{\pgfqpoint{-0.048611in}{0.000000in}}%
\pgfusepath{stroke,fill}%
}%
\begin{pgfscope}%
\pgfsys@transformshift{0.646606in}{0.427996in}%
\pgfsys@useobject{currentmarker}{}%
\end{pgfscope}%
\end{pgfscope}%
\begin{pgfscope}%
\definecolor{textcolor}{rgb}{0.000000,0.000000,0.000000}%
\pgfsetstrokecolor{textcolor}%
\pgfsetfillcolor{textcolor}%
\pgftext[x=0.194444in, y=0.379770in, left, base]{\color{textcolor}\rmfamily\fontsize{10.000000}{12.000000}\selectfont \(\displaystyle {-23.5}\)}%
\end{pgfscope}%
\begin{pgfscope}%
\pgfsetbuttcap%
\pgfsetroundjoin%
\definecolor{currentfill}{rgb}{0.000000,0.000000,0.000000}%
\pgfsetfillcolor{currentfill}%
\pgfsetlinewidth{0.803000pt}%
\definecolor{currentstroke}{rgb}{0.000000,0.000000,0.000000}%
\pgfsetstrokecolor{currentstroke}%
\pgfsetdash{}{0pt}%
\pgfsys@defobject{currentmarker}{\pgfqpoint{-0.048611in}{0.000000in}}{\pgfqpoint{-0.000000in}{0.000000in}}{%
\pgfpathmoveto{\pgfqpoint{-0.000000in}{0.000000in}}%
\pgfpathlineto{\pgfqpoint{-0.048611in}{0.000000in}}%
\pgfusepath{stroke,fill}%
}%
\begin{pgfscope}%
\pgfsys@transformshift{0.646606in}{0.881634in}%
\pgfsys@useobject{currentmarker}{}%
\end{pgfscope}%
\end{pgfscope}%
\begin{pgfscope}%
\definecolor{textcolor}{rgb}{0.000000,0.000000,0.000000}%
\pgfsetstrokecolor{textcolor}%
\pgfsetfillcolor{textcolor}%
\pgftext[x=0.194444in, y=0.833408in, left, base]{\color{textcolor}\rmfamily\fontsize{10.000000}{12.000000}\selectfont \(\displaystyle {-23.0}\)}%
\end{pgfscope}%
\begin{pgfscope}%
\pgfsetbuttcap%
\pgfsetroundjoin%
\definecolor{currentfill}{rgb}{0.000000,0.000000,0.000000}%
\pgfsetfillcolor{currentfill}%
\pgfsetlinewidth{0.803000pt}%
\definecolor{currentstroke}{rgb}{0.000000,0.000000,0.000000}%
\pgfsetstrokecolor{currentstroke}%
\pgfsetdash{}{0pt}%
\pgfsys@defobject{currentmarker}{\pgfqpoint{-0.048611in}{0.000000in}}{\pgfqpoint{-0.000000in}{0.000000in}}{%
\pgfpathmoveto{\pgfqpoint{-0.000000in}{0.000000in}}%
\pgfpathlineto{\pgfqpoint{-0.048611in}{0.000000in}}%
\pgfusepath{stroke,fill}%
}%
\begin{pgfscope}%
\pgfsys@transformshift{0.646606in}{1.335272in}%
\pgfsys@useobject{currentmarker}{}%
\end{pgfscope}%
\end{pgfscope}%
\begin{pgfscope}%
\definecolor{textcolor}{rgb}{0.000000,0.000000,0.000000}%
\pgfsetstrokecolor{textcolor}%
\pgfsetfillcolor{textcolor}%
\pgftext[x=0.194444in, y=1.287046in, left, base]{\color{textcolor}\rmfamily\fontsize{10.000000}{12.000000}\selectfont \(\displaystyle {-22.5}\)}%
\end{pgfscope}%
\begin{pgfscope}%
\pgfsetbuttcap%
\pgfsetroundjoin%
\definecolor{currentfill}{rgb}{0.000000,0.000000,0.000000}%
\pgfsetfillcolor{currentfill}%
\pgfsetlinewidth{0.803000pt}%
\definecolor{currentstroke}{rgb}{0.000000,0.000000,0.000000}%
\pgfsetstrokecolor{currentstroke}%
\pgfsetdash{}{0pt}%
\pgfsys@defobject{currentmarker}{\pgfqpoint{-0.048611in}{0.000000in}}{\pgfqpoint{-0.000000in}{0.000000in}}{%
\pgfpathmoveto{\pgfqpoint{-0.000000in}{0.000000in}}%
\pgfpathlineto{\pgfqpoint{-0.048611in}{0.000000in}}%
\pgfusepath{stroke,fill}%
}%
\begin{pgfscope}%
\pgfsys@transformshift{0.646606in}{1.788910in}%
\pgfsys@useobject{currentmarker}{}%
\end{pgfscope}%
\end{pgfscope}%
\begin{pgfscope}%
\definecolor{textcolor}{rgb}{0.000000,0.000000,0.000000}%
\pgfsetstrokecolor{textcolor}%
\pgfsetfillcolor{textcolor}%
\pgftext[x=0.194444in, y=1.740684in, left, base]{\color{textcolor}\rmfamily\fontsize{10.000000}{12.000000}\selectfont \(\displaystyle {-22.0}\)}%
\end{pgfscope}%
\begin{pgfscope}%
\pgfsetbuttcap%
\pgfsetroundjoin%
\definecolor{currentfill}{rgb}{0.000000,0.000000,0.000000}%
\pgfsetfillcolor{currentfill}%
\pgfsetlinewidth{0.803000pt}%
\definecolor{currentstroke}{rgb}{0.000000,0.000000,0.000000}%
\pgfsetstrokecolor{currentstroke}%
\pgfsetdash{}{0pt}%
\pgfsys@defobject{currentmarker}{\pgfqpoint{-0.048611in}{0.000000in}}{\pgfqpoint{-0.000000in}{0.000000in}}{%
\pgfpathmoveto{\pgfqpoint{-0.000000in}{0.000000in}}%
\pgfpathlineto{\pgfqpoint{-0.048611in}{0.000000in}}%
\pgfusepath{stroke,fill}%
}%
\begin{pgfscope}%
\pgfsys@transformshift{0.646606in}{2.242548in}%
\pgfsys@useobject{currentmarker}{}%
\end{pgfscope}%
\end{pgfscope}%
\begin{pgfscope}%
\definecolor{textcolor}{rgb}{0.000000,0.000000,0.000000}%
\pgfsetstrokecolor{textcolor}%
\pgfsetfillcolor{textcolor}%
\pgftext[x=0.194444in, y=2.194322in, left, base]{\color{textcolor}\rmfamily\fontsize{10.000000}{12.000000}\selectfont \(\displaystyle {-21.5}\)}%
\end{pgfscope}%
\begin{pgfscope}%
\pgfsetbuttcap%
\pgfsetroundjoin%
\definecolor{currentfill}{rgb}{0.000000,0.000000,0.000000}%
\pgfsetfillcolor{currentfill}%
\pgfsetlinewidth{0.803000pt}%
\definecolor{currentstroke}{rgb}{0.000000,0.000000,0.000000}%
\pgfsetstrokecolor{currentstroke}%
\pgfsetdash{}{0pt}%
\pgfsys@defobject{currentmarker}{\pgfqpoint{-0.048611in}{0.000000in}}{\pgfqpoint{-0.000000in}{0.000000in}}{%
\pgfpathmoveto{\pgfqpoint{-0.000000in}{0.000000in}}%
\pgfpathlineto{\pgfqpoint{-0.048611in}{0.000000in}}%
\pgfusepath{stroke,fill}%
}%
\begin{pgfscope}%
\pgfsys@transformshift{0.646606in}{2.696186in}%
\pgfsys@useobject{currentmarker}{}%
\end{pgfscope}%
\end{pgfscope}%
\begin{pgfscope}%
\definecolor{textcolor}{rgb}{0.000000,0.000000,0.000000}%
\pgfsetstrokecolor{textcolor}%
\pgfsetfillcolor{textcolor}%
\pgftext[x=0.194444in, y=2.647960in, left, base]{\color{textcolor}\rmfamily\fontsize{10.000000}{12.000000}\selectfont \(\displaystyle {-21.0}\)}%
\end{pgfscope}%
\begin{pgfscope}%
\pgfsetbuttcap%
\pgfsetroundjoin%
\definecolor{currentfill}{rgb}{0.000000,0.000000,0.000000}%
\pgfsetfillcolor{currentfill}%
\pgfsetlinewidth{0.803000pt}%
\definecolor{currentstroke}{rgb}{0.000000,0.000000,0.000000}%
\pgfsetstrokecolor{currentstroke}%
\pgfsetdash{}{0pt}%
\pgfsys@defobject{currentmarker}{\pgfqpoint{-0.048611in}{0.000000in}}{\pgfqpoint{-0.000000in}{0.000000in}}{%
\pgfpathmoveto{\pgfqpoint{-0.000000in}{0.000000in}}%
\pgfpathlineto{\pgfqpoint{-0.048611in}{0.000000in}}%
\pgfusepath{stroke,fill}%
}%
\begin{pgfscope}%
\pgfsys@transformshift{0.646606in}{3.149824in}%
\pgfsys@useobject{currentmarker}{}%
\end{pgfscope}%
\end{pgfscope}%
\begin{pgfscope}%
\definecolor{textcolor}{rgb}{0.000000,0.000000,0.000000}%
\pgfsetstrokecolor{textcolor}%
\pgfsetfillcolor{textcolor}%
\pgftext[x=0.194444in, y=3.101598in, left, base]{\color{textcolor}\rmfamily\fontsize{10.000000}{12.000000}\selectfont \(\displaystyle {-20.5}\)}%
\end{pgfscope}%
\begin{pgfscope}%
\pgfsetbuttcap%
\pgfsetroundjoin%
\definecolor{currentfill}{rgb}{0.000000,0.000000,0.000000}%
\pgfsetfillcolor{currentfill}%
\pgfsetlinewidth{0.803000pt}%
\definecolor{currentstroke}{rgb}{0.000000,0.000000,0.000000}%
\pgfsetstrokecolor{currentstroke}%
\pgfsetdash{}{0pt}%
\pgfsys@defobject{currentmarker}{\pgfqpoint{-0.048611in}{0.000000in}}{\pgfqpoint{-0.000000in}{0.000000in}}{%
\pgfpathmoveto{\pgfqpoint{-0.000000in}{0.000000in}}%
\pgfpathlineto{\pgfqpoint{-0.048611in}{0.000000in}}%
\pgfusepath{stroke,fill}%
}%
\begin{pgfscope}%
\pgfsys@transformshift{0.646606in}{3.603462in}%
\pgfsys@useobject{currentmarker}{}%
\end{pgfscope}%
\end{pgfscope}%
\begin{pgfscope}%
\definecolor{textcolor}{rgb}{0.000000,0.000000,0.000000}%
\pgfsetstrokecolor{textcolor}%
\pgfsetfillcolor{textcolor}%
\pgftext[x=0.194444in, y=3.555236in, left, base]{\color{textcolor}\rmfamily\fontsize{10.000000}{12.000000}\selectfont \(\displaystyle {-20.0}\)}%
\end{pgfscope}%
\begin{pgfscope}%
\pgfsetbuttcap%
\pgfsetroundjoin%
\definecolor{currentfill}{rgb}{0.000000,0.000000,0.000000}%
\pgfsetfillcolor{currentfill}%
\pgfsetlinewidth{0.803000pt}%
\definecolor{currentstroke}{rgb}{0.000000,0.000000,0.000000}%
\pgfsetstrokecolor{currentstroke}%
\pgfsetdash{}{0pt}%
\pgfsys@defobject{currentmarker}{\pgfqpoint{-0.048611in}{0.000000in}}{\pgfqpoint{-0.000000in}{0.000000in}}{%
\pgfpathmoveto{\pgfqpoint{-0.000000in}{0.000000in}}%
\pgfpathlineto{\pgfqpoint{-0.048611in}{0.000000in}}%
\pgfusepath{stroke,fill}%
}%
\begin{pgfscope}%
\pgfsys@transformshift{0.646606in}{4.057100in}%
\pgfsys@useobject{currentmarker}{}%
\end{pgfscope}%
\end{pgfscope}%
\begin{pgfscope}%
\definecolor{textcolor}{rgb}{0.000000,0.000000,0.000000}%
\pgfsetstrokecolor{textcolor}%
\pgfsetfillcolor{textcolor}%
\pgftext[x=0.194444in, y=4.008874in, left, base]{\color{textcolor}\rmfamily\fontsize{10.000000}{12.000000}\selectfont \(\displaystyle {-19.5}\)}%
\end{pgfscope}%
\begin{pgfscope}%
\definecolor{textcolor}{rgb}{0.000000,0.000000,0.000000}%
\pgfsetstrokecolor{textcolor}%
\pgfsetfillcolor{textcolor}%
\pgftext[x=0.138889in,y=2.286514in,,bottom,rotate=90.000000]{\color{textcolor}\rmfamily\fontsize{10.000000}{12.000000}\selectfont [arcsec]}%
\end{pgfscope}%
\begin{pgfscope}%
\pgfsetrectcap%
\pgfsetmiterjoin%
\pgfsetlinewidth{0.803000pt}%
\definecolor{currentstroke}{rgb}{0.000000,0.000000,0.000000}%
\pgfsetstrokecolor{currentstroke}%
\pgfsetdash{}{0pt}%
\pgfpathmoveto{\pgfqpoint{0.646606in}{0.415123in}}%
\pgfpathlineto{\pgfqpoint{0.646606in}{4.157905in}}%
\pgfusepath{stroke}%
\end{pgfscope}%
\begin{pgfscope}%
\pgfsetrectcap%
\pgfsetmiterjoin%
\pgfsetlinewidth{0.803000pt}%
\definecolor{currentstroke}{rgb}{0.000000,0.000000,0.000000}%
\pgfsetstrokecolor{currentstroke}%
\pgfsetdash{}{0pt}%
\pgfpathmoveto{\pgfqpoint{6.260779in}{0.415123in}}%
\pgfpathlineto{\pgfqpoint{6.260779in}{4.157905in}}%
\pgfusepath{stroke}%
\end{pgfscope}%
\begin{pgfscope}%
\pgfsetrectcap%
\pgfsetmiterjoin%
\pgfsetlinewidth{0.803000pt}%
\definecolor{currentstroke}{rgb}{0.000000,0.000000,0.000000}%
\pgfsetstrokecolor{currentstroke}%
\pgfsetdash{}{0pt}%
\pgfpathmoveto{\pgfqpoint{0.646606in}{0.415123in}}%
\pgfpathlineto{\pgfqpoint{6.260779in}{0.415123in}}%
\pgfusepath{stroke}%
\end{pgfscope}%
\begin{pgfscope}%
\pgfsetrectcap%
\pgfsetmiterjoin%
\pgfsetlinewidth{0.803000pt}%
\definecolor{currentstroke}{rgb}{0.000000,0.000000,0.000000}%
\pgfsetstrokecolor{currentstroke}%
\pgfsetdash{}{0pt}%
\pgfpathmoveto{\pgfqpoint{0.646606in}{4.157905in}}%
\pgfpathlineto{\pgfqpoint{6.260779in}{4.157905in}}%
\pgfusepath{stroke}%
\end{pgfscope}%
\end{pgfpicture}%
\makeatother%
\endgroup%

%% file: overview_all_masks.pgf
\begingroup%
\makeatletter%
\begin{pgfpicture}%
\pgfpathrectangle{\pgfpointorigin}{\pgfqpoint{5.854010in}{3.403173in}}%
\pgfusepath{use as bounding box, clip}%
\begin{pgfscope}%
\pgfsetbuttcap%
\pgfsetmiterjoin%
\definecolor{currentfill}{rgb}{1.000000,1.000000,1.000000}%
\pgfsetfillcolor{currentfill}%
\pgfsetlinewidth{0.000000pt}%
\definecolor{currentstroke}{rgb}{1.000000,1.000000,1.000000}%
\pgfsetstrokecolor{currentstroke}%
\pgfsetdash{}{0pt}%
\pgfpathmoveto{\pgfqpoint{0.000000in}{0.000000in}}%
\pgfpathlineto{\pgfqpoint{5.854010in}{0.000000in}}%
\pgfpathlineto{\pgfqpoint{5.854010in}{3.403173in}}%
\pgfpathlineto{\pgfqpoint{0.000000in}{3.403173in}}%
\pgfpathclose%
\pgfusepath{fill}%
\end{pgfscope}%
\begin{pgfscope}%
\pgfsetbuttcap%
\pgfsetmiterjoin%
\definecolor{currentfill}{rgb}{1.000000,1.000000,1.000000}%
\pgfsetfillcolor{currentfill}%
\pgfsetlinewidth{0.000000pt}%
\definecolor{currentstroke}{rgb}{0.000000,0.000000,0.000000}%
\pgfsetstrokecolor{currentstroke}%
\pgfsetstrokeopacity{0.000000}%
\pgfsetdash{}{0pt}%
\pgfpathmoveto{\pgfqpoint{0.538581in}{1.991501in}}%
\pgfpathlineto{\pgfqpoint{2.913253in}{1.991501in}}%
\pgfpathlineto{\pgfqpoint{2.913253in}{3.204100in}}%
\pgfpathlineto{\pgfqpoint{0.538581in}{3.204100in}}%
\pgfpathclose%
\pgfusepath{fill}%
\end{pgfscope}%
\begin{pgfscope}%
\pgfpathrectangle{\pgfqpoint{0.538581in}{1.991501in}}{\pgfqpoint{2.374672in}{1.212598in}}%
\pgfusepath{clip}%
\pgfsys@transformshift{0.538581in}{1.991501in}%
\pgftext[left,bottom]{\includegraphics[interpolate=true,width=2.380000in,height=1.220000in]{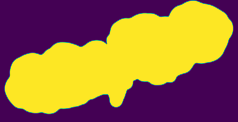}}%
\end{pgfscope}%
\begin{pgfscope}%
\pgfsetbuttcap%
\pgfsetroundjoin%
\definecolor{currentfill}{rgb}{0.000000,0.000000,0.000000}%
\pgfsetfillcolor{currentfill}%
\pgfsetlinewidth{0.803000pt}%
\definecolor{currentstroke}{rgb}{0.000000,0.000000,0.000000}%
\pgfsetstrokecolor{currentstroke}%
\pgfsetdash{}{0pt}%
\pgfsys@defobject{currentmarker}{\pgfqpoint{0.000000in}{-0.048611in}}{\pgfqpoint{0.000000in}{0.000000in}}{%
\pgfpathmoveto{\pgfqpoint{0.000000in}{0.000000in}}%
\pgfpathlineto{\pgfqpoint{0.000000in}{-0.048611in}}%
\pgfusepath{stroke,fill}%
}%
\begin{pgfscope}%
\pgfsys@transformshift{0.808047in}{1.991501in}%
\pgfsys@useobject{currentmarker}{}%
\end{pgfscope}%
\end{pgfscope}%
\begin{pgfscope}%
\pgfsetbuttcap%
\pgfsetroundjoin%
\definecolor{currentfill}{rgb}{0.000000,0.000000,0.000000}%
\pgfsetfillcolor{currentfill}%
\pgfsetlinewidth{0.803000pt}%
\definecolor{currentstroke}{rgb}{0.000000,0.000000,0.000000}%
\pgfsetstrokecolor{currentstroke}%
\pgfsetdash{}{0pt}%
\pgfsys@defobject{currentmarker}{\pgfqpoint{0.000000in}{-0.048611in}}{\pgfqpoint{0.000000in}{0.000000in}}{%
\pgfpathmoveto{\pgfqpoint{0.000000in}{0.000000in}}%
\pgfpathlineto{\pgfqpoint{0.000000in}{-0.048611in}}%
\pgfusepath{stroke,fill}%
}%
\begin{pgfscope}%
\pgfsys@transformshift{1.650130in}{1.991501in}%
\pgfsys@useobject{currentmarker}{}%
\end{pgfscope}%
\end{pgfscope}%
\begin{pgfscope}%
\pgfsetbuttcap%
\pgfsetroundjoin%
\definecolor{currentfill}{rgb}{0.000000,0.000000,0.000000}%
\pgfsetfillcolor{currentfill}%
\pgfsetlinewidth{0.803000pt}%
\definecolor{currentstroke}{rgb}{0.000000,0.000000,0.000000}%
\pgfsetstrokecolor{currentstroke}%
\pgfsetdash{}{0pt}%
\pgfsys@defobject{currentmarker}{\pgfqpoint{0.000000in}{-0.048611in}}{\pgfqpoint{0.000000in}{0.000000in}}{%
\pgfpathmoveto{\pgfqpoint{0.000000in}{0.000000in}}%
\pgfpathlineto{\pgfqpoint{0.000000in}{-0.048611in}}%
\pgfusepath{stroke,fill}%
}%
\begin{pgfscope}%
\pgfsys@transformshift{2.492212in}{1.991501in}%
\pgfsys@useobject{currentmarker}{}%
\end{pgfscope}%
\end{pgfscope}%
\begin{pgfscope}%
\pgfsetbuttcap%
\pgfsetroundjoin%
\definecolor{currentfill}{rgb}{0.000000,0.000000,0.000000}%
\pgfsetfillcolor{currentfill}%
\pgfsetlinewidth{0.803000pt}%
\definecolor{currentstroke}{rgb}{0.000000,0.000000,0.000000}%
\pgfsetstrokecolor{currentstroke}%
\pgfsetdash{}{0pt}%
\pgfsys@defobject{currentmarker}{\pgfqpoint{-0.048611in}{0.000000in}}{\pgfqpoint{-0.000000in}{0.000000in}}{%
\pgfpathmoveto{\pgfqpoint{-0.000000in}{0.000000in}}%
\pgfpathlineto{\pgfqpoint{-0.048611in}{0.000000in}}%
\pgfusepath{stroke,fill}%
}%
\begin{pgfscope}%
\pgfsys@transformshift{0.538581in}{2.260968in}%
\pgfsys@useobject{currentmarker}{}%
\end{pgfscope}%
\end{pgfscope}%
\begin{pgfscope}%
\definecolor{textcolor}{rgb}{0.000000,0.000000,0.000000}%
\pgfsetstrokecolor{textcolor}%
\pgfsetfillcolor{textcolor}%
\pgftext[x=0.194444in, y=2.212742in, left, base]{\color{textcolor}\rmfamily\fontsize{10.000000}{12.000000}\selectfont \(\displaystyle {-20}\)}%
\end{pgfscope}%
\begin{pgfscope}%
\pgfsetbuttcap%
\pgfsetroundjoin%
\definecolor{currentfill}{rgb}{0.000000,0.000000,0.000000}%
\pgfsetfillcolor{currentfill}%
\pgfsetlinewidth{0.803000pt}%
\definecolor{currentstroke}{rgb}{0.000000,0.000000,0.000000}%
\pgfsetstrokecolor{currentstroke}%
\pgfsetdash{}{0pt}%
\pgfsys@defobject{currentmarker}{\pgfqpoint{-0.048611in}{0.000000in}}{\pgfqpoint{-0.000000in}{0.000000in}}{%
\pgfpathmoveto{\pgfqpoint{-0.000000in}{0.000000in}}%
\pgfpathlineto{\pgfqpoint{-0.048611in}{0.000000in}}%
\pgfusepath{stroke,fill}%
}%
\begin{pgfscope}%
\pgfsys@transformshift{0.538581in}{2.597800in}%
\pgfsys@useobject{currentmarker}{}%
\end{pgfscope}%
\end{pgfscope}%
\begin{pgfscope}%
\definecolor{textcolor}{rgb}{0.000000,0.000000,0.000000}%
\pgfsetstrokecolor{textcolor}%
\pgfsetfillcolor{textcolor}%
\pgftext[x=0.371914in, y=2.549575in, left, base]{\color{textcolor}\rmfamily\fontsize{10.000000}{12.000000}\selectfont \(\displaystyle {0}\)}%
\end{pgfscope}%
\begin{pgfscope}%
\pgfsetbuttcap%
\pgfsetroundjoin%
\definecolor{currentfill}{rgb}{0.000000,0.000000,0.000000}%
\pgfsetfillcolor{currentfill}%
\pgfsetlinewidth{0.803000pt}%
\definecolor{currentstroke}{rgb}{0.000000,0.000000,0.000000}%
\pgfsetstrokecolor{currentstroke}%
\pgfsetdash{}{0pt}%
\pgfsys@defobject{currentmarker}{\pgfqpoint{-0.048611in}{0.000000in}}{\pgfqpoint{-0.000000in}{0.000000in}}{%
\pgfpathmoveto{\pgfqpoint{-0.000000in}{0.000000in}}%
\pgfpathlineto{\pgfqpoint{-0.048611in}{0.000000in}}%
\pgfusepath{stroke,fill}%
}%
\begin{pgfscope}%
\pgfsys@transformshift{0.538581in}{2.934633in}%
\pgfsys@useobject{currentmarker}{}%
\end{pgfscope}%
\end{pgfscope}%
\begin{pgfscope}%
\definecolor{textcolor}{rgb}{0.000000,0.000000,0.000000}%
\pgfsetstrokecolor{textcolor}%
\pgfsetfillcolor{textcolor}%
\pgftext[x=0.302469in, y=2.886408in, left, base]{\color{textcolor}\rmfamily\fontsize{10.000000}{12.000000}\selectfont \(\displaystyle {20}\)}%
\end{pgfscope}%
\begin{pgfscope}%
\definecolor{textcolor}{rgb}{0.000000,0.000000,0.000000}%
\pgfsetstrokecolor{textcolor}%
\pgfsetfillcolor{textcolor}%
\pgftext[x=0.138889in,y=2.597800in,,bottom,rotate=90.000000]{\color{textcolor}\rmfamily\fontsize{10.000000}{12.000000}\selectfont [arcsec]}%
\end{pgfscope}%
\begin{pgfscope}%
\pgfsetrectcap%
\pgfsetmiterjoin%
\pgfsetlinewidth{0.803000pt}%
\definecolor{currentstroke}{rgb}{0.000000,0.000000,0.000000}%
\pgfsetstrokecolor{currentstroke}%
\pgfsetdash{}{0pt}%
\pgfpathmoveto{\pgfqpoint{0.538581in}{1.991501in}}%
\pgfpathlineto{\pgfqpoint{0.538581in}{3.204100in}}%
\pgfusepath{stroke}%
\end{pgfscope}%
\begin{pgfscope}%
\pgfsetrectcap%
\pgfsetmiterjoin%
\pgfsetlinewidth{0.803000pt}%
\definecolor{currentstroke}{rgb}{0.000000,0.000000,0.000000}%
\pgfsetstrokecolor{currentstroke}%
\pgfsetdash{}{0pt}%
\pgfpathmoveto{\pgfqpoint{2.913253in}{1.991501in}}%
\pgfpathlineto{\pgfqpoint{2.913253in}{3.204100in}}%
\pgfusepath{stroke}%
\end{pgfscope}%
\begin{pgfscope}%
\pgfsetrectcap%
\pgfsetmiterjoin%
\pgfsetlinewidth{0.803000pt}%
\definecolor{currentstroke}{rgb}{0.000000,0.000000,0.000000}%
\pgfsetstrokecolor{currentstroke}%
\pgfsetdash{}{0pt}%
\pgfpathmoveto{\pgfqpoint{0.538581in}{1.991501in}}%
\pgfpathlineto{\pgfqpoint{2.913253in}{1.991501in}}%
\pgfusepath{stroke}%
\end{pgfscope}%
\begin{pgfscope}%
\pgfsetrectcap%
\pgfsetmiterjoin%
\pgfsetlinewidth{0.803000pt}%
\definecolor{currentstroke}{rgb}{0.000000,0.000000,0.000000}%
\pgfsetstrokecolor{currentstroke}%
\pgfsetdash{}{0pt}%
\pgfpathmoveto{\pgfqpoint{0.538581in}{3.204100in}}%
\pgfpathlineto{\pgfqpoint{2.913253in}{3.204100in}}%
\pgfusepath{stroke}%
\end{pgfscope}%
\begin{pgfscope}%
\definecolor{textcolor}{rgb}{0.000000,0.000000,0.000000}%
\pgfsetstrokecolor{textcolor}%
\pgfsetfillcolor{textcolor}%
\pgftext[x=1.725917in,y=3.287433in,,base]{\color{textcolor}\rmfamily\fontsize{12.000000}{14.400000}\selectfont 2052-2MHz}%
\end{pgfscope}%
\begin{pgfscope}%
\pgfsetbuttcap%
\pgfsetmiterjoin%
\definecolor{currentfill}{rgb}{1.000000,1.000000,1.000000}%
\pgfsetfillcolor{currentfill}%
\pgfsetlinewidth{0.000000pt}%
\definecolor{currentstroke}{rgb}{0.000000,0.000000,0.000000}%
\pgfsetstrokecolor{currentstroke}%
\pgfsetstrokeopacity{0.000000}%
\pgfsetdash{}{0pt}%
\pgfpathmoveto{\pgfqpoint{3.479338in}{1.991501in}}%
\pgfpathlineto{\pgfqpoint{5.854010in}{1.991501in}}%
\pgfpathlineto{\pgfqpoint{5.854010in}{3.204100in}}%
\pgfpathlineto{\pgfqpoint{3.479338in}{3.204100in}}%
\pgfpathclose%
\pgfusepath{fill}%
\end{pgfscope}%
\begin{pgfscope}%
\pgfpathrectangle{\pgfqpoint{3.479338in}{1.991501in}}{\pgfqpoint{2.374672in}{1.212598in}}%
\pgfusepath{clip}%
\pgfsys@transformshift{3.479338in}{1.991501in}%
\pgftext[left,bottom]{\includegraphics[interpolate=true,width=2.380000in,height=1.220000in]{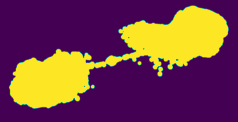}}%
\end{pgfscope}%
\begin{pgfscope}%
\pgfsetbuttcap%
\pgfsetroundjoin%
\definecolor{currentfill}{rgb}{0.000000,0.000000,0.000000}%
\pgfsetfillcolor{currentfill}%
\pgfsetlinewidth{0.803000pt}%
\definecolor{currentstroke}{rgb}{0.000000,0.000000,0.000000}%
\pgfsetstrokecolor{currentstroke}%
\pgfsetdash{}{0pt}%
\pgfsys@defobject{currentmarker}{\pgfqpoint{0.000000in}{-0.048611in}}{\pgfqpoint{0.000000in}{0.000000in}}{%
\pgfpathmoveto{\pgfqpoint{0.000000in}{0.000000in}}%
\pgfpathlineto{\pgfqpoint{0.000000in}{-0.048611in}}%
\pgfusepath{stroke,fill}%
}%
\begin{pgfscope}%
\pgfsys@transformshift{3.748805in}{1.991501in}%
\pgfsys@useobject{currentmarker}{}%
\end{pgfscope}%
\end{pgfscope}%
\begin{pgfscope}%
\pgfsetbuttcap%
\pgfsetroundjoin%
\definecolor{currentfill}{rgb}{0.000000,0.000000,0.000000}%
\pgfsetfillcolor{currentfill}%
\pgfsetlinewidth{0.803000pt}%
\definecolor{currentstroke}{rgb}{0.000000,0.000000,0.000000}%
\pgfsetstrokecolor{currentstroke}%
\pgfsetdash{}{0pt}%
\pgfsys@defobject{currentmarker}{\pgfqpoint{0.000000in}{-0.048611in}}{\pgfqpoint{0.000000in}{0.000000in}}{%
\pgfpathmoveto{\pgfqpoint{0.000000in}{0.000000in}}%
\pgfpathlineto{\pgfqpoint{0.000000in}{-0.048611in}}%
\pgfusepath{stroke,fill}%
}%
\begin{pgfscope}%
\pgfsys@transformshift{4.590887in}{1.991501in}%
\pgfsys@useobject{currentmarker}{}%
\end{pgfscope}%
\end{pgfscope}%
\begin{pgfscope}%
\pgfsetbuttcap%
\pgfsetroundjoin%
\definecolor{currentfill}{rgb}{0.000000,0.000000,0.000000}%
\pgfsetfillcolor{currentfill}%
\pgfsetlinewidth{0.803000pt}%
\definecolor{currentstroke}{rgb}{0.000000,0.000000,0.000000}%
\pgfsetstrokecolor{currentstroke}%
\pgfsetdash{}{0pt}%
\pgfsys@defobject{currentmarker}{\pgfqpoint{0.000000in}{-0.048611in}}{\pgfqpoint{0.000000in}{0.000000in}}{%
\pgfpathmoveto{\pgfqpoint{0.000000in}{0.000000in}}%
\pgfpathlineto{\pgfqpoint{0.000000in}{-0.048611in}}%
\pgfusepath{stroke,fill}%
}%
\begin{pgfscope}%
\pgfsys@transformshift{5.432969in}{1.991501in}%
\pgfsys@useobject{currentmarker}{}%
\end{pgfscope}%
\end{pgfscope}%
\begin{pgfscope}%
\pgfsetbuttcap%
\pgfsetroundjoin%
\definecolor{currentfill}{rgb}{0.000000,0.000000,0.000000}%
\pgfsetfillcolor{currentfill}%
\pgfsetlinewidth{0.803000pt}%
\definecolor{currentstroke}{rgb}{0.000000,0.000000,0.000000}%
\pgfsetstrokecolor{currentstroke}%
\pgfsetdash{}{0pt}%
\pgfsys@defobject{currentmarker}{\pgfqpoint{-0.048611in}{0.000000in}}{\pgfqpoint{-0.000000in}{0.000000in}}{%
\pgfpathmoveto{\pgfqpoint{-0.000000in}{0.000000in}}%
\pgfpathlineto{\pgfqpoint{-0.048611in}{0.000000in}}%
\pgfusepath{stroke,fill}%
}%
\begin{pgfscope}%
\pgfsys@transformshift{3.479338in}{2.260968in}%
\pgfsys@useobject{currentmarker}{}%
\end{pgfscope}%
\end{pgfscope}%
\begin{pgfscope}%
\pgfsetbuttcap%
\pgfsetroundjoin%
\definecolor{currentfill}{rgb}{0.000000,0.000000,0.000000}%
\pgfsetfillcolor{currentfill}%
\pgfsetlinewidth{0.803000pt}%
\definecolor{currentstroke}{rgb}{0.000000,0.000000,0.000000}%
\pgfsetstrokecolor{currentstroke}%
\pgfsetdash{}{0pt}%
\pgfsys@defobject{currentmarker}{\pgfqpoint{-0.048611in}{0.000000in}}{\pgfqpoint{-0.000000in}{0.000000in}}{%
\pgfpathmoveto{\pgfqpoint{-0.000000in}{0.000000in}}%
\pgfpathlineto{\pgfqpoint{-0.048611in}{0.000000in}}%
\pgfusepath{stroke,fill}%
}%
\begin{pgfscope}%
\pgfsys@transformshift{3.479338in}{2.597800in}%
\pgfsys@useobject{currentmarker}{}%
\end{pgfscope}%
\end{pgfscope}%
\begin{pgfscope}%
\pgfsetbuttcap%
\pgfsetroundjoin%
\definecolor{currentfill}{rgb}{0.000000,0.000000,0.000000}%
\pgfsetfillcolor{currentfill}%
\pgfsetlinewidth{0.803000pt}%
\definecolor{currentstroke}{rgb}{0.000000,0.000000,0.000000}%
\pgfsetstrokecolor{currentstroke}%
\pgfsetdash{}{0pt}%
\pgfsys@defobject{currentmarker}{\pgfqpoint{-0.048611in}{0.000000in}}{\pgfqpoint{-0.000000in}{0.000000in}}{%
\pgfpathmoveto{\pgfqpoint{-0.000000in}{0.000000in}}%
\pgfpathlineto{\pgfqpoint{-0.048611in}{0.000000in}}%
\pgfusepath{stroke,fill}%
}%
\begin{pgfscope}%
\pgfsys@transformshift{3.479338in}{2.934633in}%
\pgfsys@useobject{currentmarker}{}%
\end{pgfscope}%
\end{pgfscope}%
\begin{pgfscope}%
\pgfsetrectcap%
\pgfsetmiterjoin%
\pgfsetlinewidth{0.803000pt}%
\definecolor{currentstroke}{rgb}{0.000000,0.000000,0.000000}%
\pgfsetstrokecolor{currentstroke}%
\pgfsetdash{}{0pt}%
\pgfpathmoveto{\pgfqpoint{3.479338in}{1.991501in}}%
\pgfpathlineto{\pgfqpoint{3.479338in}{3.204100in}}%
\pgfusepath{stroke}%
\end{pgfscope}%
\begin{pgfscope}%
\pgfsetrectcap%
\pgfsetmiterjoin%
\pgfsetlinewidth{0.803000pt}%
\definecolor{currentstroke}{rgb}{0.000000,0.000000,0.000000}%
\pgfsetstrokecolor{currentstroke}%
\pgfsetdash{}{0pt}%
\pgfpathmoveto{\pgfqpoint{5.854010in}{1.991501in}}%
\pgfpathlineto{\pgfqpoint{5.854010in}{3.204100in}}%
\pgfusepath{stroke}%
\end{pgfscope}%
\begin{pgfscope}%
\pgfsetrectcap%
\pgfsetmiterjoin%
\pgfsetlinewidth{0.803000pt}%
\definecolor{currentstroke}{rgb}{0.000000,0.000000,0.000000}%
\pgfsetstrokecolor{currentstroke}%
\pgfsetdash{}{0pt}%
\pgfpathmoveto{\pgfqpoint{3.479338in}{1.991501in}}%
\pgfpathlineto{\pgfqpoint{5.854010in}{1.991501in}}%
\pgfusepath{stroke}%
\end{pgfscope}%
\begin{pgfscope}%
\pgfsetrectcap%
\pgfsetmiterjoin%
\pgfsetlinewidth{0.803000pt}%
\definecolor{currentstroke}{rgb}{0.000000,0.000000,0.000000}%
\pgfsetstrokecolor{currentstroke}%
\pgfsetdash{}{0pt}%
\pgfpathmoveto{\pgfqpoint{3.479338in}{3.204100in}}%
\pgfpathlineto{\pgfqpoint{5.854010in}{3.204100in}}%
\pgfusepath{stroke}%
\end{pgfscope}%
\begin{pgfscope}%
\definecolor{textcolor}{rgb}{0.000000,0.000000,0.000000}%
\pgfsetstrokecolor{textcolor}%
\pgfsetfillcolor{textcolor}%
\pgftext[x=4.666674in,y=3.287433in,,base]{\color{textcolor}\rmfamily\fontsize{12.000000}{14.400000}\selectfont 4811-8MHz}%
\end{pgfscope}%
\begin{pgfscope}%
\pgfsetbuttcap%
\pgfsetmiterjoin%
\definecolor{currentfill}{rgb}{1.000000,1.000000,1.000000}%
\pgfsetfillcolor{currentfill}%
\pgfsetlinewidth{0.000000pt}%
\definecolor{currentstroke}{rgb}{0.000000,0.000000,0.000000}%
\pgfsetstrokecolor{currentstroke}%
\pgfsetstrokeopacity{0.000000}%
\pgfsetdash{}{0pt}%
\pgfpathmoveto{\pgfqpoint{0.538581in}{0.415123in}}%
\pgfpathlineto{\pgfqpoint{2.913253in}{0.415123in}}%
\pgfpathlineto{\pgfqpoint{2.913253in}{1.627722in}}%
\pgfpathlineto{\pgfqpoint{0.538581in}{1.627722in}}%
\pgfpathclose%
\pgfusepath{fill}%
\end{pgfscope}%
\begin{pgfscope}%
\pgfpathrectangle{\pgfqpoint{0.538581in}{0.415123in}}{\pgfqpoint{2.374672in}{1.212598in}}%
\pgfusepath{clip}%
\pgfsys@transformshift{0.538581in}{0.415123in}%
\pgftext[left,bottom]{\includegraphics[interpolate=true,width=2.380000in,height=1.220000in]{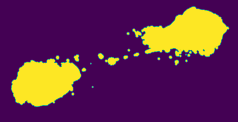}}%
\end{pgfscope}%
\begin{pgfscope}%
\pgfsetbuttcap%
\pgfsetroundjoin%
\definecolor{currentfill}{rgb}{0.000000,0.000000,0.000000}%
\pgfsetfillcolor{currentfill}%
\pgfsetlinewidth{0.803000pt}%
\definecolor{currentstroke}{rgb}{0.000000,0.000000,0.000000}%
\pgfsetstrokecolor{currentstroke}%
\pgfsetdash{}{0pt}%
\pgfsys@defobject{currentmarker}{\pgfqpoint{0.000000in}{-0.048611in}}{\pgfqpoint{0.000000in}{0.000000in}}{%
\pgfpathmoveto{\pgfqpoint{0.000000in}{0.000000in}}%
\pgfpathlineto{\pgfqpoint{0.000000in}{-0.048611in}}%
\pgfusepath{stroke,fill}%
}%
\begin{pgfscope}%
\pgfsys@transformshift{0.808047in}{0.415123in}%
\pgfsys@useobject{currentmarker}{}%
\end{pgfscope}%
\end{pgfscope}%
\begin{pgfscope}%
\definecolor{textcolor}{rgb}{0.000000,0.000000,0.000000}%
\pgfsetstrokecolor{textcolor}%
\pgfsetfillcolor{textcolor}%
\pgftext[x=0.808047in,y=0.317901in,,top]{\color{textcolor}\rmfamily\fontsize{10.000000}{12.000000}\selectfont \(\displaystyle {-50}\)}%
\end{pgfscope}%
\begin{pgfscope}%
\pgfsetbuttcap%
\pgfsetroundjoin%
\definecolor{currentfill}{rgb}{0.000000,0.000000,0.000000}%
\pgfsetfillcolor{currentfill}%
\pgfsetlinewidth{0.803000pt}%
\definecolor{currentstroke}{rgb}{0.000000,0.000000,0.000000}%
\pgfsetstrokecolor{currentstroke}%
\pgfsetdash{}{0pt}%
\pgfsys@defobject{currentmarker}{\pgfqpoint{0.000000in}{-0.048611in}}{\pgfqpoint{0.000000in}{0.000000in}}{%
\pgfpathmoveto{\pgfqpoint{0.000000in}{0.000000in}}%
\pgfpathlineto{\pgfqpoint{0.000000in}{-0.048611in}}%
\pgfusepath{stroke,fill}%
}%
\begin{pgfscope}%
\pgfsys@transformshift{1.650130in}{0.415123in}%
\pgfsys@useobject{currentmarker}{}%
\end{pgfscope}%
\end{pgfscope}%
\begin{pgfscope}%
\definecolor{textcolor}{rgb}{0.000000,0.000000,0.000000}%
\pgfsetstrokecolor{textcolor}%
\pgfsetfillcolor{textcolor}%
\pgftext[x=1.650130in,y=0.317901in,,top]{\color{textcolor}\rmfamily\fontsize{10.000000}{12.000000}\selectfont \(\displaystyle {0}\)}%
\end{pgfscope}%
\begin{pgfscope}%
\pgfsetbuttcap%
\pgfsetroundjoin%
\definecolor{currentfill}{rgb}{0.000000,0.000000,0.000000}%
\pgfsetfillcolor{currentfill}%
\pgfsetlinewidth{0.803000pt}%
\definecolor{currentstroke}{rgb}{0.000000,0.000000,0.000000}%
\pgfsetstrokecolor{currentstroke}%
\pgfsetdash{}{0pt}%
\pgfsys@defobject{currentmarker}{\pgfqpoint{0.000000in}{-0.048611in}}{\pgfqpoint{0.000000in}{0.000000in}}{%
\pgfpathmoveto{\pgfqpoint{0.000000in}{0.000000in}}%
\pgfpathlineto{\pgfqpoint{0.000000in}{-0.048611in}}%
\pgfusepath{stroke,fill}%
}%
\begin{pgfscope}%
\pgfsys@transformshift{2.492212in}{0.415123in}%
\pgfsys@useobject{currentmarker}{}%
\end{pgfscope}%
\end{pgfscope}%
\begin{pgfscope}%
\definecolor{textcolor}{rgb}{0.000000,0.000000,0.000000}%
\pgfsetstrokecolor{textcolor}%
\pgfsetfillcolor{textcolor}%
\pgftext[x=2.492212in,y=0.317901in,,top]{\color{textcolor}\rmfamily\fontsize{10.000000}{12.000000}\selectfont \(\displaystyle {50}\)}%
\end{pgfscope}%
\begin{pgfscope}%
\definecolor{textcolor}{rgb}{0.000000,0.000000,0.000000}%
\pgfsetstrokecolor{textcolor}%
\pgfsetfillcolor{textcolor}%
\pgftext[x=1.725917in,y=0.138889in,,top]{\color{textcolor}\rmfamily\fontsize{10.000000}{12.000000}\selectfont [arcsec]}%
\end{pgfscope}%
\begin{pgfscope}%
\pgfsetbuttcap%
\pgfsetroundjoin%
\definecolor{currentfill}{rgb}{0.000000,0.000000,0.000000}%
\pgfsetfillcolor{currentfill}%
\pgfsetlinewidth{0.803000pt}%
\definecolor{currentstroke}{rgb}{0.000000,0.000000,0.000000}%
\pgfsetstrokecolor{currentstroke}%
\pgfsetdash{}{0pt}%
\pgfsys@defobject{currentmarker}{\pgfqpoint{-0.048611in}{0.000000in}}{\pgfqpoint{-0.000000in}{0.000000in}}{%
\pgfpathmoveto{\pgfqpoint{-0.000000in}{0.000000in}}%
\pgfpathlineto{\pgfqpoint{-0.048611in}{0.000000in}}%
\pgfusepath{stroke,fill}%
}%
\begin{pgfscope}%
\pgfsys@transformshift{0.538581in}{0.684590in}%
\pgfsys@useobject{currentmarker}{}%
\end{pgfscope}%
\end{pgfscope}%
\begin{pgfscope}%
\definecolor{textcolor}{rgb}{0.000000,0.000000,0.000000}%
\pgfsetstrokecolor{textcolor}%
\pgfsetfillcolor{textcolor}%
\pgftext[x=0.194444in, y=0.636364in, left, base]{\color{textcolor}\rmfamily\fontsize{10.000000}{12.000000}\selectfont \(\displaystyle {-20}\)}%
\end{pgfscope}%
\begin{pgfscope}%
\pgfsetbuttcap%
\pgfsetroundjoin%
\definecolor{currentfill}{rgb}{0.000000,0.000000,0.000000}%
\pgfsetfillcolor{currentfill}%
\pgfsetlinewidth{0.803000pt}%
\definecolor{currentstroke}{rgb}{0.000000,0.000000,0.000000}%
\pgfsetstrokecolor{currentstroke}%
\pgfsetdash{}{0pt}%
\pgfsys@defobject{currentmarker}{\pgfqpoint{-0.048611in}{0.000000in}}{\pgfqpoint{-0.000000in}{0.000000in}}{%
\pgfpathmoveto{\pgfqpoint{-0.000000in}{0.000000in}}%
\pgfpathlineto{\pgfqpoint{-0.048611in}{0.000000in}}%
\pgfusepath{stroke,fill}%
}%
\begin{pgfscope}%
\pgfsys@transformshift{0.538581in}{1.021423in}%
\pgfsys@useobject{currentmarker}{}%
\end{pgfscope}%
\end{pgfscope}%
\begin{pgfscope}%
\definecolor{textcolor}{rgb}{0.000000,0.000000,0.000000}%
\pgfsetstrokecolor{textcolor}%
\pgfsetfillcolor{textcolor}%
\pgftext[x=0.371914in, y=0.973197in, left, base]{\color{textcolor}\rmfamily\fontsize{10.000000}{12.000000}\selectfont \(\displaystyle {0}\)}%
\end{pgfscope}%
\begin{pgfscope}%
\pgfsetbuttcap%
\pgfsetroundjoin%
\definecolor{currentfill}{rgb}{0.000000,0.000000,0.000000}%
\pgfsetfillcolor{currentfill}%
\pgfsetlinewidth{0.803000pt}%
\definecolor{currentstroke}{rgb}{0.000000,0.000000,0.000000}%
\pgfsetstrokecolor{currentstroke}%
\pgfsetdash{}{0pt}%
\pgfsys@defobject{currentmarker}{\pgfqpoint{-0.048611in}{0.000000in}}{\pgfqpoint{-0.000000in}{0.000000in}}{%
\pgfpathmoveto{\pgfqpoint{-0.000000in}{0.000000in}}%
\pgfpathlineto{\pgfqpoint{-0.048611in}{0.000000in}}%
\pgfusepath{stroke,fill}%
}%
\begin{pgfscope}%
\pgfsys@transformshift{0.538581in}{1.358255in}%
\pgfsys@useobject{currentmarker}{}%
\end{pgfscope}%
\end{pgfscope}%
\begin{pgfscope}%
\definecolor{textcolor}{rgb}{0.000000,0.000000,0.000000}%
\pgfsetstrokecolor{textcolor}%
\pgfsetfillcolor{textcolor}%
\pgftext[x=0.302469in, y=1.310030in, left, base]{\color{textcolor}\rmfamily\fontsize{10.000000}{12.000000}\selectfont \(\displaystyle {20}\)}%
\end{pgfscope}%
\begin{pgfscope}%
\definecolor{textcolor}{rgb}{0.000000,0.000000,0.000000}%
\pgfsetstrokecolor{textcolor}%
\pgfsetfillcolor{textcolor}%
\pgftext[x=0.138889in,y=1.021423in,,bottom,rotate=90.000000]{\color{textcolor}\rmfamily\fontsize{10.000000}{12.000000}\selectfont [arcsec]}%
\end{pgfscope}%
\begin{pgfscope}%
\pgfsetrectcap%
\pgfsetmiterjoin%
\pgfsetlinewidth{0.803000pt}%
\definecolor{currentstroke}{rgb}{0.000000,0.000000,0.000000}%
\pgfsetstrokecolor{currentstroke}%
\pgfsetdash{}{0pt}%
\pgfpathmoveto{\pgfqpoint{0.538581in}{0.415123in}}%
\pgfpathlineto{\pgfqpoint{0.538581in}{1.627722in}}%
\pgfusepath{stroke}%
\end{pgfscope}%
\begin{pgfscope}%
\pgfsetrectcap%
\pgfsetmiterjoin%
\pgfsetlinewidth{0.803000pt}%
\definecolor{currentstroke}{rgb}{0.000000,0.000000,0.000000}%
\pgfsetstrokecolor{currentstroke}%
\pgfsetdash{}{0pt}%
\pgfpathmoveto{\pgfqpoint{2.913253in}{0.415123in}}%
\pgfpathlineto{\pgfqpoint{2.913253in}{1.627722in}}%
\pgfusepath{stroke}%
\end{pgfscope}%
\begin{pgfscope}%
\pgfsetrectcap%
\pgfsetmiterjoin%
\pgfsetlinewidth{0.803000pt}%
\definecolor{currentstroke}{rgb}{0.000000,0.000000,0.000000}%
\pgfsetstrokecolor{currentstroke}%
\pgfsetdash{}{0pt}%
\pgfpathmoveto{\pgfqpoint{0.538581in}{0.415123in}}%
\pgfpathlineto{\pgfqpoint{2.913253in}{0.415123in}}%
\pgfusepath{stroke}%
\end{pgfscope}%
\begin{pgfscope}%
\pgfsetrectcap%
\pgfsetmiterjoin%
\pgfsetlinewidth{0.803000pt}%
\definecolor{currentstroke}{rgb}{0.000000,0.000000,0.000000}%
\pgfsetstrokecolor{currentstroke}%
\pgfsetdash{}{0pt}%
\pgfpathmoveto{\pgfqpoint{0.538581in}{1.627722in}}%
\pgfpathlineto{\pgfqpoint{2.913253in}{1.627722in}}%
\pgfusepath{stroke}%
\end{pgfscope}%
\begin{pgfscope}%
\definecolor{textcolor}{rgb}{0.000000,0.000000,0.000000}%
\pgfsetstrokecolor{textcolor}%
\pgfsetfillcolor{textcolor}%
\pgftext[x=1.725917in,y=1.711055in,,base]{\color{textcolor}\rmfamily\fontsize{12.000000}{14.400000}\selectfont 8427-8MHz}%
\end{pgfscope}%
\begin{pgfscope}%
\pgfsetbuttcap%
\pgfsetmiterjoin%
\definecolor{currentfill}{rgb}{1.000000,1.000000,1.000000}%
\pgfsetfillcolor{currentfill}%
\pgfsetlinewidth{0.000000pt}%
\definecolor{currentstroke}{rgb}{0.000000,0.000000,0.000000}%
\pgfsetstrokecolor{currentstroke}%
\pgfsetstrokeopacity{0.000000}%
\pgfsetdash{}{0pt}%
\pgfpathmoveto{\pgfqpoint{3.479338in}{0.415123in}}%
\pgfpathlineto{\pgfqpoint{5.854010in}{0.415123in}}%
\pgfpathlineto{\pgfqpoint{5.854010in}{1.627722in}}%
\pgfpathlineto{\pgfqpoint{3.479338in}{1.627722in}}%
\pgfpathclose%
\pgfusepath{fill}%
\end{pgfscope}%
\begin{pgfscope}%
\pgfpathrectangle{\pgfqpoint{3.479338in}{0.415123in}}{\pgfqpoint{2.374672in}{1.212598in}}%
\pgfusepath{clip}%
\pgfsys@transformshift{3.479338in}{0.415123in}%
\pgftext[left,bottom]{\includegraphics[interpolate=true,width=2.380000in,height=1.220000in]{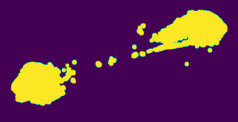}}%
\end{pgfscope}%
\begin{pgfscope}%
\pgfsetbuttcap%
\pgfsetroundjoin%
\definecolor{currentfill}{rgb}{0.000000,0.000000,0.000000}%
\pgfsetfillcolor{currentfill}%
\pgfsetlinewidth{0.803000pt}%
\definecolor{currentstroke}{rgb}{0.000000,0.000000,0.000000}%
\pgfsetstrokecolor{currentstroke}%
\pgfsetdash{}{0pt}%
\pgfsys@defobject{currentmarker}{\pgfqpoint{0.000000in}{-0.048611in}}{\pgfqpoint{0.000000in}{0.000000in}}{%
\pgfpathmoveto{\pgfqpoint{0.000000in}{0.000000in}}%
\pgfpathlineto{\pgfqpoint{0.000000in}{-0.048611in}}%
\pgfusepath{stroke,fill}%
}%
\begin{pgfscope}%
\pgfsys@transformshift{3.748805in}{0.415123in}%
\pgfsys@useobject{currentmarker}{}%
\end{pgfscope}%
\end{pgfscope}%
\begin{pgfscope}%
\definecolor{textcolor}{rgb}{0.000000,0.000000,0.000000}%
\pgfsetstrokecolor{textcolor}%
\pgfsetfillcolor{textcolor}%
\pgftext[x=3.748805in,y=0.317901in,,top]{\color{textcolor}\rmfamily\fontsize{10.000000}{12.000000}\selectfont \(\displaystyle {-50}\)}%
\end{pgfscope}%
\begin{pgfscope}%
\pgfsetbuttcap%
\pgfsetroundjoin%
\definecolor{currentfill}{rgb}{0.000000,0.000000,0.000000}%
\pgfsetfillcolor{currentfill}%
\pgfsetlinewidth{0.803000pt}%
\definecolor{currentstroke}{rgb}{0.000000,0.000000,0.000000}%
\pgfsetstrokecolor{currentstroke}%
\pgfsetdash{}{0pt}%
\pgfsys@defobject{currentmarker}{\pgfqpoint{0.000000in}{-0.048611in}}{\pgfqpoint{0.000000in}{0.000000in}}{%
\pgfpathmoveto{\pgfqpoint{0.000000in}{0.000000in}}%
\pgfpathlineto{\pgfqpoint{0.000000in}{-0.048611in}}%
\pgfusepath{stroke,fill}%
}%
\begin{pgfscope}%
\pgfsys@transformshift{4.590887in}{0.415123in}%
\pgfsys@useobject{currentmarker}{}%
\end{pgfscope}%
\end{pgfscope}%
\begin{pgfscope}%
\definecolor{textcolor}{rgb}{0.000000,0.000000,0.000000}%
\pgfsetstrokecolor{textcolor}%
\pgfsetfillcolor{textcolor}%
\pgftext[x=4.590887in,y=0.317901in,,top]{\color{textcolor}\rmfamily\fontsize{10.000000}{12.000000}\selectfont \(\displaystyle {0}\)}%
\end{pgfscope}%
\begin{pgfscope}%
\pgfsetbuttcap%
\pgfsetroundjoin%
\definecolor{currentfill}{rgb}{0.000000,0.000000,0.000000}%
\pgfsetfillcolor{currentfill}%
\pgfsetlinewidth{0.803000pt}%
\definecolor{currentstroke}{rgb}{0.000000,0.000000,0.000000}%
\pgfsetstrokecolor{currentstroke}%
\pgfsetdash{}{0pt}%
\pgfsys@defobject{currentmarker}{\pgfqpoint{0.000000in}{-0.048611in}}{\pgfqpoint{0.000000in}{0.000000in}}{%
\pgfpathmoveto{\pgfqpoint{0.000000in}{0.000000in}}%
\pgfpathlineto{\pgfqpoint{0.000000in}{-0.048611in}}%
\pgfusepath{stroke,fill}%
}%
\begin{pgfscope}%
\pgfsys@transformshift{5.432969in}{0.415123in}%
\pgfsys@useobject{currentmarker}{}%
\end{pgfscope}%
\end{pgfscope}%
\begin{pgfscope}%
\definecolor{textcolor}{rgb}{0.000000,0.000000,0.000000}%
\pgfsetstrokecolor{textcolor}%
\pgfsetfillcolor{textcolor}%
\pgftext[x=5.432969in,y=0.317901in,,top]{\color{textcolor}\rmfamily\fontsize{10.000000}{12.000000}\selectfont \(\displaystyle {50}\)}%
\end{pgfscope}%
\begin{pgfscope}%
\definecolor{textcolor}{rgb}{0.000000,0.000000,0.000000}%
\pgfsetstrokecolor{textcolor}%
\pgfsetfillcolor{textcolor}%
\pgftext[x=4.666674in,y=0.138889in,,top]{\color{textcolor}\rmfamily\fontsize{10.000000}{12.000000}\selectfont [arcsec]}%
\end{pgfscope}%
\begin{pgfscope}%
\pgfsetbuttcap%
\pgfsetroundjoin%
\definecolor{currentfill}{rgb}{0.000000,0.000000,0.000000}%
\pgfsetfillcolor{currentfill}%
\pgfsetlinewidth{0.803000pt}%
\definecolor{currentstroke}{rgb}{0.000000,0.000000,0.000000}%
\pgfsetstrokecolor{currentstroke}%
\pgfsetdash{}{0pt}%
\pgfsys@defobject{currentmarker}{\pgfqpoint{-0.048611in}{0.000000in}}{\pgfqpoint{-0.000000in}{0.000000in}}{%
\pgfpathmoveto{\pgfqpoint{-0.000000in}{0.000000in}}%
\pgfpathlineto{\pgfqpoint{-0.048611in}{0.000000in}}%
\pgfusepath{stroke,fill}%
}%
\begin{pgfscope}%
\pgfsys@transformshift{3.479338in}{0.684590in}%
\pgfsys@useobject{currentmarker}{}%
\end{pgfscope}%
\end{pgfscope}%
\begin{pgfscope}%
\pgfsetbuttcap%
\pgfsetroundjoin%
\definecolor{currentfill}{rgb}{0.000000,0.000000,0.000000}%
\pgfsetfillcolor{currentfill}%
\pgfsetlinewidth{0.803000pt}%
\definecolor{currentstroke}{rgb}{0.000000,0.000000,0.000000}%
\pgfsetstrokecolor{currentstroke}%
\pgfsetdash{}{0pt}%
\pgfsys@defobject{currentmarker}{\pgfqpoint{-0.048611in}{0.000000in}}{\pgfqpoint{-0.000000in}{0.000000in}}{%
\pgfpathmoveto{\pgfqpoint{-0.000000in}{0.000000in}}%
\pgfpathlineto{\pgfqpoint{-0.048611in}{0.000000in}}%
\pgfusepath{stroke,fill}%
}%
\begin{pgfscope}%
\pgfsys@transformshift{3.479338in}{1.021423in}%
\pgfsys@useobject{currentmarker}{}%
\end{pgfscope}%
\end{pgfscope}%
\begin{pgfscope}%
\pgfsetbuttcap%
\pgfsetroundjoin%
\definecolor{currentfill}{rgb}{0.000000,0.000000,0.000000}%
\pgfsetfillcolor{currentfill}%
\pgfsetlinewidth{0.803000pt}%
\definecolor{currentstroke}{rgb}{0.000000,0.000000,0.000000}%
\pgfsetstrokecolor{currentstroke}%
\pgfsetdash{}{0pt}%
\pgfsys@defobject{currentmarker}{\pgfqpoint{-0.048611in}{0.000000in}}{\pgfqpoint{-0.000000in}{0.000000in}}{%
\pgfpathmoveto{\pgfqpoint{-0.000000in}{0.000000in}}%
\pgfpathlineto{\pgfqpoint{-0.048611in}{0.000000in}}%
\pgfusepath{stroke,fill}%
}%
\begin{pgfscope}%
\pgfsys@transformshift{3.479338in}{1.358255in}%
\pgfsys@useobject{currentmarker}{}%
\end{pgfscope}%
\end{pgfscope}%
\begin{pgfscope}%
\pgfsetrectcap%
\pgfsetmiterjoin%
\pgfsetlinewidth{0.803000pt}%
\definecolor{currentstroke}{rgb}{0.000000,0.000000,0.000000}%
\pgfsetstrokecolor{currentstroke}%
\pgfsetdash{}{0pt}%
\pgfpathmoveto{\pgfqpoint{3.479338in}{0.415123in}}%
\pgfpathlineto{\pgfqpoint{3.479338in}{1.627722in}}%
\pgfusepath{stroke}%
\end{pgfscope}%
\begin{pgfscope}%
\pgfsetrectcap%
\pgfsetmiterjoin%
\pgfsetlinewidth{0.803000pt}%
\definecolor{currentstroke}{rgb}{0.000000,0.000000,0.000000}%
\pgfsetstrokecolor{currentstroke}%
\pgfsetdash{}{0pt}%
\pgfpathmoveto{\pgfqpoint{5.854010in}{0.415123in}}%
\pgfpathlineto{\pgfqpoint{5.854010in}{1.627722in}}%
\pgfusepath{stroke}%
\end{pgfscope}%
\begin{pgfscope}%
\pgfsetrectcap%
\pgfsetmiterjoin%
\pgfsetlinewidth{0.803000pt}%
\definecolor{currentstroke}{rgb}{0.000000,0.000000,0.000000}%
\pgfsetstrokecolor{currentstroke}%
\pgfsetdash{}{0pt}%
\pgfpathmoveto{\pgfqpoint{3.479338in}{0.415123in}}%
\pgfpathlineto{\pgfqpoint{5.854010in}{0.415123in}}%
\pgfusepath{stroke}%
\end{pgfscope}%
\begin{pgfscope}%
\pgfsetrectcap%
\pgfsetmiterjoin%
\pgfsetlinewidth{0.803000pt}%
\definecolor{currentstroke}{rgb}{0.000000,0.000000,0.000000}%
\pgfsetstrokecolor{currentstroke}%
\pgfsetdash{}{0pt}%
\pgfpathmoveto{\pgfqpoint{3.479338in}{1.627722in}}%
\pgfpathlineto{\pgfqpoint{5.854010in}{1.627722in}}%
\pgfusepath{stroke}%
\end{pgfscope}%
\begin{pgfscope}%
\definecolor{textcolor}{rgb}{0.000000,0.000000,0.000000}%
\pgfsetstrokecolor{textcolor}%
\pgfsetfillcolor{textcolor}%
\pgftext[x=4.666674in,y=1.711055in,,base]{\color{textcolor}\rmfamily\fontsize{12.000000}{14.400000}\selectfont 13360-8MHz}%
\end{pgfscope}%
\end{pgfpicture}%
\makeatother%
\endgroup%

%% file: uncertainty0.pgf
\begingroup%
\makeatletter%
\begin{pgfpicture}%
\pgfpathrectangle{\pgfpointorigin}{\pgfqpoint{5.710824in}{3.102304in}}%
\pgfusepath{use as bounding box, clip}%
\begin{pgfscope}%
\pgfsetbuttcap%
\pgfsetmiterjoin%
\definecolor{currentfill}{rgb}{1.000000,1.000000,1.000000}%
\pgfsetfillcolor{currentfill}%
\pgfsetlinewidth{0.000000pt}%
\definecolor{currentstroke}{rgb}{1.000000,1.000000,1.000000}%
\pgfsetstrokecolor{currentstroke}%
\pgfsetdash{}{0pt}%
\pgfpathmoveto{\pgfqpoint{0.000000in}{0.000000in}}%
\pgfpathlineto{\pgfqpoint{5.710824in}{0.000000in}}%
\pgfpathlineto{\pgfqpoint{5.710824in}{3.102304in}}%
\pgfpathlineto{\pgfqpoint{0.000000in}{3.102304in}}%
\pgfpathclose%
\pgfusepath{fill}%
\end{pgfscope}%
\begin{pgfscope}%
\pgfsetbuttcap%
\pgfsetmiterjoin%
\definecolor{currentfill}{rgb}{1.000000,1.000000,1.000000}%
\pgfsetfillcolor{currentfill}%
\pgfsetlinewidth{0.000000pt}%
\definecolor{currentstroke}{rgb}{0.000000,0.000000,0.000000}%
\pgfsetstrokecolor{currentstroke}%
\pgfsetstrokeopacity{0.000000}%
\pgfsetdash{}{0pt}%
\pgfpathmoveto{\pgfqpoint{0.538581in}{1.833863in}}%
\pgfpathlineto{\pgfqpoint{2.677314in}{1.833863in}}%
\pgfpathlineto{\pgfqpoint{2.677314in}{2.903230in}}%
\pgfpathlineto{\pgfqpoint{0.538581in}{2.903230in}}%
\pgfpathclose%
\pgfusepath{fill}%
\end{pgfscope}%
\begin{pgfscope}%
\pgfpathrectangle{\pgfqpoint{0.538581in}{1.833863in}}{\pgfqpoint{2.138733in}{1.069366in}}%
\pgfusepath{clip}%
\pgfsys@transformshift{0.538581in}{1.833863in}%
\pgftext[left,bottom]{\includegraphics[interpolate=true,width=2.140000in,height=1.070000in]{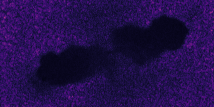}}%
\end{pgfscope}%
\begin{pgfscope}%
\pgfsetbuttcap%
\pgfsetroundjoin%
\definecolor{currentfill}{rgb}{0.000000,0.000000,0.000000}%
\pgfsetfillcolor{currentfill}%
\pgfsetlinewidth{0.803000pt}%
\definecolor{currentstroke}{rgb}{0.000000,0.000000,0.000000}%
\pgfsetstrokecolor{currentstroke}%
\pgfsetdash{}{0pt}%
\pgfsys@defobject{currentmarker}{\pgfqpoint{0.000000in}{-0.048611in}}{\pgfqpoint{0.000000in}{0.000000in}}{%
\pgfpathmoveto{\pgfqpoint{0.000000in}{0.000000in}}%
\pgfpathlineto{\pgfqpoint{0.000000in}{-0.048611in}}%
\pgfusepath{stroke,fill}%
}%
\begin{pgfscope}%
\pgfsys@transformshift{1.013855in}{1.833863in}%
\pgfsys@useobject{currentmarker}{}%
\end{pgfscope}%
\end{pgfscope}%
\begin{pgfscope}%
\pgfsetbuttcap%
\pgfsetroundjoin%
\definecolor{currentfill}{rgb}{0.000000,0.000000,0.000000}%
\pgfsetfillcolor{currentfill}%
\pgfsetlinewidth{0.803000pt}%
\definecolor{currentstroke}{rgb}{0.000000,0.000000,0.000000}%
\pgfsetstrokecolor{currentstroke}%
\pgfsetdash{}{0pt}%
\pgfsys@defobject{currentmarker}{\pgfqpoint{0.000000in}{-0.048611in}}{\pgfqpoint{0.000000in}{0.000000in}}{%
\pgfpathmoveto{\pgfqpoint{0.000000in}{0.000000in}}%
\pgfpathlineto{\pgfqpoint{0.000000in}{-0.048611in}}%
\pgfusepath{stroke,fill}%
}%
\begin{pgfscope}%
\pgfsys@transformshift{1.607947in}{1.833863in}%
\pgfsys@useobject{currentmarker}{}%
\end{pgfscope}%
\end{pgfscope}%
\begin{pgfscope}%
\pgfsetbuttcap%
\pgfsetroundjoin%
\definecolor{currentfill}{rgb}{0.000000,0.000000,0.000000}%
\pgfsetfillcolor{currentfill}%
\pgfsetlinewidth{0.803000pt}%
\definecolor{currentstroke}{rgb}{0.000000,0.000000,0.000000}%
\pgfsetstrokecolor{currentstroke}%
\pgfsetdash{}{0pt}%
\pgfsys@defobject{currentmarker}{\pgfqpoint{0.000000in}{-0.048611in}}{\pgfqpoint{0.000000in}{0.000000in}}{%
\pgfpathmoveto{\pgfqpoint{0.000000in}{0.000000in}}%
\pgfpathlineto{\pgfqpoint{0.000000in}{-0.048611in}}%
\pgfusepath{stroke,fill}%
}%
\begin{pgfscope}%
\pgfsys@transformshift{2.202040in}{1.833863in}%
\pgfsys@useobject{currentmarker}{}%
\end{pgfscope}%
\end{pgfscope}%
\begin{pgfscope}%
\pgfsetbuttcap%
\pgfsetroundjoin%
\definecolor{currentfill}{rgb}{0.000000,0.000000,0.000000}%
\pgfsetfillcolor{currentfill}%
\pgfsetlinewidth{0.803000pt}%
\definecolor{currentstroke}{rgb}{0.000000,0.000000,0.000000}%
\pgfsetstrokecolor{currentstroke}%
\pgfsetdash{}{0pt}%
\pgfsys@defobject{currentmarker}{\pgfqpoint{-0.048611in}{0.000000in}}{\pgfqpoint{-0.000000in}{0.000000in}}{%
\pgfpathmoveto{\pgfqpoint{-0.000000in}{0.000000in}}%
\pgfpathlineto{\pgfqpoint{-0.048611in}{0.000000in}}%
\pgfusepath{stroke,fill}%
}%
\begin{pgfscope}%
\pgfsys@transformshift{0.538581in}{2.071500in}%
\pgfsys@useobject{currentmarker}{}%
\end{pgfscope}%
\end{pgfscope}%
\begin{pgfscope}%
\definecolor{textcolor}{rgb}{0.000000,0.000000,0.000000}%
\pgfsetstrokecolor{textcolor}%
\pgfsetfillcolor{textcolor}%
\pgftext[x=0.194444in, y=2.023275in, left, base]{\color{textcolor}\rmfamily\fontsize{10.000000}{12.000000}\selectfont \(\displaystyle {-25}\)}%
\end{pgfscope}%
\begin{pgfscope}%
\pgfsetbuttcap%
\pgfsetroundjoin%
\definecolor{currentfill}{rgb}{0.000000,0.000000,0.000000}%
\pgfsetfillcolor{currentfill}%
\pgfsetlinewidth{0.803000pt}%
\definecolor{currentstroke}{rgb}{0.000000,0.000000,0.000000}%
\pgfsetstrokecolor{currentstroke}%
\pgfsetdash{}{0pt}%
\pgfsys@defobject{currentmarker}{\pgfqpoint{-0.048611in}{0.000000in}}{\pgfqpoint{-0.000000in}{0.000000in}}{%
\pgfpathmoveto{\pgfqpoint{-0.000000in}{0.000000in}}%
\pgfpathlineto{\pgfqpoint{-0.048611in}{0.000000in}}%
\pgfusepath{stroke,fill}%
}%
\begin{pgfscope}%
\pgfsys@transformshift{0.538581in}{2.368547in}%
\pgfsys@useobject{currentmarker}{}%
\end{pgfscope}%
\end{pgfscope}%
\begin{pgfscope}%
\definecolor{textcolor}{rgb}{0.000000,0.000000,0.000000}%
\pgfsetstrokecolor{textcolor}%
\pgfsetfillcolor{textcolor}%
\pgftext[x=0.371914in, y=2.320321in, left, base]{\color{textcolor}\rmfamily\fontsize{10.000000}{12.000000}\selectfont \(\displaystyle {0}\)}%
\end{pgfscope}%
\begin{pgfscope}%
\pgfsetbuttcap%
\pgfsetroundjoin%
\definecolor{currentfill}{rgb}{0.000000,0.000000,0.000000}%
\pgfsetfillcolor{currentfill}%
\pgfsetlinewidth{0.803000pt}%
\definecolor{currentstroke}{rgb}{0.000000,0.000000,0.000000}%
\pgfsetstrokecolor{currentstroke}%
\pgfsetdash{}{0pt}%
\pgfsys@defobject{currentmarker}{\pgfqpoint{-0.048611in}{0.000000in}}{\pgfqpoint{-0.000000in}{0.000000in}}{%
\pgfpathmoveto{\pgfqpoint{-0.000000in}{0.000000in}}%
\pgfpathlineto{\pgfqpoint{-0.048611in}{0.000000in}}%
\pgfusepath{stroke,fill}%
}%
\begin{pgfscope}%
\pgfsys@transformshift{0.538581in}{2.665593in}%
\pgfsys@useobject{currentmarker}{}%
\end{pgfscope}%
\end{pgfscope}%
\begin{pgfscope}%
\definecolor{textcolor}{rgb}{0.000000,0.000000,0.000000}%
\pgfsetstrokecolor{textcolor}%
\pgfsetfillcolor{textcolor}%
\pgftext[x=0.302469in, y=2.617368in, left, base]{\color{textcolor}\rmfamily\fontsize{10.000000}{12.000000}\selectfont \(\displaystyle {25}\)}%
\end{pgfscope}%
\begin{pgfscope}%
\definecolor{textcolor}{rgb}{0.000000,0.000000,0.000000}%
\pgfsetstrokecolor{textcolor}%
\pgfsetfillcolor{textcolor}%
\pgftext[x=0.138889in,y=2.368547in,,bottom,rotate=90.000000]{\color{textcolor}\rmfamily\fontsize{10.000000}{12.000000}\selectfont [arcsec]}%
\end{pgfscope}%
\begin{pgfscope}%
\pgfsetrectcap%
\pgfsetmiterjoin%
\pgfsetlinewidth{0.803000pt}%
\definecolor{currentstroke}{rgb}{0.000000,0.000000,0.000000}%
\pgfsetstrokecolor{currentstroke}%
\pgfsetdash{}{0pt}%
\pgfpathmoveto{\pgfqpoint{0.538581in}{1.833863in}}%
\pgfpathlineto{\pgfqpoint{0.538581in}{2.903230in}}%
\pgfusepath{stroke}%
\end{pgfscope}%
\begin{pgfscope}%
\pgfsetrectcap%
\pgfsetmiterjoin%
\pgfsetlinewidth{0.803000pt}%
\definecolor{currentstroke}{rgb}{0.000000,0.000000,0.000000}%
\pgfsetstrokecolor{currentstroke}%
\pgfsetdash{}{0pt}%
\pgfpathmoveto{\pgfqpoint{2.677314in}{1.833863in}}%
\pgfpathlineto{\pgfqpoint{2.677314in}{2.903230in}}%
\pgfusepath{stroke}%
\end{pgfscope}%
\begin{pgfscope}%
\pgfsetrectcap%
\pgfsetmiterjoin%
\pgfsetlinewidth{0.803000pt}%
\definecolor{currentstroke}{rgb}{0.000000,0.000000,0.000000}%
\pgfsetstrokecolor{currentstroke}%
\pgfsetdash{}{0pt}%
\pgfpathmoveto{\pgfqpoint{0.538581in}{1.833863in}}%
\pgfpathlineto{\pgfqpoint{2.677314in}{1.833863in}}%
\pgfusepath{stroke}%
\end{pgfscope}%
\begin{pgfscope}%
\pgfsetrectcap%
\pgfsetmiterjoin%
\pgfsetlinewidth{0.803000pt}%
\definecolor{currentstroke}{rgb}{0.000000,0.000000,0.000000}%
\pgfsetstrokecolor{currentstroke}%
\pgfsetdash{}{0pt}%
\pgfpathmoveto{\pgfqpoint{0.538581in}{2.903230in}}%
\pgfpathlineto{\pgfqpoint{2.677314in}{2.903230in}}%
\pgfusepath{stroke}%
\end{pgfscope}%
\begin{pgfscope}%
\definecolor{textcolor}{rgb}{0.000000,0.000000,0.000000}%
\pgfsetstrokecolor{textcolor}%
\pgfsetfillcolor{textcolor}%
\pgftext[x=1.607947in,y=2.986563in,,base]{\color{textcolor}\rmfamily\fontsize{12.000000}{14.400000}\selectfont 2052-2MHz}%
\end{pgfscope}%
\begin{pgfscope}%
\pgfsetbuttcap%
\pgfsetmiterjoin%
\definecolor{currentfill}{rgb}{1.000000,1.000000,1.000000}%
\pgfsetfillcolor{currentfill}%
\pgfsetlinewidth{0.000000pt}%
\definecolor{currentstroke}{rgb}{0.000000,0.000000,0.000000}%
\pgfsetstrokecolor{currentstroke}%
\pgfsetstrokeopacity{0.000000}%
\pgfsetdash{}{0pt}%
\pgfpathmoveto{\pgfqpoint{2.891187in}{1.833863in}}%
\pgfpathlineto{\pgfqpoint{5.029919in}{1.833863in}}%
\pgfpathlineto{\pgfqpoint{5.029919in}{2.903230in}}%
\pgfpathlineto{\pgfqpoint{2.891187in}{2.903230in}}%
\pgfpathclose%
\pgfusepath{fill}%
\end{pgfscope}%
\begin{pgfscope}%
\pgfpathrectangle{\pgfqpoint{2.891187in}{1.833863in}}{\pgfqpoint{2.138733in}{1.069366in}}%
\pgfusepath{clip}%
\pgfsys@transformshift{2.891187in}{1.833863in}%
\pgftext[left,bottom]{\includegraphics[interpolate=true,width=2.140000in,height=1.070000in]{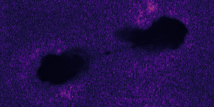}}%
\end{pgfscope}%
\begin{pgfscope}%
\pgfsetbuttcap%
\pgfsetroundjoin%
\definecolor{currentfill}{rgb}{0.000000,0.000000,0.000000}%
\pgfsetfillcolor{currentfill}%
\pgfsetlinewidth{0.803000pt}%
\definecolor{currentstroke}{rgb}{0.000000,0.000000,0.000000}%
\pgfsetstrokecolor{currentstroke}%
\pgfsetdash{}{0pt}%
\pgfsys@defobject{currentmarker}{\pgfqpoint{0.000000in}{-0.048611in}}{\pgfqpoint{0.000000in}{0.000000in}}{%
\pgfpathmoveto{\pgfqpoint{0.000000in}{0.000000in}}%
\pgfpathlineto{\pgfqpoint{0.000000in}{-0.048611in}}%
\pgfusepath{stroke,fill}%
}%
\begin{pgfscope}%
\pgfsys@transformshift{3.366461in}{1.833863in}%
\pgfsys@useobject{currentmarker}{}%
\end{pgfscope}%
\end{pgfscope}%
\begin{pgfscope}%
\pgfsetbuttcap%
\pgfsetroundjoin%
\definecolor{currentfill}{rgb}{0.000000,0.000000,0.000000}%
\pgfsetfillcolor{currentfill}%
\pgfsetlinewidth{0.803000pt}%
\definecolor{currentstroke}{rgb}{0.000000,0.000000,0.000000}%
\pgfsetstrokecolor{currentstroke}%
\pgfsetdash{}{0pt}%
\pgfsys@defobject{currentmarker}{\pgfqpoint{0.000000in}{-0.048611in}}{\pgfqpoint{0.000000in}{0.000000in}}{%
\pgfpathmoveto{\pgfqpoint{0.000000in}{0.000000in}}%
\pgfpathlineto{\pgfqpoint{0.000000in}{-0.048611in}}%
\pgfusepath{stroke,fill}%
}%
\begin{pgfscope}%
\pgfsys@transformshift{3.960553in}{1.833863in}%
\pgfsys@useobject{currentmarker}{}%
\end{pgfscope}%
\end{pgfscope}%
\begin{pgfscope}%
\pgfsetbuttcap%
\pgfsetroundjoin%
\definecolor{currentfill}{rgb}{0.000000,0.000000,0.000000}%
\pgfsetfillcolor{currentfill}%
\pgfsetlinewidth{0.803000pt}%
\definecolor{currentstroke}{rgb}{0.000000,0.000000,0.000000}%
\pgfsetstrokecolor{currentstroke}%
\pgfsetdash{}{0pt}%
\pgfsys@defobject{currentmarker}{\pgfqpoint{0.000000in}{-0.048611in}}{\pgfqpoint{0.000000in}{0.000000in}}{%
\pgfpathmoveto{\pgfqpoint{0.000000in}{0.000000in}}%
\pgfpathlineto{\pgfqpoint{0.000000in}{-0.048611in}}%
\pgfusepath{stroke,fill}%
}%
\begin{pgfscope}%
\pgfsys@transformshift{4.554646in}{1.833863in}%
\pgfsys@useobject{currentmarker}{}%
\end{pgfscope}%
\end{pgfscope}%
\begin{pgfscope}%
\pgfsetbuttcap%
\pgfsetroundjoin%
\definecolor{currentfill}{rgb}{0.000000,0.000000,0.000000}%
\pgfsetfillcolor{currentfill}%
\pgfsetlinewidth{0.803000pt}%
\definecolor{currentstroke}{rgb}{0.000000,0.000000,0.000000}%
\pgfsetstrokecolor{currentstroke}%
\pgfsetdash{}{0pt}%
\pgfsys@defobject{currentmarker}{\pgfqpoint{-0.048611in}{0.000000in}}{\pgfqpoint{-0.000000in}{0.000000in}}{%
\pgfpathmoveto{\pgfqpoint{-0.000000in}{0.000000in}}%
\pgfpathlineto{\pgfqpoint{-0.048611in}{0.000000in}}%
\pgfusepath{stroke,fill}%
}%
\begin{pgfscope}%
\pgfsys@transformshift{2.891187in}{2.071500in}%
\pgfsys@useobject{currentmarker}{}%
\end{pgfscope}%
\end{pgfscope}%
\begin{pgfscope}%
\pgfsetbuttcap%
\pgfsetroundjoin%
\definecolor{currentfill}{rgb}{0.000000,0.000000,0.000000}%
\pgfsetfillcolor{currentfill}%
\pgfsetlinewidth{0.803000pt}%
\definecolor{currentstroke}{rgb}{0.000000,0.000000,0.000000}%
\pgfsetstrokecolor{currentstroke}%
\pgfsetdash{}{0pt}%
\pgfsys@defobject{currentmarker}{\pgfqpoint{-0.048611in}{0.000000in}}{\pgfqpoint{-0.000000in}{0.000000in}}{%
\pgfpathmoveto{\pgfqpoint{-0.000000in}{0.000000in}}%
\pgfpathlineto{\pgfqpoint{-0.048611in}{0.000000in}}%
\pgfusepath{stroke,fill}%
}%
\begin{pgfscope}%
\pgfsys@transformshift{2.891187in}{2.368547in}%
\pgfsys@useobject{currentmarker}{}%
\end{pgfscope}%
\end{pgfscope}%
\begin{pgfscope}%
\pgfsetbuttcap%
\pgfsetroundjoin%
\definecolor{currentfill}{rgb}{0.000000,0.000000,0.000000}%
\pgfsetfillcolor{currentfill}%
\pgfsetlinewidth{0.803000pt}%
\definecolor{currentstroke}{rgb}{0.000000,0.000000,0.000000}%
\pgfsetstrokecolor{currentstroke}%
\pgfsetdash{}{0pt}%
\pgfsys@defobject{currentmarker}{\pgfqpoint{-0.048611in}{0.000000in}}{\pgfqpoint{-0.000000in}{0.000000in}}{%
\pgfpathmoveto{\pgfqpoint{-0.000000in}{0.000000in}}%
\pgfpathlineto{\pgfqpoint{-0.048611in}{0.000000in}}%
\pgfusepath{stroke,fill}%
}%
\begin{pgfscope}%
\pgfsys@transformshift{2.891187in}{2.665593in}%
\pgfsys@useobject{currentmarker}{}%
\end{pgfscope}%
\end{pgfscope}%
\begin{pgfscope}%
\pgfsetrectcap%
\pgfsetmiterjoin%
\pgfsetlinewidth{0.803000pt}%
\definecolor{currentstroke}{rgb}{0.000000,0.000000,0.000000}%
\pgfsetstrokecolor{currentstroke}%
\pgfsetdash{}{0pt}%
\pgfpathmoveto{\pgfqpoint{2.891187in}{1.833863in}}%
\pgfpathlineto{\pgfqpoint{2.891187in}{2.903230in}}%
\pgfusepath{stroke}%
\end{pgfscope}%
\begin{pgfscope}%
\pgfsetrectcap%
\pgfsetmiterjoin%
\pgfsetlinewidth{0.803000pt}%
\definecolor{currentstroke}{rgb}{0.000000,0.000000,0.000000}%
\pgfsetstrokecolor{currentstroke}%
\pgfsetdash{}{0pt}%
\pgfpathmoveto{\pgfqpoint{5.029919in}{1.833863in}}%
\pgfpathlineto{\pgfqpoint{5.029919in}{2.903230in}}%
\pgfusepath{stroke}%
\end{pgfscope}%
\begin{pgfscope}%
\pgfsetrectcap%
\pgfsetmiterjoin%
\pgfsetlinewidth{0.803000pt}%
\definecolor{currentstroke}{rgb}{0.000000,0.000000,0.000000}%
\pgfsetstrokecolor{currentstroke}%
\pgfsetdash{}{0pt}%
\pgfpathmoveto{\pgfqpoint{2.891187in}{1.833863in}}%
\pgfpathlineto{\pgfqpoint{5.029919in}{1.833863in}}%
\pgfusepath{stroke}%
\end{pgfscope}%
\begin{pgfscope}%
\pgfsetrectcap%
\pgfsetmiterjoin%
\pgfsetlinewidth{0.803000pt}%
\definecolor{currentstroke}{rgb}{0.000000,0.000000,0.000000}%
\pgfsetstrokecolor{currentstroke}%
\pgfsetdash{}{0pt}%
\pgfpathmoveto{\pgfqpoint{2.891187in}{2.903230in}}%
\pgfpathlineto{\pgfqpoint{5.029919in}{2.903230in}}%
\pgfusepath{stroke}%
\end{pgfscope}%
\begin{pgfscope}%
\definecolor{textcolor}{rgb}{0.000000,0.000000,0.000000}%
\pgfsetstrokecolor{textcolor}%
\pgfsetfillcolor{textcolor}%
\pgftext[x=3.960553in,y=2.986563in,,base]{\color{textcolor}\rmfamily\fontsize{12.000000}{14.400000}\selectfont 4811-8MHz}%
\end{pgfscope}%
\begin{pgfscope}%
\pgfsetbuttcap%
\pgfsetmiterjoin%
\definecolor{currentfill}{rgb}{1.000000,1.000000,1.000000}%
\pgfsetfillcolor{currentfill}%
\pgfsetlinewidth{0.000000pt}%
\definecolor{currentstroke}{rgb}{0.000000,0.000000,0.000000}%
\pgfsetstrokecolor{currentstroke}%
\pgfsetstrokeopacity{0.000000}%
\pgfsetdash{}{0pt}%
\pgfpathmoveto{\pgfqpoint{0.538581in}{0.415123in}}%
\pgfpathlineto{\pgfqpoint{2.677314in}{0.415123in}}%
\pgfpathlineto{\pgfqpoint{2.677314in}{1.484490in}}%
\pgfpathlineto{\pgfqpoint{0.538581in}{1.484490in}}%
\pgfpathclose%
\pgfusepath{fill}%
\end{pgfscope}%
\begin{pgfscope}%
\pgfpathrectangle{\pgfqpoint{0.538581in}{0.415123in}}{\pgfqpoint{2.138733in}{1.069366in}}%
\pgfusepath{clip}%
\pgfsys@transformshift{0.538581in}{0.415123in}%
\pgftext[left,bottom]{\includegraphics[interpolate=true,width=2.140000in,height=1.070000in]{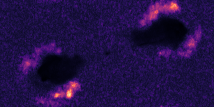}}%
\end{pgfscope}%
\begin{pgfscope}%
\pgfsetbuttcap%
\pgfsetroundjoin%
\definecolor{currentfill}{rgb}{0.000000,0.000000,0.000000}%
\pgfsetfillcolor{currentfill}%
\pgfsetlinewidth{0.803000pt}%
\definecolor{currentstroke}{rgb}{0.000000,0.000000,0.000000}%
\pgfsetstrokecolor{currentstroke}%
\pgfsetdash{}{0pt}%
\pgfsys@defobject{currentmarker}{\pgfqpoint{0.000000in}{-0.048611in}}{\pgfqpoint{0.000000in}{0.000000in}}{%
\pgfpathmoveto{\pgfqpoint{0.000000in}{0.000000in}}%
\pgfpathlineto{\pgfqpoint{0.000000in}{-0.048611in}}%
\pgfusepath{stroke,fill}%
}%
\begin{pgfscope}%
\pgfsys@transformshift{1.013855in}{0.415123in}%
\pgfsys@useobject{currentmarker}{}%
\end{pgfscope}%
\end{pgfscope}%
\begin{pgfscope}%
\definecolor{textcolor}{rgb}{0.000000,0.000000,0.000000}%
\pgfsetstrokecolor{textcolor}%
\pgfsetfillcolor{textcolor}%
\pgftext[x=1.013855in,y=0.317901in,,top]{\color{textcolor}\rmfamily\fontsize{10.000000}{12.000000}\selectfont \(\displaystyle {-50}\)}%
\end{pgfscope}%
\begin{pgfscope}%
\pgfsetbuttcap%
\pgfsetroundjoin%
\definecolor{currentfill}{rgb}{0.000000,0.000000,0.000000}%
\pgfsetfillcolor{currentfill}%
\pgfsetlinewidth{0.803000pt}%
\definecolor{currentstroke}{rgb}{0.000000,0.000000,0.000000}%
\pgfsetstrokecolor{currentstroke}%
\pgfsetdash{}{0pt}%
\pgfsys@defobject{currentmarker}{\pgfqpoint{0.000000in}{-0.048611in}}{\pgfqpoint{0.000000in}{0.000000in}}{%
\pgfpathmoveto{\pgfqpoint{0.000000in}{0.000000in}}%
\pgfpathlineto{\pgfqpoint{0.000000in}{-0.048611in}}%
\pgfusepath{stroke,fill}%
}%
\begin{pgfscope}%
\pgfsys@transformshift{1.607947in}{0.415123in}%
\pgfsys@useobject{currentmarker}{}%
\end{pgfscope}%
\end{pgfscope}%
\begin{pgfscope}%
\definecolor{textcolor}{rgb}{0.000000,0.000000,0.000000}%
\pgfsetstrokecolor{textcolor}%
\pgfsetfillcolor{textcolor}%
\pgftext[x=1.607947in,y=0.317901in,,top]{\color{textcolor}\rmfamily\fontsize{10.000000}{12.000000}\selectfont \(\displaystyle {0}\)}%
\end{pgfscope}%
\begin{pgfscope}%
\pgfsetbuttcap%
\pgfsetroundjoin%
\definecolor{currentfill}{rgb}{0.000000,0.000000,0.000000}%
\pgfsetfillcolor{currentfill}%
\pgfsetlinewidth{0.803000pt}%
\definecolor{currentstroke}{rgb}{0.000000,0.000000,0.000000}%
\pgfsetstrokecolor{currentstroke}%
\pgfsetdash{}{0pt}%
\pgfsys@defobject{currentmarker}{\pgfqpoint{0.000000in}{-0.048611in}}{\pgfqpoint{0.000000in}{0.000000in}}{%
\pgfpathmoveto{\pgfqpoint{0.000000in}{0.000000in}}%
\pgfpathlineto{\pgfqpoint{0.000000in}{-0.048611in}}%
\pgfusepath{stroke,fill}%
}%
\begin{pgfscope}%
\pgfsys@transformshift{2.202040in}{0.415123in}%
\pgfsys@useobject{currentmarker}{}%
\end{pgfscope}%
\end{pgfscope}%
\begin{pgfscope}%
\definecolor{textcolor}{rgb}{0.000000,0.000000,0.000000}%
\pgfsetstrokecolor{textcolor}%
\pgfsetfillcolor{textcolor}%
\pgftext[x=2.202040in,y=0.317901in,,top]{\color{textcolor}\rmfamily\fontsize{10.000000}{12.000000}\selectfont \(\displaystyle {50}\)}%
\end{pgfscope}%
\begin{pgfscope}%
\definecolor{textcolor}{rgb}{0.000000,0.000000,0.000000}%
\pgfsetstrokecolor{textcolor}%
\pgfsetfillcolor{textcolor}%
\pgftext[x=1.607947in,y=0.138889in,,top]{\color{textcolor}\rmfamily\fontsize{10.000000}{12.000000}\selectfont [arcsec]}%
\end{pgfscope}%
\begin{pgfscope}%
\pgfsetbuttcap%
\pgfsetroundjoin%
\definecolor{currentfill}{rgb}{0.000000,0.000000,0.000000}%
\pgfsetfillcolor{currentfill}%
\pgfsetlinewidth{0.803000pt}%
\definecolor{currentstroke}{rgb}{0.000000,0.000000,0.000000}%
\pgfsetstrokecolor{currentstroke}%
\pgfsetdash{}{0pt}%
\pgfsys@defobject{currentmarker}{\pgfqpoint{-0.048611in}{0.000000in}}{\pgfqpoint{-0.000000in}{0.000000in}}{%
\pgfpathmoveto{\pgfqpoint{-0.000000in}{0.000000in}}%
\pgfpathlineto{\pgfqpoint{-0.048611in}{0.000000in}}%
\pgfusepath{stroke,fill}%
}%
\begin{pgfscope}%
\pgfsys@transformshift{0.538581in}{0.652760in}%
\pgfsys@useobject{currentmarker}{}%
\end{pgfscope}%
\end{pgfscope}%
\begin{pgfscope}%
\definecolor{textcolor}{rgb}{0.000000,0.000000,0.000000}%
\pgfsetstrokecolor{textcolor}%
\pgfsetfillcolor{textcolor}%
\pgftext[x=0.194444in, y=0.604535in, left, base]{\color{textcolor}\rmfamily\fontsize{10.000000}{12.000000}\selectfont \(\displaystyle {-25}\)}%
\end{pgfscope}%
\begin{pgfscope}%
\pgfsetbuttcap%
\pgfsetroundjoin%
\definecolor{currentfill}{rgb}{0.000000,0.000000,0.000000}%
\pgfsetfillcolor{currentfill}%
\pgfsetlinewidth{0.803000pt}%
\definecolor{currentstroke}{rgb}{0.000000,0.000000,0.000000}%
\pgfsetstrokecolor{currentstroke}%
\pgfsetdash{}{0pt}%
\pgfsys@defobject{currentmarker}{\pgfqpoint{-0.048611in}{0.000000in}}{\pgfqpoint{-0.000000in}{0.000000in}}{%
\pgfpathmoveto{\pgfqpoint{-0.000000in}{0.000000in}}%
\pgfpathlineto{\pgfqpoint{-0.048611in}{0.000000in}}%
\pgfusepath{stroke,fill}%
}%
\begin{pgfscope}%
\pgfsys@transformshift{0.538581in}{0.949806in}%
\pgfsys@useobject{currentmarker}{}%
\end{pgfscope}%
\end{pgfscope}%
\begin{pgfscope}%
\definecolor{textcolor}{rgb}{0.000000,0.000000,0.000000}%
\pgfsetstrokecolor{textcolor}%
\pgfsetfillcolor{textcolor}%
\pgftext[x=0.371914in, y=0.901581in, left, base]{\color{textcolor}\rmfamily\fontsize{10.000000}{12.000000}\selectfont \(\displaystyle {0}\)}%
\end{pgfscope}%
\begin{pgfscope}%
\pgfsetbuttcap%
\pgfsetroundjoin%
\definecolor{currentfill}{rgb}{0.000000,0.000000,0.000000}%
\pgfsetfillcolor{currentfill}%
\pgfsetlinewidth{0.803000pt}%
\definecolor{currentstroke}{rgb}{0.000000,0.000000,0.000000}%
\pgfsetstrokecolor{currentstroke}%
\pgfsetdash{}{0pt}%
\pgfsys@defobject{currentmarker}{\pgfqpoint{-0.048611in}{0.000000in}}{\pgfqpoint{-0.000000in}{0.000000in}}{%
\pgfpathmoveto{\pgfqpoint{-0.000000in}{0.000000in}}%
\pgfpathlineto{\pgfqpoint{-0.048611in}{0.000000in}}%
\pgfusepath{stroke,fill}%
}%
\begin{pgfscope}%
\pgfsys@transformshift{0.538581in}{1.246853in}%
\pgfsys@useobject{currentmarker}{}%
\end{pgfscope}%
\end{pgfscope}%
\begin{pgfscope}%
\definecolor{textcolor}{rgb}{0.000000,0.000000,0.000000}%
\pgfsetstrokecolor{textcolor}%
\pgfsetfillcolor{textcolor}%
\pgftext[x=0.302469in, y=1.198627in, left, base]{\color{textcolor}\rmfamily\fontsize{10.000000}{12.000000}\selectfont \(\displaystyle {25}\)}%
\end{pgfscope}%
\begin{pgfscope}%
\definecolor{textcolor}{rgb}{0.000000,0.000000,0.000000}%
\pgfsetstrokecolor{textcolor}%
\pgfsetfillcolor{textcolor}%
\pgftext[x=0.138889in,y=0.949806in,,bottom,rotate=90.000000]{\color{textcolor}\rmfamily\fontsize{10.000000}{12.000000}\selectfont [arcsec]}%
\end{pgfscope}%
\begin{pgfscope}%
\pgfsetrectcap%
\pgfsetmiterjoin%
\pgfsetlinewidth{0.803000pt}%
\definecolor{currentstroke}{rgb}{0.000000,0.000000,0.000000}%
\pgfsetstrokecolor{currentstroke}%
\pgfsetdash{}{0pt}%
\pgfpathmoveto{\pgfqpoint{0.538581in}{0.415123in}}%
\pgfpathlineto{\pgfqpoint{0.538581in}{1.484490in}}%
\pgfusepath{stroke}%
\end{pgfscope}%
\begin{pgfscope}%
\pgfsetrectcap%
\pgfsetmiterjoin%
\pgfsetlinewidth{0.803000pt}%
\definecolor{currentstroke}{rgb}{0.000000,0.000000,0.000000}%
\pgfsetstrokecolor{currentstroke}%
\pgfsetdash{}{0pt}%
\pgfpathmoveto{\pgfqpoint{2.677314in}{0.415123in}}%
\pgfpathlineto{\pgfqpoint{2.677314in}{1.484490in}}%
\pgfusepath{stroke}%
\end{pgfscope}%
\begin{pgfscope}%
\pgfsetrectcap%
\pgfsetmiterjoin%
\pgfsetlinewidth{0.803000pt}%
\definecolor{currentstroke}{rgb}{0.000000,0.000000,0.000000}%
\pgfsetstrokecolor{currentstroke}%
\pgfsetdash{}{0pt}%
\pgfpathmoveto{\pgfqpoint{0.538581in}{0.415123in}}%
\pgfpathlineto{\pgfqpoint{2.677314in}{0.415123in}}%
\pgfusepath{stroke}%
\end{pgfscope}%
\begin{pgfscope}%
\pgfsetrectcap%
\pgfsetmiterjoin%
\pgfsetlinewidth{0.803000pt}%
\definecolor{currentstroke}{rgb}{0.000000,0.000000,0.000000}%
\pgfsetstrokecolor{currentstroke}%
\pgfsetdash{}{0pt}%
\pgfpathmoveto{\pgfqpoint{0.538581in}{1.484490in}}%
\pgfpathlineto{\pgfqpoint{2.677314in}{1.484490in}}%
\pgfusepath{stroke}%
\end{pgfscope}%
\begin{pgfscope}%
\definecolor{textcolor}{rgb}{0.000000,0.000000,0.000000}%
\pgfsetstrokecolor{textcolor}%
\pgfsetfillcolor{textcolor}%
\pgftext[x=1.607947in,y=1.567823in,,base]{\color{textcolor}\rmfamily\fontsize{12.000000}{14.400000}\selectfont 8427-8MHz}%
\end{pgfscope}%
\begin{pgfscope}%
\pgfsetbuttcap%
\pgfsetmiterjoin%
\definecolor{currentfill}{rgb}{1.000000,1.000000,1.000000}%
\pgfsetfillcolor{currentfill}%
\pgfsetlinewidth{0.000000pt}%
\definecolor{currentstroke}{rgb}{0.000000,0.000000,0.000000}%
\pgfsetstrokecolor{currentstroke}%
\pgfsetstrokeopacity{0.000000}%
\pgfsetdash{}{0pt}%
\pgfpathmoveto{\pgfqpoint{2.891187in}{0.415123in}}%
\pgfpathlineto{\pgfqpoint{5.029919in}{0.415123in}}%
\pgfpathlineto{\pgfqpoint{5.029919in}{1.484490in}}%
\pgfpathlineto{\pgfqpoint{2.891187in}{1.484490in}}%
\pgfpathclose%
\pgfusepath{fill}%
\end{pgfscope}%
\begin{pgfscope}%
\pgfpathrectangle{\pgfqpoint{2.891187in}{0.415123in}}{\pgfqpoint{2.138733in}{1.069366in}}%
\pgfusepath{clip}%
\pgfsys@transformshift{2.891187in}{0.415123in}%
\pgftext[left,bottom]{\includegraphics[interpolate=true,width=2.140000in,height=1.070000in]{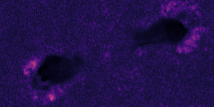}}%
\end{pgfscope}%
\begin{pgfscope}%
\pgfsetbuttcap%
\pgfsetroundjoin%
\definecolor{currentfill}{rgb}{0.000000,0.000000,0.000000}%
\pgfsetfillcolor{currentfill}%
\pgfsetlinewidth{0.803000pt}%
\definecolor{currentstroke}{rgb}{0.000000,0.000000,0.000000}%
\pgfsetstrokecolor{currentstroke}%
\pgfsetdash{}{0pt}%
\pgfsys@defobject{currentmarker}{\pgfqpoint{0.000000in}{-0.048611in}}{\pgfqpoint{0.000000in}{0.000000in}}{%
\pgfpathmoveto{\pgfqpoint{0.000000in}{0.000000in}}%
\pgfpathlineto{\pgfqpoint{0.000000in}{-0.048611in}}%
\pgfusepath{stroke,fill}%
}%
\begin{pgfscope}%
\pgfsys@transformshift{3.366461in}{0.415123in}%
\pgfsys@useobject{currentmarker}{}%
\end{pgfscope}%
\end{pgfscope}%
\begin{pgfscope}%
\definecolor{textcolor}{rgb}{0.000000,0.000000,0.000000}%
\pgfsetstrokecolor{textcolor}%
\pgfsetfillcolor{textcolor}%
\pgftext[x=3.366461in,y=0.317901in,,top]{\color{textcolor}\rmfamily\fontsize{10.000000}{12.000000}\selectfont \(\displaystyle {-50}\)}%
\end{pgfscope}%
\begin{pgfscope}%
\pgfsetbuttcap%
\pgfsetroundjoin%
\definecolor{currentfill}{rgb}{0.000000,0.000000,0.000000}%
\pgfsetfillcolor{currentfill}%
\pgfsetlinewidth{0.803000pt}%
\definecolor{currentstroke}{rgb}{0.000000,0.000000,0.000000}%
\pgfsetstrokecolor{currentstroke}%
\pgfsetdash{}{0pt}%
\pgfsys@defobject{currentmarker}{\pgfqpoint{0.000000in}{-0.048611in}}{\pgfqpoint{0.000000in}{0.000000in}}{%
\pgfpathmoveto{\pgfqpoint{0.000000in}{0.000000in}}%
\pgfpathlineto{\pgfqpoint{0.000000in}{-0.048611in}}%
\pgfusepath{stroke,fill}%
}%
\begin{pgfscope}%
\pgfsys@transformshift{3.960553in}{0.415123in}%
\pgfsys@useobject{currentmarker}{}%
\end{pgfscope}%
\end{pgfscope}%
\begin{pgfscope}%
\definecolor{textcolor}{rgb}{0.000000,0.000000,0.000000}%
\pgfsetstrokecolor{textcolor}%
\pgfsetfillcolor{textcolor}%
\pgftext[x=3.960553in,y=0.317901in,,top]{\color{textcolor}\rmfamily\fontsize{10.000000}{12.000000}\selectfont \(\displaystyle {0}\)}%
\end{pgfscope}%
\begin{pgfscope}%
\pgfsetbuttcap%
\pgfsetroundjoin%
\definecolor{currentfill}{rgb}{0.000000,0.000000,0.000000}%
\pgfsetfillcolor{currentfill}%
\pgfsetlinewidth{0.803000pt}%
\definecolor{currentstroke}{rgb}{0.000000,0.000000,0.000000}%
\pgfsetstrokecolor{currentstroke}%
\pgfsetdash{}{0pt}%
\pgfsys@defobject{currentmarker}{\pgfqpoint{0.000000in}{-0.048611in}}{\pgfqpoint{0.000000in}{0.000000in}}{%
\pgfpathmoveto{\pgfqpoint{0.000000in}{0.000000in}}%
\pgfpathlineto{\pgfqpoint{0.000000in}{-0.048611in}}%
\pgfusepath{stroke,fill}%
}%
\begin{pgfscope}%
\pgfsys@transformshift{4.554646in}{0.415123in}%
\pgfsys@useobject{currentmarker}{}%
\end{pgfscope}%
\end{pgfscope}%
\begin{pgfscope}%
\definecolor{textcolor}{rgb}{0.000000,0.000000,0.000000}%
\pgfsetstrokecolor{textcolor}%
\pgfsetfillcolor{textcolor}%
\pgftext[x=4.554646in,y=0.317901in,,top]{\color{textcolor}\rmfamily\fontsize{10.000000}{12.000000}\selectfont \(\displaystyle {50}\)}%
\end{pgfscope}%
\begin{pgfscope}%
\definecolor{textcolor}{rgb}{0.000000,0.000000,0.000000}%
\pgfsetstrokecolor{textcolor}%
\pgfsetfillcolor{textcolor}%
\pgftext[x=3.960553in,y=0.138889in,,top]{\color{textcolor}\rmfamily\fontsize{10.000000}{12.000000}\selectfont [arcsec]}%
\end{pgfscope}%
\begin{pgfscope}%
\pgfsetbuttcap%
\pgfsetroundjoin%
\definecolor{currentfill}{rgb}{0.000000,0.000000,0.000000}%
\pgfsetfillcolor{currentfill}%
\pgfsetlinewidth{0.803000pt}%
\definecolor{currentstroke}{rgb}{0.000000,0.000000,0.000000}%
\pgfsetstrokecolor{currentstroke}%
\pgfsetdash{}{0pt}%
\pgfsys@defobject{currentmarker}{\pgfqpoint{-0.048611in}{0.000000in}}{\pgfqpoint{-0.000000in}{0.000000in}}{%
\pgfpathmoveto{\pgfqpoint{-0.000000in}{0.000000in}}%
\pgfpathlineto{\pgfqpoint{-0.048611in}{0.000000in}}%
\pgfusepath{stroke,fill}%
}%
\begin{pgfscope}%
\pgfsys@transformshift{2.891187in}{0.652760in}%
\pgfsys@useobject{currentmarker}{}%
\end{pgfscope}%
\end{pgfscope}%
\begin{pgfscope}%
\pgfsetbuttcap%
\pgfsetroundjoin%
\definecolor{currentfill}{rgb}{0.000000,0.000000,0.000000}%
\pgfsetfillcolor{currentfill}%
\pgfsetlinewidth{0.803000pt}%
\definecolor{currentstroke}{rgb}{0.000000,0.000000,0.000000}%
\pgfsetstrokecolor{currentstroke}%
\pgfsetdash{}{0pt}%
\pgfsys@defobject{currentmarker}{\pgfqpoint{-0.048611in}{0.000000in}}{\pgfqpoint{-0.000000in}{0.000000in}}{%
\pgfpathmoveto{\pgfqpoint{-0.000000in}{0.000000in}}%
\pgfpathlineto{\pgfqpoint{-0.048611in}{0.000000in}}%
\pgfusepath{stroke,fill}%
}%
\begin{pgfscope}%
\pgfsys@transformshift{2.891187in}{0.949806in}%
\pgfsys@useobject{currentmarker}{}%
\end{pgfscope}%
\end{pgfscope}%
\begin{pgfscope}%
\pgfsetbuttcap%
\pgfsetroundjoin%
\definecolor{currentfill}{rgb}{0.000000,0.000000,0.000000}%
\pgfsetfillcolor{currentfill}%
\pgfsetlinewidth{0.803000pt}%
\definecolor{currentstroke}{rgb}{0.000000,0.000000,0.000000}%
\pgfsetstrokecolor{currentstroke}%
\pgfsetdash{}{0pt}%
\pgfsys@defobject{currentmarker}{\pgfqpoint{-0.048611in}{0.000000in}}{\pgfqpoint{-0.000000in}{0.000000in}}{%
\pgfpathmoveto{\pgfqpoint{-0.000000in}{0.000000in}}%
\pgfpathlineto{\pgfqpoint{-0.048611in}{0.000000in}}%
\pgfusepath{stroke,fill}%
}%
\begin{pgfscope}%
\pgfsys@transformshift{2.891187in}{1.246853in}%
\pgfsys@useobject{currentmarker}{}%
\end{pgfscope}%
\end{pgfscope}%
\begin{pgfscope}%
\pgfsetrectcap%
\pgfsetmiterjoin%
\pgfsetlinewidth{0.803000pt}%
\definecolor{currentstroke}{rgb}{0.000000,0.000000,0.000000}%
\pgfsetstrokecolor{currentstroke}%
\pgfsetdash{}{0pt}%
\pgfpathmoveto{\pgfqpoint{2.891187in}{0.415123in}}%
\pgfpathlineto{\pgfqpoint{2.891187in}{1.484490in}}%
\pgfusepath{stroke}%
\end{pgfscope}%
\begin{pgfscope}%
\pgfsetrectcap%
\pgfsetmiterjoin%
\pgfsetlinewidth{0.803000pt}%
\definecolor{currentstroke}{rgb}{0.000000,0.000000,0.000000}%
\pgfsetstrokecolor{currentstroke}%
\pgfsetdash{}{0pt}%
\pgfpathmoveto{\pgfqpoint{5.029919in}{0.415123in}}%
\pgfpathlineto{\pgfqpoint{5.029919in}{1.484490in}}%
\pgfusepath{stroke}%
\end{pgfscope}%
\begin{pgfscope}%
\pgfsetrectcap%
\pgfsetmiterjoin%
\pgfsetlinewidth{0.803000pt}%
\definecolor{currentstroke}{rgb}{0.000000,0.000000,0.000000}%
\pgfsetstrokecolor{currentstroke}%
\pgfsetdash{}{0pt}%
\pgfpathmoveto{\pgfqpoint{2.891187in}{0.415123in}}%
\pgfpathlineto{\pgfqpoint{5.029919in}{0.415123in}}%
\pgfusepath{stroke}%
\end{pgfscope}%
\begin{pgfscope}%
\pgfsetrectcap%
\pgfsetmiterjoin%
\pgfsetlinewidth{0.803000pt}%
\definecolor{currentstroke}{rgb}{0.000000,0.000000,0.000000}%
\pgfsetstrokecolor{currentstroke}%
\pgfsetdash{}{0pt}%
\pgfpathmoveto{\pgfqpoint{2.891187in}{1.484490in}}%
\pgfpathlineto{\pgfqpoint{5.029919in}{1.484490in}}%
\pgfusepath{stroke}%
\end{pgfscope}%
\begin{pgfscope}%
\definecolor{textcolor}{rgb}{0.000000,0.000000,0.000000}%
\pgfsetstrokecolor{textcolor}%
\pgfsetfillcolor{textcolor}%
\pgftext[x=3.960553in,y=1.567823in,,base]{\color{textcolor}\rmfamily\fontsize{12.000000}{14.400000}\selectfont 13360-8MHz}%
\end{pgfscope}%
\begin{pgfscope}%
\pgfpathrectangle{\pgfqpoint{5.310628in}{0.404137in}}{\pgfqpoint{0.125504in}{2.510079in}}%
\pgfusepath{clip}%
\pgfsetbuttcap%
\pgfsetmiterjoin%
\definecolor{currentfill}{rgb}{1.000000,1.000000,1.000000}%
\pgfsetfillcolor{currentfill}%
\pgfsetlinewidth{0.010037pt}%
\definecolor{currentstroke}{rgb}{1.000000,1.000000,1.000000}%
\pgfsetstrokecolor{currentstroke}%
\pgfsetdash{}{0pt}%
\pgfpathmoveto{\pgfqpoint{5.310628in}{0.404137in}}%
\pgfpathlineto{\pgfqpoint{5.310628in}{0.413942in}}%
\pgfpathlineto{\pgfqpoint{5.310628in}{2.904411in}}%
\pgfpathlineto{\pgfqpoint{5.310628in}{2.914216in}}%
\pgfpathlineto{\pgfqpoint{5.436132in}{2.914216in}}%
\pgfpathlineto{\pgfqpoint{5.436132in}{2.904411in}}%
\pgfpathlineto{\pgfqpoint{5.436132in}{0.413942in}}%
\pgfpathlineto{\pgfqpoint{5.436132in}{0.404137in}}%
\pgfpathclose%
\pgfusepath{stroke,fill}%
\end{pgfscope}%
\begin{pgfscope}%
\pgfsys@transformshift{5.310000in}{0.402304in}%
\pgftext[left,bottom]{\includegraphics[interpolate=true,width=0.130000in,height=2.510000in]{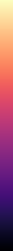}}%
\end{pgfscope}%
\begin{pgfscope}%
\pgfsetbuttcap%
\pgfsetroundjoin%
\definecolor{currentfill}{rgb}{0.000000,0.000000,0.000000}%
\pgfsetfillcolor{currentfill}%
\pgfsetlinewidth{0.803000pt}%
\definecolor{currentstroke}{rgb}{0.000000,0.000000,0.000000}%
\pgfsetstrokecolor{currentstroke}%
\pgfsetdash{}{0pt}%
\pgfsys@defobject{currentmarker}{\pgfqpoint{0.000000in}{0.000000in}}{\pgfqpoint{0.048611in}{0.000000in}}{%
\pgfpathmoveto{\pgfqpoint{0.000000in}{0.000000in}}%
\pgfpathlineto{\pgfqpoint{0.048611in}{0.000000in}}%
\pgfusepath{stroke,fill}%
}%
\begin{pgfscope}%
\pgfsys@transformshift{5.436132in}{0.404137in}%
\pgfsys@useobject{currentmarker}{}%
\end{pgfscope}%
\end{pgfscope}%
\begin{pgfscope}%
\definecolor{textcolor}{rgb}{0.000000,0.000000,0.000000}%
\pgfsetstrokecolor{textcolor}%
\pgfsetfillcolor{textcolor}%
\pgftext[x=5.533354in, y=0.355912in, left, base]{\color{textcolor}\rmfamily\fontsize{10.000000}{12.000000}\selectfont \(\displaystyle {0.0}\)}%
\end{pgfscope}%
\begin{pgfscope}%
\pgfsetbuttcap%
\pgfsetroundjoin%
\definecolor{currentfill}{rgb}{0.000000,0.000000,0.000000}%
\pgfsetfillcolor{currentfill}%
\pgfsetlinewidth{0.803000pt}%
\definecolor{currentstroke}{rgb}{0.000000,0.000000,0.000000}%
\pgfsetstrokecolor{currentstroke}%
\pgfsetdash{}{0pt}%
\pgfsys@defobject{currentmarker}{\pgfqpoint{0.000000in}{0.000000in}}{\pgfqpoint{0.048611in}{0.000000in}}{%
\pgfpathmoveto{\pgfqpoint{0.000000in}{0.000000in}}%
\pgfpathlineto{\pgfqpoint{0.048611in}{0.000000in}}%
\pgfusepath{stroke,fill}%
}%
\begin{pgfscope}%
\pgfsys@transformshift{5.436132in}{0.837281in}%
\pgfsys@useobject{currentmarker}{}%
\end{pgfscope}%
\end{pgfscope}%
\begin{pgfscope}%
\definecolor{textcolor}{rgb}{0.000000,0.000000,0.000000}%
\pgfsetstrokecolor{textcolor}%
\pgfsetfillcolor{textcolor}%
\pgftext[x=5.533354in, y=0.789056in, left, base]{\color{textcolor}\rmfamily\fontsize{10.000000}{12.000000}\selectfont \(\displaystyle {0.5}\)}%
\end{pgfscope}%
\begin{pgfscope}%
\pgfsetbuttcap%
\pgfsetroundjoin%
\definecolor{currentfill}{rgb}{0.000000,0.000000,0.000000}%
\pgfsetfillcolor{currentfill}%
\pgfsetlinewidth{0.803000pt}%
\definecolor{currentstroke}{rgb}{0.000000,0.000000,0.000000}%
\pgfsetstrokecolor{currentstroke}%
\pgfsetdash{}{0pt}%
\pgfsys@defobject{currentmarker}{\pgfqpoint{0.000000in}{0.000000in}}{\pgfqpoint{0.048611in}{0.000000in}}{%
\pgfpathmoveto{\pgfqpoint{0.000000in}{0.000000in}}%
\pgfpathlineto{\pgfqpoint{0.048611in}{0.000000in}}%
\pgfusepath{stroke,fill}%
}%
\begin{pgfscope}%
\pgfsys@transformshift{5.436132in}{1.270424in}%
\pgfsys@useobject{currentmarker}{}%
\end{pgfscope}%
\end{pgfscope}%
\begin{pgfscope}%
\definecolor{textcolor}{rgb}{0.000000,0.000000,0.000000}%
\pgfsetstrokecolor{textcolor}%
\pgfsetfillcolor{textcolor}%
\pgftext[x=5.533354in, y=1.222199in, left, base]{\color{textcolor}\rmfamily\fontsize{10.000000}{12.000000}\selectfont \(\displaystyle {1.0}\)}%
\end{pgfscope}%
\begin{pgfscope}%
\pgfsetbuttcap%
\pgfsetroundjoin%
\definecolor{currentfill}{rgb}{0.000000,0.000000,0.000000}%
\pgfsetfillcolor{currentfill}%
\pgfsetlinewidth{0.803000pt}%
\definecolor{currentstroke}{rgb}{0.000000,0.000000,0.000000}%
\pgfsetstrokecolor{currentstroke}%
\pgfsetdash{}{0pt}%
\pgfsys@defobject{currentmarker}{\pgfqpoint{0.000000in}{0.000000in}}{\pgfqpoint{0.048611in}{0.000000in}}{%
\pgfpathmoveto{\pgfqpoint{0.000000in}{0.000000in}}%
\pgfpathlineto{\pgfqpoint{0.048611in}{0.000000in}}%
\pgfusepath{stroke,fill}%
}%
\begin{pgfscope}%
\pgfsys@transformshift{5.436132in}{1.703568in}%
\pgfsys@useobject{currentmarker}{}%
\end{pgfscope}%
\end{pgfscope}%
\begin{pgfscope}%
\definecolor{textcolor}{rgb}{0.000000,0.000000,0.000000}%
\pgfsetstrokecolor{textcolor}%
\pgfsetfillcolor{textcolor}%
\pgftext[x=5.533354in, y=1.655343in, left, base]{\color{textcolor}\rmfamily\fontsize{10.000000}{12.000000}\selectfont \(\displaystyle {1.5}\)}%
\end{pgfscope}%
\begin{pgfscope}%
\pgfsetbuttcap%
\pgfsetroundjoin%
\definecolor{currentfill}{rgb}{0.000000,0.000000,0.000000}%
\pgfsetfillcolor{currentfill}%
\pgfsetlinewidth{0.803000pt}%
\definecolor{currentstroke}{rgb}{0.000000,0.000000,0.000000}%
\pgfsetstrokecolor{currentstroke}%
\pgfsetdash{}{0pt}%
\pgfsys@defobject{currentmarker}{\pgfqpoint{0.000000in}{0.000000in}}{\pgfqpoint{0.048611in}{0.000000in}}{%
\pgfpathmoveto{\pgfqpoint{0.000000in}{0.000000in}}%
\pgfpathlineto{\pgfqpoint{0.048611in}{0.000000in}}%
\pgfusepath{stroke,fill}%
}%
\begin{pgfscope}%
\pgfsys@transformshift{5.436132in}{2.136712in}%
\pgfsys@useobject{currentmarker}{}%
\end{pgfscope}%
\end{pgfscope}%
\begin{pgfscope}%
\definecolor{textcolor}{rgb}{0.000000,0.000000,0.000000}%
\pgfsetstrokecolor{textcolor}%
\pgfsetfillcolor{textcolor}%
\pgftext[x=5.533354in, y=2.088486in, left, base]{\color{textcolor}\rmfamily\fontsize{10.000000}{12.000000}\selectfont \(\displaystyle {2.0}\)}%
\end{pgfscope}%
\begin{pgfscope}%
\pgfsetbuttcap%
\pgfsetroundjoin%
\definecolor{currentfill}{rgb}{0.000000,0.000000,0.000000}%
\pgfsetfillcolor{currentfill}%
\pgfsetlinewidth{0.803000pt}%
\definecolor{currentstroke}{rgb}{0.000000,0.000000,0.000000}%
\pgfsetstrokecolor{currentstroke}%
\pgfsetdash{}{0pt}%
\pgfsys@defobject{currentmarker}{\pgfqpoint{0.000000in}{0.000000in}}{\pgfqpoint{0.048611in}{0.000000in}}{%
\pgfpathmoveto{\pgfqpoint{0.000000in}{0.000000in}}%
\pgfpathlineto{\pgfqpoint{0.048611in}{0.000000in}}%
\pgfusepath{stroke,fill}%
}%
\begin{pgfscope}%
\pgfsys@transformshift{5.436132in}{2.569855in}%
\pgfsys@useobject{currentmarker}{}%
\end{pgfscope}%
\end{pgfscope}%
\begin{pgfscope}%
\definecolor{textcolor}{rgb}{0.000000,0.000000,0.000000}%
\pgfsetstrokecolor{textcolor}%
\pgfsetfillcolor{textcolor}%
\pgftext[x=5.533354in, y=2.521630in, left, base]{\color{textcolor}\rmfamily\fontsize{10.000000}{12.000000}\selectfont \(\displaystyle {2.5}\)}%
\end{pgfscope}%
\begin{pgfscope}%
\pgfsetbuttcap%
\pgfsetmiterjoin%
\pgfsetlinewidth{0.803000pt}%
\definecolor{currentstroke}{rgb}{0.000000,0.000000,0.000000}%
\pgfsetstrokecolor{currentstroke}%
\pgfsetdash{}{0pt}%
\pgfpathmoveto{\pgfqpoint{5.310628in}{0.404137in}}%
\pgfpathlineto{\pgfqpoint{5.310628in}{0.413942in}}%
\pgfpathlineto{\pgfqpoint{5.310628in}{2.904411in}}%
\pgfpathlineto{\pgfqpoint{5.310628in}{2.914216in}}%
\pgfpathlineto{\pgfqpoint{5.436132in}{2.914216in}}%
\pgfpathlineto{\pgfqpoint{5.436132in}{2.904411in}}%
\pgfpathlineto{\pgfqpoint{5.436132in}{0.413942in}}%
\pgfpathlineto{\pgfqpoint{5.436132in}{0.404137in}}%
\pgfpathclose%
\pgfusepath{stroke}%
\end{pgfscope}%
\end{pgfpicture}%
\makeatother%
\endgroup%

%% file: chisq.tex
\begin{tabular}{llrrrr}
\hline\hline Data set & Weighting & \texttt{Resolve} & Multi-scale \texttt{CLEAN} model & Multi-scale \texttt{CLEAN} & Single-scale \texttt{CLEAN}\\\hline
2052-2MHz & Bayesian & 1.4, 1.1 & \textcolor{gray}{0.5, 0.5} & 210.3, 207.7 & 379.9, 390.7\\
& \texttt{wsclean} & \textcolor{gray}{3.6, 3.6} & 0.1, 0.1 & 79.7, 78.6 & 119.2, 120.8\\
4811-8MHz & Bayesian & 1.6, 1.4 & \textcolor{gray}{0.7, 0.7} & 79.2, 49.1 & 110.8, 84.3\\
& \texttt{wsclean} & \textcolor{gray}{3.5, 5.6} & 0.2, 0.2 & 31.2, 18.1 & 38.4, 26.0\\
8427-8MHz & Bayesian & 1.1, 1.0 & \textcolor{gray}{0.8, 0.8} & 233.4, 19.2 & 216.3, 46.1\\
& \texttt{wsclean} & \textcolor{gray}{7.3, 36.3} & 0.2, 0.2 & 82.3, 5.5 & 76.4, 12.5\\
13360-8MHz & Bayesian & 1.0, 0.9 & \textcolor{gray}{0.8, 0.8} & 199.4, 3.4 & 211.7, 49.8\\
& \texttt{wsclean} & \textcolor{gray}{26.9, 73.9} & 0.2, 0.2 & 97.7, 0.9 & 101.8, 16.9\\
\hline \end{tabular}

%% file: overview_contour_2052_all.pgf
\begingroup%
\makeatletter%
\begin{pgfpicture}%
\pgfpathrectangle{\pgfpointorigin}{\pgfqpoint{6.152754in}{3.281935in}}%
\pgfusepath{use as bounding box, clip}%
\begin{pgfscope}%
\pgfsetbuttcap%
\pgfsetmiterjoin%
\definecolor{currentfill}{rgb}{1.000000,1.000000,1.000000}%
\pgfsetfillcolor{currentfill}%
\pgfsetlinewidth{0.000000pt}%
\definecolor{currentstroke}{rgb}{1.000000,1.000000,1.000000}%
\pgfsetstrokecolor{currentstroke}%
\pgfsetdash{}{0pt}%
\pgfpathmoveto{\pgfqpoint{0.000000in}{0.000000in}}%
\pgfpathlineto{\pgfqpoint{6.152754in}{0.000000in}}%
\pgfpathlineto{\pgfqpoint{6.152754in}{3.281935in}}%
\pgfpathlineto{\pgfqpoint{0.000000in}{3.281935in}}%
\pgfpathclose%
\pgfusepath{fill}%
\end{pgfscope}%
\begin{pgfscope}%
\pgfsys@transformshift{0.540000in}{0.421935in}%
\pgftext[left,bottom]{\includegraphics[interpolate=true,width=5.610000in,height=2.860000in]{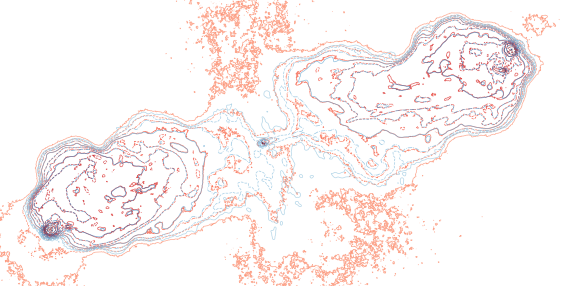}}%
\end{pgfscope}%
\begin{pgfscope}%
\pgfpathrectangle{\pgfqpoint{0.538581in}{0.415123in}}{\pgfqpoint{5.614173in}{2.866812in}}%
\pgfusepath{clip}%
\pgfsys@transformshift{0.538581in}{0.415123in}%
\pgftext[left,bottom]{\includegraphics[interpolate=true,width=5.620000in,height=2.870000in]{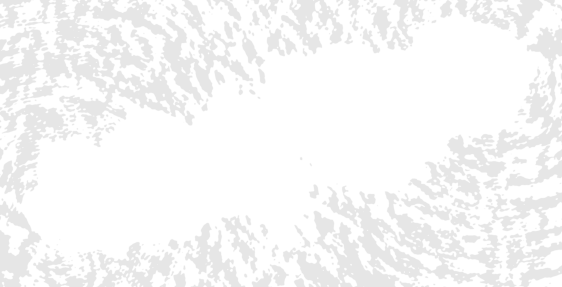}}%
\end{pgfscope}%
\begin{pgfscope}%
\pgfsetbuttcap%
\pgfsetroundjoin%
\definecolor{currentfill}{rgb}{0.000000,0.000000,0.000000}%
\pgfsetfillcolor{currentfill}%
\pgfsetlinewidth{0.803000pt}%
\definecolor{currentstroke}{rgb}{0.000000,0.000000,0.000000}%
\pgfsetstrokecolor{currentstroke}%
\pgfsetdash{}{0pt}%
\pgfsys@defobject{currentmarker}{\pgfqpoint{0.000000in}{-0.048611in}}{\pgfqpoint{0.000000in}{0.000000in}}{%
\pgfpathmoveto{\pgfqpoint{0.000000in}{0.000000in}}%
\pgfpathlineto{\pgfqpoint{0.000000in}{-0.048611in}}%
\pgfusepath{stroke,fill}%
}%
\begin{pgfscope}%
\pgfsys@transformshift{0.777482in}{0.415123in}%
\pgfsys@useobject{currentmarker}{}%
\end{pgfscope}%
\end{pgfscope}%
\begin{pgfscope}%
\definecolor{textcolor}{rgb}{0.000000,0.000000,0.000000}%
\pgfsetstrokecolor{textcolor}%
\pgfsetfillcolor{textcolor}%
\pgftext[x=0.777482in,y=0.317901in,,top]{\color{textcolor}\rmfamily\fontsize{10.000000}{12.000000}\selectfont \(\displaystyle {-60}\)}%
\end{pgfscope}%
\begin{pgfscope}%
\pgfsetbuttcap%
\pgfsetroundjoin%
\definecolor{currentfill}{rgb}{0.000000,0.000000,0.000000}%
\pgfsetfillcolor{currentfill}%
\pgfsetlinewidth{0.803000pt}%
\definecolor{currentstroke}{rgb}{0.000000,0.000000,0.000000}%
\pgfsetstrokecolor{currentstroke}%
\pgfsetdash{}{0pt}%
\pgfsys@defobject{currentmarker}{\pgfqpoint{0.000000in}{-0.048611in}}{\pgfqpoint{0.000000in}{0.000000in}}{%
\pgfpathmoveto{\pgfqpoint{0.000000in}{0.000000in}}%
\pgfpathlineto{\pgfqpoint{0.000000in}{-0.048611in}}%
\pgfusepath{stroke,fill}%
}%
\begin{pgfscope}%
\pgfsys@transformshift{1.573819in}{0.415123in}%
\pgfsys@useobject{currentmarker}{}%
\end{pgfscope}%
\end{pgfscope}%
\begin{pgfscope}%
\definecolor{textcolor}{rgb}{0.000000,0.000000,0.000000}%
\pgfsetstrokecolor{textcolor}%
\pgfsetfillcolor{textcolor}%
\pgftext[x=1.573819in,y=0.317901in,,top]{\color{textcolor}\rmfamily\fontsize{10.000000}{12.000000}\selectfont \(\displaystyle {-40}\)}%
\end{pgfscope}%
\begin{pgfscope}%
\pgfsetbuttcap%
\pgfsetroundjoin%
\definecolor{currentfill}{rgb}{0.000000,0.000000,0.000000}%
\pgfsetfillcolor{currentfill}%
\pgfsetlinewidth{0.803000pt}%
\definecolor{currentstroke}{rgb}{0.000000,0.000000,0.000000}%
\pgfsetstrokecolor{currentstroke}%
\pgfsetdash{}{0pt}%
\pgfsys@defobject{currentmarker}{\pgfqpoint{0.000000in}{-0.048611in}}{\pgfqpoint{0.000000in}{0.000000in}}{%
\pgfpathmoveto{\pgfqpoint{0.000000in}{0.000000in}}%
\pgfpathlineto{\pgfqpoint{0.000000in}{-0.048611in}}%
\pgfusepath{stroke,fill}%
}%
\begin{pgfscope}%
\pgfsys@transformshift{2.370155in}{0.415123in}%
\pgfsys@useobject{currentmarker}{}%
\end{pgfscope}%
\end{pgfscope}%
\begin{pgfscope}%
\definecolor{textcolor}{rgb}{0.000000,0.000000,0.000000}%
\pgfsetstrokecolor{textcolor}%
\pgfsetfillcolor{textcolor}%
\pgftext[x=2.370155in,y=0.317901in,,top]{\color{textcolor}\rmfamily\fontsize{10.000000}{12.000000}\selectfont \(\displaystyle {-20}\)}%
\end{pgfscope}%
\begin{pgfscope}%
\pgfsetbuttcap%
\pgfsetroundjoin%
\definecolor{currentfill}{rgb}{0.000000,0.000000,0.000000}%
\pgfsetfillcolor{currentfill}%
\pgfsetlinewidth{0.803000pt}%
\definecolor{currentstroke}{rgb}{0.000000,0.000000,0.000000}%
\pgfsetstrokecolor{currentstroke}%
\pgfsetdash{}{0pt}%
\pgfsys@defobject{currentmarker}{\pgfqpoint{0.000000in}{-0.048611in}}{\pgfqpoint{0.000000in}{0.000000in}}{%
\pgfpathmoveto{\pgfqpoint{0.000000in}{0.000000in}}%
\pgfpathlineto{\pgfqpoint{0.000000in}{-0.048611in}}%
\pgfusepath{stroke,fill}%
}%
\begin{pgfscope}%
\pgfsys@transformshift{3.166492in}{0.415123in}%
\pgfsys@useobject{currentmarker}{}%
\end{pgfscope}%
\end{pgfscope}%
\begin{pgfscope}%
\definecolor{textcolor}{rgb}{0.000000,0.000000,0.000000}%
\pgfsetstrokecolor{textcolor}%
\pgfsetfillcolor{textcolor}%
\pgftext[x=3.166492in,y=0.317901in,,top]{\color{textcolor}\rmfamily\fontsize{10.000000}{12.000000}\selectfont \(\displaystyle {0}\)}%
\end{pgfscope}%
\begin{pgfscope}%
\pgfsetbuttcap%
\pgfsetroundjoin%
\definecolor{currentfill}{rgb}{0.000000,0.000000,0.000000}%
\pgfsetfillcolor{currentfill}%
\pgfsetlinewidth{0.803000pt}%
\definecolor{currentstroke}{rgb}{0.000000,0.000000,0.000000}%
\pgfsetstrokecolor{currentstroke}%
\pgfsetdash{}{0pt}%
\pgfsys@defobject{currentmarker}{\pgfqpoint{0.000000in}{-0.048611in}}{\pgfqpoint{0.000000in}{0.000000in}}{%
\pgfpathmoveto{\pgfqpoint{0.000000in}{0.000000in}}%
\pgfpathlineto{\pgfqpoint{0.000000in}{-0.048611in}}%
\pgfusepath{stroke,fill}%
}%
\begin{pgfscope}%
\pgfsys@transformshift{3.962828in}{0.415123in}%
\pgfsys@useobject{currentmarker}{}%
\end{pgfscope}%
\end{pgfscope}%
\begin{pgfscope}%
\definecolor{textcolor}{rgb}{0.000000,0.000000,0.000000}%
\pgfsetstrokecolor{textcolor}%
\pgfsetfillcolor{textcolor}%
\pgftext[x=3.962828in,y=0.317901in,,top]{\color{textcolor}\rmfamily\fontsize{10.000000}{12.000000}\selectfont \(\displaystyle {20}\)}%
\end{pgfscope}%
\begin{pgfscope}%
\pgfsetbuttcap%
\pgfsetroundjoin%
\definecolor{currentfill}{rgb}{0.000000,0.000000,0.000000}%
\pgfsetfillcolor{currentfill}%
\pgfsetlinewidth{0.803000pt}%
\definecolor{currentstroke}{rgb}{0.000000,0.000000,0.000000}%
\pgfsetstrokecolor{currentstroke}%
\pgfsetdash{}{0pt}%
\pgfsys@defobject{currentmarker}{\pgfqpoint{0.000000in}{-0.048611in}}{\pgfqpoint{0.000000in}{0.000000in}}{%
\pgfpathmoveto{\pgfqpoint{0.000000in}{0.000000in}}%
\pgfpathlineto{\pgfqpoint{0.000000in}{-0.048611in}}%
\pgfusepath{stroke,fill}%
}%
\begin{pgfscope}%
\pgfsys@transformshift{4.759165in}{0.415123in}%
\pgfsys@useobject{currentmarker}{}%
\end{pgfscope}%
\end{pgfscope}%
\begin{pgfscope}%
\definecolor{textcolor}{rgb}{0.000000,0.000000,0.000000}%
\pgfsetstrokecolor{textcolor}%
\pgfsetfillcolor{textcolor}%
\pgftext[x=4.759165in,y=0.317901in,,top]{\color{textcolor}\rmfamily\fontsize{10.000000}{12.000000}\selectfont \(\displaystyle {40}\)}%
\end{pgfscope}%
\begin{pgfscope}%
\pgfsetbuttcap%
\pgfsetroundjoin%
\definecolor{currentfill}{rgb}{0.000000,0.000000,0.000000}%
\pgfsetfillcolor{currentfill}%
\pgfsetlinewidth{0.803000pt}%
\definecolor{currentstroke}{rgb}{0.000000,0.000000,0.000000}%
\pgfsetstrokecolor{currentstroke}%
\pgfsetdash{}{0pt}%
\pgfsys@defobject{currentmarker}{\pgfqpoint{0.000000in}{-0.048611in}}{\pgfqpoint{0.000000in}{0.000000in}}{%
\pgfpathmoveto{\pgfqpoint{0.000000in}{0.000000in}}%
\pgfpathlineto{\pgfqpoint{0.000000in}{-0.048611in}}%
\pgfusepath{stroke,fill}%
}%
\begin{pgfscope}%
\pgfsys@transformshift{5.555502in}{0.415123in}%
\pgfsys@useobject{currentmarker}{}%
\end{pgfscope}%
\end{pgfscope}%
\begin{pgfscope}%
\definecolor{textcolor}{rgb}{0.000000,0.000000,0.000000}%
\pgfsetstrokecolor{textcolor}%
\pgfsetfillcolor{textcolor}%
\pgftext[x=5.555502in,y=0.317901in,,top]{\color{textcolor}\rmfamily\fontsize{10.000000}{12.000000}\selectfont \(\displaystyle {60}\)}%
\end{pgfscope}%
\begin{pgfscope}%
\definecolor{textcolor}{rgb}{0.000000,0.000000,0.000000}%
\pgfsetstrokecolor{textcolor}%
\pgfsetfillcolor{textcolor}%
\pgftext[x=3.345668in,y=0.138889in,,top]{\color{textcolor}\rmfamily\fontsize{10.000000}{12.000000}\selectfont [arcsec]}%
\end{pgfscope}%
\begin{pgfscope}%
\pgfsetbuttcap%
\pgfsetroundjoin%
\definecolor{currentfill}{rgb}{0.000000,0.000000,0.000000}%
\pgfsetfillcolor{currentfill}%
\pgfsetlinewidth{0.803000pt}%
\definecolor{currentstroke}{rgb}{0.000000,0.000000,0.000000}%
\pgfsetstrokecolor{currentstroke}%
\pgfsetdash{}{0pt}%
\pgfsys@defobject{currentmarker}{\pgfqpoint{-0.048611in}{0.000000in}}{\pgfqpoint{-0.000000in}{0.000000in}}{%
\pgfpathmoveto{\pgfqpoint{-0.000000in}{0.000000in}}%
\pgfpathlineto{\pgfqpoint{-0.048611in}{0.000000in}}%
\pgfusepath{stroke,fill}%
}%
\begin{pgfscope}%
\pgfsys@transformshift{0.538581in}{0.654024in}%
\pgfsys@useobject{currentmarker}{}%
\end{pgfscope}%
\end{pgfscope}%
\begin{pgfscope}%
\definecolor{textcolor}{rgb}{0.000000,0.000000,0.000000}%
\pgfsetstrokecolor{textcolor}%
\pgfsetfillcolor{textcolor}%
\pgftext[x=0.194444in, y=0.605799in, left, base]{\color{textcolor}\rmfamily\fontsize{10.000000}{12.000000}\selectfont \(\displaystyle {-30}\)}%
\end{pgfscope}%
\begin{pgfscope}%
\pgfsetbuttcap%
\pgfsetroundjoin%
\definecolor{currentfill}{rgb}{0.000000,0.000000,0.000000}%
\pgfsetfillcolor{currentfill}%
\pgfsetlinewidth{0.803000pt}%
\definecolor{currentstroke}{rgb}{0.000000,0.000000,0.000000}%
\pgfsetstrokecolor{currentstroke}%
\pgfsetdash{}{0pt}%
\pgfsys@defobject{currentmarker}{\pgfqpoint{-0.048611in}{0.000000in}}{\pgfqpoint{-0.000000in}{0.000000in}}{%
\pgfpathmoveto{\pgfqpoint{-0.000000in}{0.000000in}}%
\pgfpathlineto{\pgfqpoint{-0.048611in}{0.000000in}}%
\pgfusepath{stroke,fill}%
}%
\begin{pgfscope}%
\pgfsys@transformshift{0.538581in}{1.052193in}%
\pgfsys@useobject{currentmarker}{}%
\end{pgfscope}%
\end{pgfscope}%
\begin{pgfscope}%
\definecolor{textcolor}{rgb}{0.000000,0.000000,0.000000}%
\pgfsetstrokecolor{textcolor}%
\pgfsetfillcolor{textcolor}%
\pgftext[x=0.194444in, y=1.003967in, left, base]{\color{textcolor}\rmfamily\fontsize{10.000000}{12.000000}\selectfont \(\displaystyle {-20}\)}%
\end{pgfscope}%
\begin{pgfscope}%
\pgfsetbuttcap%
\pgfsetroundjoin%
\definecolor{currentfill}{rgb}{0.000000,0.000000,0.000000}%
\pgfsetfillcolor{currentfill}%
\pgfsetlinewidth{0.803000pt}%
\definecolor{currentstroke}{rgb}{0.000000,0.000000,0.000000}%
\pgfsetstrokecolor{currentstroke}%
\pgfsetdash{}{0pt}%
\pgfsys@defobject{currentmarker}{\pgfqpoint{-0.048611in}{0.000000in}}{\pgfqpoint{-0.000000in}{0.000000in}}{%
\pgfpathmoveto{\pgfqpoint{-0.000000in}{0.000000in}}%
\pgfpathlineto{\pgfqpoint{-0.048611in}{0.000000in}}%
\pgfusepath{stroke,fill}%
}%
\begin{pgfscope}%
\pgfsys@transformshift{0.538581in}{1.450361in}%
\pgfsys@useobject{currentmarker}{}%
\end{pgfscope}%
\end{pgfscope}%
\begin{pgfscope}%
\definecolor{textcolor}{rgb}{0.000000,0.000000,0.000000}%
\pgfsetstrokecolor{textcolor}%
\pgfsetfillcolor{textcolor}%
\pgftext[x=0.194444in, y=1.402136in, left, base]{\color{textcolor}\rmfamily\fontsize{10.000000}{12.000000}\selectfont \(\displaystyle {-10}\)}%
\end{pgfscope}%
\begin{pgfscope}%
\pgfsetbuttcap%
\pgfsetroundjoin%
\definecolor{currentfill}{rgb}{0.000000,0.000000,0.000000}%
\pgfsetfillcolor{currentfill}%
\pgfsetlinewidth{0.803000pt}%
\definecolor{currentstroke}{rgb}{0.000000,0.000000,0.000000}%
\pgfsetstrokecolor{currentstroke}%
\pgfsetdash{}{0pt}%
\pgfsys@defobject{currentmarker}{\pgfqpoint{-0.048611in}{0.000000in}}{\pgfqpoint{-0.000000in}{0.000000in}}{%
\pgfpathmoveto{\pgfqpoint{-0.000000in}{0.000000in}}%
\pgfpathlineto{\pgfqpoint{-0.048611in}{0.000000in}}%
\pgfusepath{stroke,fill}%
}%
\begin{pgfscope}%
\pgfsys@transformshift{0.538581in}{1.848529in}%
\pgfsys@useobject{currentmarker}{}%
\end{pgfscope}%
\end{pgfscope}%
\begin{pgfscope}%
\definecolor{textcolor}{rgb}{0.000000,0.000000,0.000000}%
\pgfsetstrokecolor{textcolor}%
\pgfsetfillcolor{textcolor}%
\pgftext[x=0.371914in, y=1.800304in, left, base]{\color{textcolor}\rmfamily\fontsize{10.000000}{12.000000}\selectfont \(\displaystyle {0}\)}%
\end{pgfscope}%
\begin{pgfscope}%
\pgfsetbuttcap%
\pgfsetroundjoin%
\definecolor{currentfill}{rgb}{0.000000,0.000000,0.000000}%
\pgfsetfillcolor{currentfill}%
\pgfsetlinewidth{0.803000pt}%
\definecolor{currentstroke}{rgb}{0.000000,0.000000,0.000000}%
\pgfsetstrokecolor{currentstroke}%
\pgfsetdash{}{0pt}%
\pgfsys@defobject{currentmarker}{\pgfqpoint{-0.048611in}{0.000000in}}{\pgfqpoint{-0.000000in}{0.000000in}}{%
\pgfpathmoveto{\pgfqpoint{-0.000000in}{0.000000in}}%
\pgfpathlineto{\pgfqpoint{-0.048611in}{0.000000in}}%
\pgfusepath{stroke,fill}%
}%
\begin{pgfscope}%
\pgfsys@transformshift{0.538581in}{2.246698in}%
\pgfsys@useobject{currentmarker}{}%
\end{pgfscope}%
\end{pgfscope}%
\begin{pgfscope}%
\definecolor{textcolor}{rgb}{0.000000,0.000000,0.000000}%
\pgfsetstrokecolor{textcolor}%
\pgfsetfillcolor{textcolor}%
\pgftext[x=0.302469in, y=2.198472in, left, base]{\color{textcolor}\rmfamily\fontsize{10.000000}{12.000000}\selectfont \(\displaystyle {10}\)}%
\end{pgfscope}%
\begin{pgfscope}%
\pgfsetbuttcap%
\pgfsetroundjoin%
\definecolor{currentfill}{rgb}{0.000000,0.000000,0.000000}%
\pgfsetfillcolor{currentfill}%
\pgfsetlinewidth{0.803000pt}%
\definecolor{currentstroke}{rgb}{0.000000,0.000000,0.000000}%
\pgfsetstrokecolor{currentstroke}%
\pgfsetdash{}{0pt}%
\pgfsys@defobject{currentmarker}{\pgfqpoint{-0.048611in}{0.000000in}}{\pgfqpoint{-0.000000in}{0.000000in}}{%
\pgfpathmoveto{\pgfqpoint{-0.000000in}{0.000000in}}%
\pgfpathlineto{\pgfqpoint{-0.048611in}{0.000000in}}%
\pgfusepath{stroke,fill}%
}%
\begin{pgfscope}%
\pgfsys@transformshift{0.538581in}{2.644866in}%
\pgfsys@useobject{currentmarker}{}%
\end{pgfscope}%
\end{pgfscope}%
\begin{pgfscope}%
\definecolor{textcolor}{rgb}{0.000000,0.000000,0.000000}%
\pgfsetstrokecolor{textcolor}%
\pgfsetfillcolor{textcolor}%
\pgftext[x=0.302469in, y=2.596641in, left, base]{\color{textcolor}\rmfamily\fontsize{10.000000}{12.000000}\selectfont \(\displaystyle {20}\)}%
\end{pgfscope}%
\begin{pgfscope}%
\pgfsetbuttcap%
\pgfsetroundjoin%
\definecolor{currentfill}{rgb}{0.000000,0.000000,0.000000}%
\pgfsetfillcolor{currentfill}%
\pgfsetlinewidth{0.803000pt}%
\definecolor{currentstroke}{rgb}{0.000000,0.000000,0.000000}%
\pgfsetstrokecolor{currentstroke}%
\pgfsetdash{}{0pt}%
\pgfsys@defobject{currentmarker}{\pgfqpoint{-0.048611in}{0.000000in}}{\pgfqpoint{-0.000000in}{0.000000in}}{%
\pgfpathmoveto{\pgfqpoint{-0.000000in}{0.000000in}}%
\pgfpathlineto{\pgfqpoint{-0.048611in}{0.000000in}}%
\pgfusepath{stroke,fill}%
}%
\begin{pgfscope}%
\pgfsys@transformshift{0.538581in}{3.043034in}%
\pgfsys@useobject{currentmarker}{}%
\end{pgfscope}%
\end{pgfscope}%
\begin{pgfscope}%
\definecolor{textcolor}{rgb}{0.000000,0.000000,0.000000}%
\pgfsetstrokecolor{textcolor}%
\pgfsetfillcolor{textcolor}%
\pgftext[x=0.302469in, y=2.994809in, left, base]{\color{textcolor}\rmfamily\fontsize{10.000000}{12.000000}\selectfont \(\displaystyle {30}\)}%
\end{pgfscope}%
\begin{pgfscope}%
\definecolor{textcolor}{rgb}{0.000000,0.000000,0.000000}%
\pgfsetstrokecolor{textcolor}%
\pgfsetfillcolor{textcolor}%
\pgftext[x=0.138889in,y=1.848529in,,bottom,rotate=90.000000]{\color{textcolor}\rmfamily\fontsize{10.000000}{12.000000}\selectfont [arcsec]}%
\end{pgfscope}%
\begin{pgfscope}%
\pgfsetrectcap%
\pgfsetmiterjoin%
\pgfsetlinewidth{0.803000pt}%
\definecolor{currentstroke}{rgb}{0.000000,0.000000,0.000000}%
\pgfsetstrokecolor{currentstroke}%
\pgfsetdash{}{0pt}%
\pgfpathmoveto{\pgfqpoint{0.538581in}{0.415123in}}%
\pgfpathlineto{\pgfqpoint{0.538581in}{3.281935in}}%
\pgfusepath{stroke}%
\end{pgfscope}%
\begin{pgfscope}%
\pgfsetrectcap%
\pgfsetmiterjoin%
\pgfsetlinewidth{0.803000pt}%
\definecolor{currentstroke}{rgb}{0.000000,0.000000,0.000000}%
\pgfsetstrokecolor{currentstroke}%
\pgfsetdash{}{0pt}%
\pgfpathmoveto{\pgfqpoint{6.152754in}{0.415123in}}%
\pgfpathlineto{\pgfqpoint{6.152754in}{3.281935in}}%
\pgfusepath{stroke}%
\end{pgfscope}%
\begin{pgfscope}%
\pgfsetrectcap%
\pgfsetmiterjoin%
\pgfsetlinewidth{0.803000pt}%
\definecolor{currentstroke}{rgb}{0.000000,0.000000,0.000000}%
\pgfsetstrokecolor{currentstroke}%
\pgfsetdash{}{0pt}%
\pgfpathmoveto{\pgfqpoint{0.538581in}{0.415123in}}%
\pgfpathlineto{\pgfqpoint{6.152754in}{0.415123in}}%
\pgfusepath{stroke}%
\end{pgfscope}%
\begin{pgfscope}%
\pgfsetrectcap%
\pgfsetmiterjoin%
\pgfsetlinewidth{0.803000pt}%
\definecolor{currentstroke}{rgb}{0.000000,0.000000,0.000000}%
\pgfsetstrokecolor{currentstroke}%
\pgfsetdash{}{0pt}%
\pgfpathmoveto{\pgfqpoint{0.538581in}{3.281935in}}%
\pgfpathlineto{\pgfqpoint{6.152754in}{3.281935in}}%
\pgfusepath{stroke}%
\end{pgfscope}%
\end{pgfpicture}%
\makeatother%
\endgroup%

%% file: main.bbl
\begin{thebibliography}{40}
\expandafter\ifx\csname natexlab\endcsname\relax\def\natexlab#1{#1}\fi

\bibitem[{Abdulaziz {et~al.}(2019)Abdulaziz, Dabbech, \& Wiaux}]{hypersara}
Abdulaziz, A., Dabbech, A., \& Wiaux, Y. 2019, Monthly Notices of the Royal
  Astronomical Society, 489, 1230

\bibitem[{Arras {et~al.}(2019{\natexlab{a}})Arras, Baltac, Ensslin, Frank,
  Hutschenreuter, Knollmueller, Leike, Newrzella, Platz, Reinecke,
  {et~al.}}]{nifty5}
Arras, P., Baltac, M., Ensslin, T.~A., {et~al.} 2019{\natexlab{a}},
  Astrophysics Source Code Library

\bibitem[{Arras {et~al.}(2020{\natexlab{a}})Arras, Bester, Perley, Leike,
  Smirnov, Westermann, \& Enßlin}]{zenodo}
Arras, P., Bester, H.~L., Perley, R.~A., {et~al.} 2020{\natexlab{a}}, Zenodo
  [\eprint[zenodo]{4267057}]

\bibitem[{Arras {et~al.}(2020{\natexlab{b}})Arras, Frank, Haim,
  Knollm{\"u}ller, Leike, Reinecke, \& En{\ss}lin}]{ifteht}
Arras, P., Frank, P., Haim, P., {et~al.} 2020{\natexlab{b}}, arXiv preprint
  arXiv:2002.05218

\bibitem[{Arras {et~al.}(2019{\natexlab{b}})Arras, Frank, Leike, Westermann, \&
  En\ss{}lin}]{resolve19}
Arras, P., Frank, P., Leike, R., Westermann, R., \& En\ss{}lin, T.~A.
  2019{\natexlab{b}}, A\&A, 627, A134

\bibitem[{Arras {et~al.}(2018)Arras, Knollm{\"u}ller, Junklewitz, \&
  En{\ss}lin}]{resolve18}
Arras, P., Knollm{\"u}ller, J., Junklewitz, H., \& En{\ss}lin, T.~A. 2018, in
  2018 26th European Signal Processing Conference (EUSIPCO), IEEE, 2683--2687

\bibitem[{Cai {et~al.}(2018)Cai, Pereyra, \& McEwen}]{cai2018uncertaintyI}
Cai, X., Pereyra, M., \& McEwen, J.~D. 2018, Monthly Notices of the Royal
  Astronomical Society, 480, 4154

\bibitem[{Cand{\`e}s {et~al.}(2006)}]{csoriginal}
Cand{\`e}s, E.~J. {et~al.} 2006, in Proceedings of the international congress
  of mathematicians, Vol.~3, Madrid, Spain, 1433--1452

\bibitem[{Carrillo {et~al.}(2012)Carrillo, McEwen, \& Wiaux}]{sara}
Carrillo, R.~E., McEwen, J.~D., \& Wiaux, Y. 2012, Monthly Notices of the Royal
  Astronomical Society, 426, 1223

\bibitem[{Carrillo {et~al.}(2014)Carrillo, McEwen, \& Wiaux}]{purify}
Carrillo, R.~E., McEwen, J.~D., \& Wiaux, Y. 2014, Monthly Notices of the Royal
  Astronomical Society, 439, 3591

\bibitem[{{Clark}(1980)}]{clarkclean1980}
{Clark}, B.~G. 1980, \aap, 89, 377

\bibitem[{Cornwell(2008)}]{cornwell2008}
Cornwell, T.~J. 2008, IEEE Journal of Selected Topics in Signal Processing, 2,
  793

\bibitem[{Cornwell \& Evans(1985)}]{cornwellmaxent}
Cornwell, T.~J. \& Evans, K. 1985, Astronomy and Astrophysics, 143, 77

\bibitem[{Cornwell {et~al.}(2008)Cornwell, Golap, \& Bhatnagar}]{wprojection}
Cornwell, T.~J., Golap, K., \& Bhatnagar, S. 2008, IEEE Journal of Selected
  Topics in Signal Processing, 2, 647

\bibitem[{Cox(1946)}]{cox}
Cox, R.~T. 1946, American journal of physics, 14, 1

\bibitem[{Dabbech {et~al.}(2018)Dabbech, Onose, Abdulaziz, Perley, Smirnov, \&
  Wiaux}]{convex}
Dabbech, A., Onose, A., Abdulaziz, A., {et~al.} 2018, Monthly Notices of the
  Royal Astronomical Society, 476, 2853

\bibitem[{En{\ss}lin(2018)}]{ensslin18}
En{\ss}lin, T.~A. 2018, Annalen der Physik, 1800127

\bibitem[{En{\ss}lin \& Frommert(2011)}]{ensslin11}
En{\ss}lin, T.~A. \& Frommert, M. 2011, Physical Review D, 83, 105014

\bibitem[{En{\ss}lin {et~al.}(2009)En{\ss}lin, Frommert, \&
  Kitaura}]{ensslin09}
En{\ss}lin, T.~A., Frommert, M., \& Kitaura, F.~S. 2009, Physical Review D, 80,
  105005

\bibitem[{Greiner {et~al.}(2016)Greiner, Vacca, Junklewitz, \&
  En{\ss}lin}]{fastresolve}
Greiner, M., Vacca, V., Junklewitz, H., \& En{\ss}lin, T.~A. 2016, arXiv
  preprint arXiv:1605.04317

\bibitem[{Gull \& Skilling(1984)}]{memgullskilling}
Gull, S.~F. \& Skilling, J. 1984, in IEE Proceedings F (Communications, Radar
  and Signal Processing), Vol. 131, IET, 646--659

\bibitem[{H{\"o}gbom(1974)}]{hogbom1974aperture}
H{\"o}gbom, J. 1974, Astronomy and Astrophysics Supplement Series, 15, 417

\bibitem[{Honma {et~al.}(2014)Honma, Akiyama, Uemura, \&
  Ikeda}]{superresolution}
Honma, M., Akiyama, K., Uemura, M., \& Ikeda, S. 2014, Publications of the
  Astronomical Society of Japan, 66, 95

\bibitem[{Junklewitz {et~al.}(2015)Junklewitz, Bell, \& En{\ss}lin}]{resolve15}
Junklewitz, H., Bell, M., \& En{\ss}lin, T. 2015, Astronomy \& Astrophysics,
  581, A59

\bibitem[{Junklewitz {et~al.}(2016)Junklewitz, Bell, Selig, \&
  En{\ss}lin}]{resolve16}
Junklewitz, H., Bell, M., Selig, M., \& En{\ss}lin, T.~A. 2016, Astronomy \&
  Astrophysics, 586, A76

\bibitem[{Khintchin(1934)}]{Khintchin}
Khintchin, A. 1934, Mathematische Annalen, 109, 604

\bibitem[{Kingma {et~al.}(2015)Kingma, Salimans, \& Welling}]{reparatrick}
Kingma, D.~P., Salimans, T., \& Welling, M. 2015, in Advances in neural
  information processing systems, 2575--2583

\bibitem[{Knollm{\"u}ller \& En{\ss}lin(2019)}]{mgvi}
Knollm{\"u}ller, J. \& En{\ss}lin, T.~A. 2019, arXiv preprint arXiv:1901.11033

\bibitem[{Offringa {et~al.}(2014)Offringa, McKinley, Hurley-Walker, Briggs,
  Wayth, Kaplan, Bell, Feng, Neben, Hughes, {et~al.}}]{wstacking}
Offringa, A., McKinley, B., Hurley-Walker, N., {et~al.} 2014, Monthly Notices
  of the Royal Astronomical Society, 444, 606

\bibitem[{Offringa \& Smirnov(2017)}]{offringamsclean}
Offringa, A.~R. \& Smirnov, O. 2017, \mnras, 471, 301

\bibitem[{Perley {et~al.}(2017)Perley, Perley, Dhawan, \& Carilli}]{discovery}
Perley, D.~A., Perley, R.~A., Dhawan, V., \& Carilli, C.~L. 2017, The
  Astrophysical Journal, 841, 117

\bibitem[{Rau \& Cornwell(2011)}]{rau2011}
Rau, U. \& Cornwell, T.~J. 2011, \aap, 532, A71

\bibitem[{Repetti {et~al.}(2019)Repetti, Pereyra, \& Wiaux}]{uncertaintycs}
Repetti, A., Pereyra, M., \& Wiaux, Y. 2019, SIAM Journal on Imaging Sciences,
  12, 87

\bibitem[{{Schwab} \& {Cotton}(1983)}]{cottonschwab}
{Schwab}, F.~R. \& {Cotton}, W.~D. 1983, \aj, 88, 688

\bibitem[{Sebokolodi {et~al.}(2020)Sebokolodi, Perley, Eilek, Carilli, Smirnov,
  Laing, Greisen, \& Wise}]{datapaper}
Sebokolodi, M.~L., Perley, R., Eilek, J., {et~al.} 2020, arXiv preprint
  arXiv:2009.06554

\bibitem[{Selig {et~al.}(2015)Selig, Vacca, Oppermann, \& En{\ss}lin}]{d3po}
Selig, M., Vacca, V., Oppermann, N., \& En{\ss}lin, T.~A. 2015, Astronomy \&
  Astrophysics, 581, A126

\bibitem[{Sutter {et~al.}(2014)Sutter, Wandelt, McEwen, Bunn, Karakci,
  Korotkov, Timbie, Tucker, \& Zhang}]{sutter2014}
Sutter, P.~M., Wandelt, B.~D., McEwen, J.~D., {et~al.} 2014, Monthly Notices of
  the Royal Astronomical Society, 438, 768

\bibitem[{Sutton \& Wandelt(2006)}]{optimalimaging}
Sutton, E.~C. \& Wandelt, B.~D. 2006, The Astrophysical Journal Supplement
  Series, 162, 401

\bibitem[{Thompson {et~al.}(2017)Thompson, Moran, \& Swenson~Jr}]{book}
Thompson, R.~A., Moran, J.~M., \& Swenson~Jr, G.~W. 2017, Interferometry and
  synthesis in radio astronomy (Springer Nature)

\bibitem[{Wiener(1949)}]{Wiener}
Wiener, N. 1949, Extrapolation, interpolation, and smoothing of stationary time
  series, Vol.~2 (MIT press Cambridge)

\end{thebibliography}
